\documentclass[universe,article,accept,moreauthors,pdftex]{Definitions/mdpi} 

\usepackage{upgreek}
\usepackage{textcomp}

\firstpage{1} 
\pubvolume{1}
\issuenum{1}
\articlenumber{0}
\pubyear{2021}
\copyrightyear{2021}
\externaleditor{{Academic Editor: Gonzalo Olmo}}
\datereceived{} 
\dateaccepted{} 
\datepublished{} 
\hreflink{https://doi.org/} 
\pdfoutput=1

\usepackage{multicol}
\usepackage{caption}
\usepackage[labelformat=simple]{subcaption} 

\DeclareCaptionLabelFormat{subcaptionlabel}{\normalfont(\textbf{#2}\normalfont)}
\usepackage{changes}
\definechangesauthor[name={second revision}, color=red]{SR}

\def \beq  {\begin{equation}}
\def \eeq  {\end{equation}}
\newcommand{\be}{\begin{equation}}
\newcommand{\ee}{\end{equation}}
\newcommand{\newc}{\newcommand}
\newc{\etal}{{\it et al.}}
\newc{\lcdm}{$\Lambda$CDM }
\newc{\lcdmnospace}{$\Lambda$CDM}
\newc{\wcdm}{$w$CDM }
\newc{\plcdm}{Planck/$\Lambda$CDM }
\newc{\plcdmnospace}{Planck/$\Lambda$CDM}
\newc{\wlcdm}{WMAP7/$\Lambda$CDM }
\newc{\wlcdmnospace}{WMAP7/$\Lambda$CDM}

\newc{\ra}{\Rightarrow}
\newc{\fs}{$f\sigma_8$}
\newc{\fsz}{$f\sigma_8(z)$}

\newc{\bea}{\begin{eqnarray*}}
\newc{\eea}{\end{eqnarray*}}
\extrafloats{100}

\Title{The Hubble Tension, The $M$ Crisis of Late Time $H(z)$ Deformation Models and the Reconstruction of \mbox{Quintessence Lagrangians$^{\dagger}$}
}

\TitleCitation{The Hubble Tension, The $M$ Crisis of Late Time $H(z)$ Deformation Models and the Reconstruction of {Quintessence Lagrangians}}

\Author{Anastasios Theodoropoulos \orcidB{} and Leandros Perivolaropoulos $^{}$*\orcidA{}}

\AuthorNames{Anastasios Theodoropoulos,Leandros Perivolaropoulos}

\AuthorCitation{Theodoropoulos, A.; Perivolaropoulos, L.}

\address[1]{Department of Physics, University of Ioannina, {45110, Ioannina,Greece}; theodorotasos@gmail.com }
\corres{\hangafter=1 \hangindent=1.0em \hspace{-.82em}
Correspondence: leandros@uoi.gr}
\firstnote{This paper is an extended version from the proceeding paper: {George Alestas, George V. Kraniotis and Leandros Perivolaropoulos. An overview of nonstandard signals in cosmological data}. In Proceedings of the 1st Electronic Conference on Universe, online, 22–28 February 2021.} 

\abstract{We present a detailed and pedagogical analysis of recent cosmological  data, including CMB, BAO,  SnIa and the recent local measurement of $H_0$. We thus obtain constraints on the parameters of these standard dark energy parameterizations,
including $\Lambda CDM$, and $H(z)$ deformation models such as $wCDM$ (constant equation of state $w$ of  dark energy), and the CPL model (corresponding to the evolving dark energy equation-of-state parameter $w(z) = w_0 + w_a \frac{z}{1+z}$). The fitted parameters include the dark matter density $\Omega_{0m}$, the SnIa absolute magnitude $M$, the Hubble constant $H_0$  and the dark energy parameters (e.g., $w$ for $wCDM$). All models considered  lead to a best-fit value of $M$ that is inconsistent with the locally determined value obtained by Cepheid calibrators ($M$ tension). We then use the best-fit dark energy parameters to reconstruct the quintessence Lagrangian that would be able to reproduce these best-fit parameterizations. Due to the derived late phantom behavior of the best-fit dark energy equation-of-state parameter $w(z)$, the reconstructed quintessence models have a negative kinetic term and are therefore plagued with instabilities. 
}

\begin{document}

\section{Introduction}
The success of the current standard cosmological model $\Lambda$CDM has been challenged by the mismatch of the value of the Hubble constant obtained by different cosmological~probes.

The latest  constraint as obtained by the Planck Collaboration~\cite{Aghanim:2018eyx} is:
\begin{align} \label{P18}
H^{\rm P18}_0 = 67.36 \pm 0.54 \text{ km s}^{-1} {\rm Mpc}^{-1} \,,
\end{align}
while the latest local determination as obtained by the SH0ES collaboration~\cite{Riess:2020fzl} is:
\begin{align}
H^{\rm R20}_0 = 73.2 \pm 1.3 \text{ km s}^{-1} {\rm Mpc}^{-1} \,.
\label{hr20}
\end{align}

This $9\%$ mismatch corresponding to a tension of more than $4\sigma$ constitutes the well-known Hubble~crisis.

This mismatch is equivalent to the mismatch of the Pantheon SnIa  absolute magnitudes, which when calibrated using the CMB sound horizon and propagated via BAO measurements to low $z$ (inverse distance ladder, $z\simeq 1100$)  have a value~\cite{Camarena:2019rmj}
\begin{align} \label{P18M}
M^{\rm P18} = - 19.401 \pm 0.027 \text{ mag} \,,
\end{align}
while when calibrated using the Cepheid stars have a value~\cite{Camarena:2019moy}
\begin{align} \label{R20M}
M^{\rm R20} = -19.244 \pm 0.037 \text{ mag} \,.
\end{align}

\begin{figure}[H]
\includegraphics[width = 0.95\columnwidth]{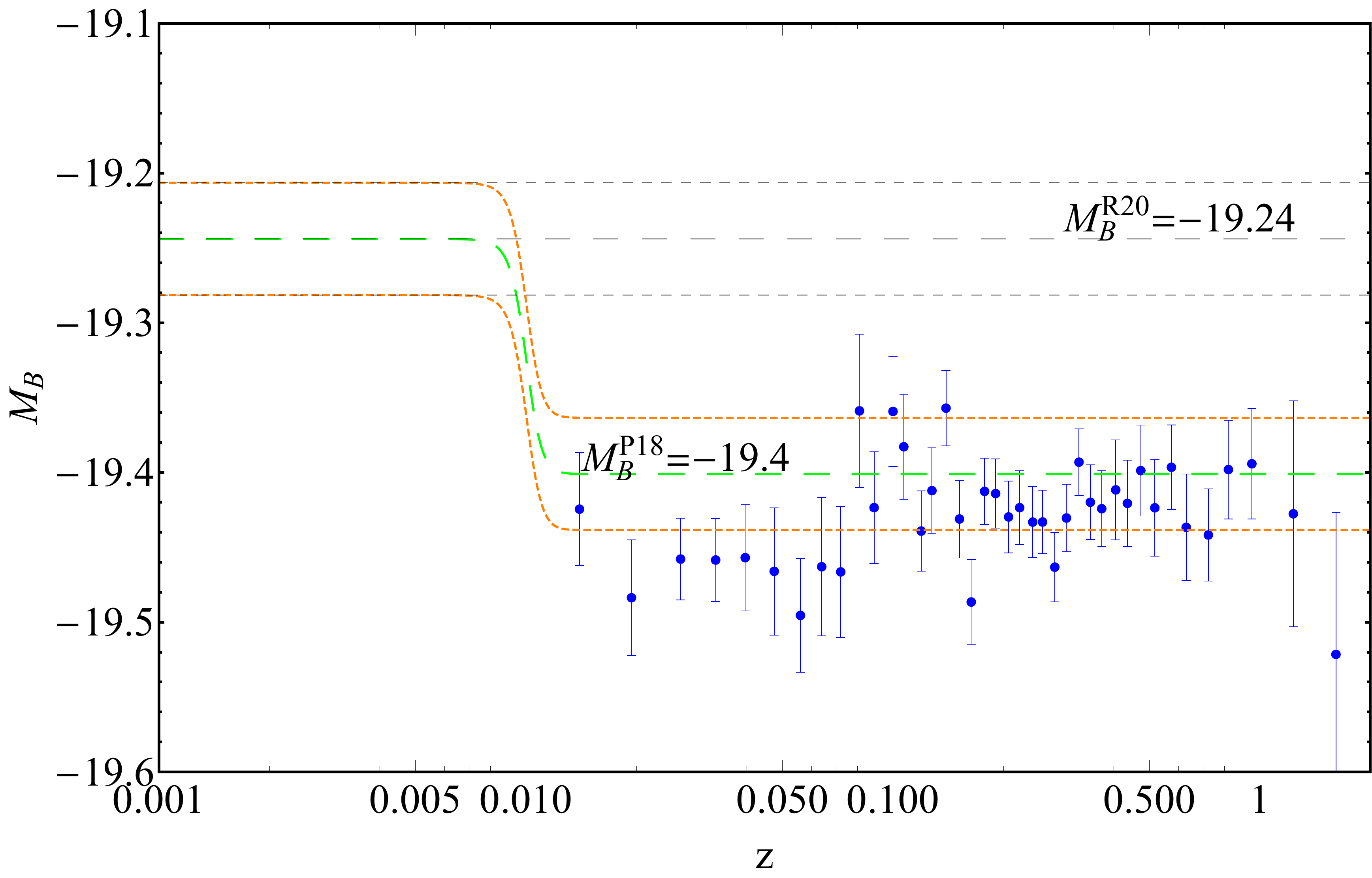}
\caption{
The SnIa absolute magnitudes obtained as $M_{i}= m_{i}-\mu(z_i)$ of the binned Pantheon sample~\cite{Scolnic:2017caz} assuming \plcdm luminosity distance.
The data are inconsistent with the $M^{\rm R20}$ of Equation~(\ref{R20M}), but they become consistent if there is a transition in the absolute magnitude with amplitude $\Delta M \approx-0.2$. Note that in this model there is no transition in the Hubble function $H(z)$.}
\label{fig1}
\end{figure}

These measurements are in tension at a level of about 3.5  $\upsigma$, and this is illustrated in\mbox{ Figure~\ref{fig1}}.
Note: It is also common to denote M as $M_B$.

The SnIa absolute magnitudes $M$ obtained from the distance modulus equation
\begin{align} \label{muu}
\mu(z) \equiv m(z) - M = 5\log_{10}\left[\frac{d_L(z)}{10 \text{pc}} \right]   \,.
\end{align}
with a luminosity distance obtained from the \plcdm Hubble expansion
\be 
H(z) =H_0^{\rm P18} \sqrt{\Omega_{0m}(1+z)^3 +1-\Omega_{0m}} \,.
\label{eq:hzlcdm} 
\ee 
($z>0.01$) are in tension with the SnIa absolute magnitudes obtained from the Cepheid calibrators at $z<0.01$ (Figure \ref{fig1}). However, since the measurements are made at different redshift ranges, the~discrepancy can be reconciled by assuming a transition of the absolute magnitude $M$ at $z\lesssim 0.01$ by $\Delta M \simeq 0.02$, as shown in Figure~\ref{fig1}.

This transition corresponds to brighter SnIa at high z (early times) compared to low z (late times) since the SnIa absolute Luminosity $L$ is connected with the absolute \mbox{magnitude as} 
\be
L\sim 10^{-2M/5}
\label{lvsm}
\ee

$H(z)$ deformations obtained by dynamical dark energy have been used as possible approaches to the Hubble tension. In~this class of models, $H(z)$ becomes deformed from its \lcdm form (\ref{eq:hzlcdm}) in such a way that the CMB anisotropy spectrum remains practically  invariant  while the Hubble parameter $H_0$ is shifted to the $H_0^{R20}$ value (e.g.,~\cite{Alestas:2020mvb}). This class of models has been shown to require the presence of phantom dark energy at least at late times and has three important~problems: 
\begin{itemize}
\item
It worsens the growth tension of the \lcdm model as it indicates larger values of the parameters $\sigma_8$ and $\Omega_{0m}$ than indicated by dynamical cosmological probes~\cite{Alestas:2020mvb,Alestas:2021xes}
\item
It provides a worse fit than \lcdm to low $z$ geometric probes such as SnIa and BAO~\cite{Alestas:2020mvb}.
\item
As in the case of \lcdm, it favors a lower value of the SnIa absolute magnitude $M$ than the local Cepheid calibrators.
\end{itemize}

One of the goals of the present analysis is to investigate in some more detail the last problem of the $H(z)$ deformation models. In~addition to \lcdm, we consider a generic class of $H(z)$ deformation models with dynamical dark energy and identify their best-fit parameter values, including the SnIa absolute magnitude $M$ and the Hubble parameter $H_0$. The~cosmological data used include CMB shift parameters, BAO, SnIa Pantheon data and the local determination of $H_0$ data point (Equation \eqref{hr20}).  We thus compare these models with \lcdm with respect to the quality of fit and the best-fit parameter values with emphasis on the best-fit values of $H_0$ and $M$. We thus find the extent to which these models suffer from the $M$ and $H_0$ tensions and to what extent they imply the presence of phantom dark energy at late~times.

Another goal of the present analysis is to identify possible quintessence Lagrangians that are able to reproduce the identified best-fit parametrizations in the context of a physical field theory model. We also test such quintessence models for instabilities and unphysical~properties.

The structure of this analysis is the following: In the next section, we provide a pedagogical review of the cosmological data used in our analysis and the statistical techniques utilized. All of our codes are based on Mathematica and are publicly available. In~Section~\ref{sec3}, we describe the data analysis performed. In~Section~\ref{sec4}, we present the dark energy models considered, including field theoretic quintessence and dark energy parameterizations. We also present the likelihood contours of the parameters of the models considered and the quality of their fit to the data. In~Section~\ref{sec5}, we reconstruct a scalar field quintessence Lagrangian that can potentially reproduce the best-fit forms of the parameterizations considered in Section~\ref{sec4}. Finally, in Section~\ref{sec6}, we conclude, summarize our main results and discuss possible extensions of the present analysis. In~the appendices, we provide a pedagogical review of basic cosmological concepts and describe our notation. We also present some derivations of equations used in our~analysis.

\section{Cosmological Data---Parameters}
Cosmological models are described by parameters whose number for most models ranges between 4 and 20~\cite{Frieman:2008sn}. The~most common parameters used and~the ones involved in the present analysis~are:
\begin{itemize}
\item $h$: The dimensionless Hubble parameter defined as: $H_0 = 100\,\rm h \cdot \rm kms^{-1}\,Mpc^{-1}$.
\item $\Omega_{m,0}$: Present value of the matter density parameter.
\item $\Omega_{b}$: The baryon density parameter.
\item $w(z)$: The equation-of-state parameter. It is also common to use two or more parameters $(w_0, w_a,...)$ to define it. For~example, in~the CPL model~\cite{Chevallier:2000qy,Linder:2002et} $w(z) = w_0 + w_a/(1+z)$. From~the $w(z)$ parametrization, it is straightforward to obtain the dark energy density parameter $\Omega_{DE}(z)$ (see Appendix \ref{section:notation}).
\item $M$: In the present analysis, we also consider the SnIa absolute magnitude $M$. This parameter can be  constrained using either a combination of cosmological data (SnIa, BAO and CMB) at $z>0.01$ or Cepheid calibrators at $z<0.01$. The~root of the Hubble crisis lies in the mismatch of the values of $M$ obtained by the above two distinct approaches, as discussed below.
\end{itemize}

Our goal is to impose constraints on the values of the parameters of these models using observational data and identify the implications of these values for cosmology in general and for quintessence models in particular. The~types of cosmological data considered {are Type Ia supernovae, the~CMB shift parameters and BAO measurements, which} are discussed in what follows. { We could also use other data such as cosmic chronometers, i.e.,~measurements of the Hubble parameter at different redshifts, but~they have big error bars, and we do not think they have much constraining power. Thus, we use the data that we think are the most important.}

\subsection{Supernovae as Distance~Indicators}
Recent data coming especially from distant supernovae indicate that the expansion of the Universe is accelerating. These data come from distance modulus measurements of a certain type of supernovae: Type Ia or SnIa. They can be used as distance indicators (standard candles).

A supernova is a very energetic explosion, which releases vast amounts of energy, as~electromagnetic radiation, in~a relatively short period of time. Supernovae are extremely luminous and due to the burst of radiation they emit, they can light up their whole galaxy for weeks. A~scale to compare the energy that one supernova can unleash is approximately the energy that our Sun will radiate in its whole lifetime ($10^{44} J$). \\
A supernova can only emerge  by two~mechanisms:
\begin{itemize}
\item The collapse of the core of a massive star. Such a star has a core mass higher than the Chandrasekhar limit, which is 1.4 solar masses ($1.4$ $M_\odot$).
\item The abrupt re-ignition of nuclear fusion in a compact star (white dwarfs, neutron stars and black holes). In~order to have re-ignition, additional energy is required to raise the temperature in the stars core. The~star can obtain this energy either by a merger or \mbox{by accretion.}
\end{itemize}

Supernovae are classified according to their light curves and their absorption line of different chemical elements in their spectra~\cite{Farooq:2013syn,Amendola:2015ksp}. If~the spectrum of a supernova includes a spectral line of hydrogen, it is classified as Type II. Otherwise, it is classified as Type I. Now, if~the spectrum of a Type I supernova contains a single ionized silicon at 615 nm, it is called a Type Ia, while if it contains a line of non-ionized helium at 587.6 nm, it is called a Type Ib. Otherwise, if~it lacks both of these lines, it is called a Type Ic. There is also a way to subdivide the type II supernovae on the basis of their light curves, but it will not be of our interest. Lastly, only Type Ia supernovae emerge from the second mechanism, while every other type emerges from the~first.

\begin{figure}[H]
     \begin{subfigure}[b]{5cm}\captionsetup{justification=centering}
         \includegraphics[width=5cm]{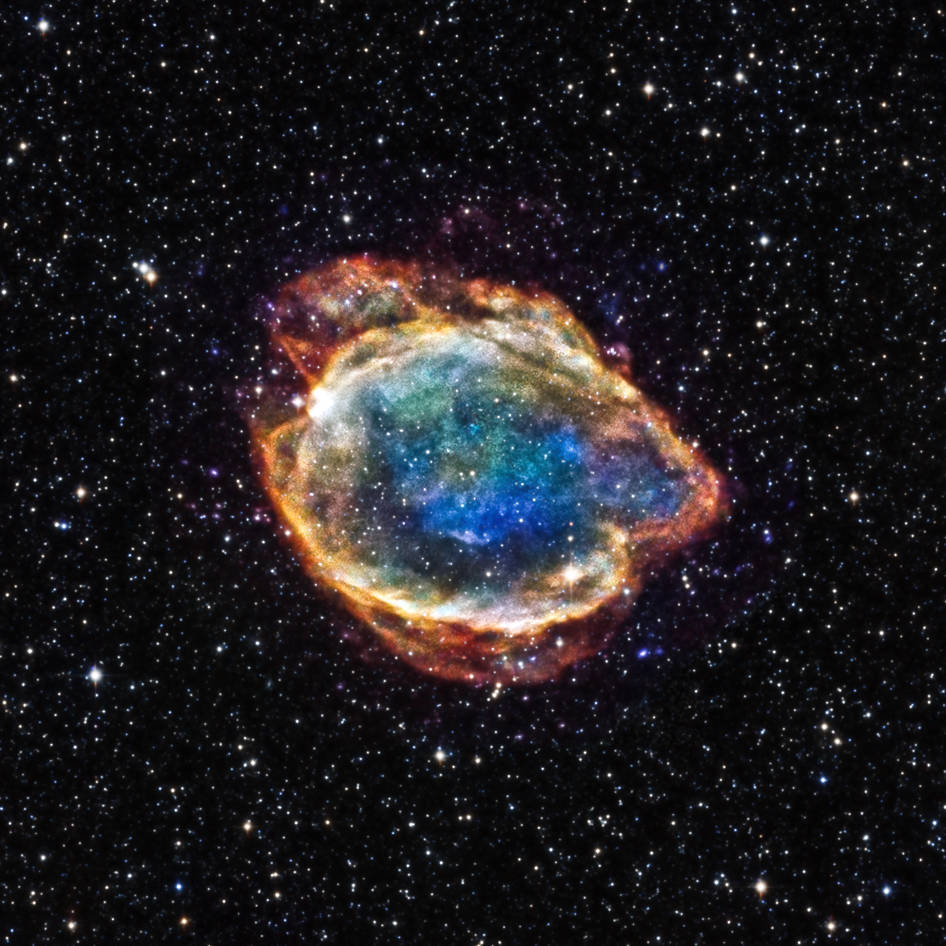}
         \caption{}
         \label{fig:G299-Remnants}
     \end{subfigure}
     \begin{subfigure}[b]{7cm}\captionsetup{justification=centering}
         \includegraphics[width=7cm]{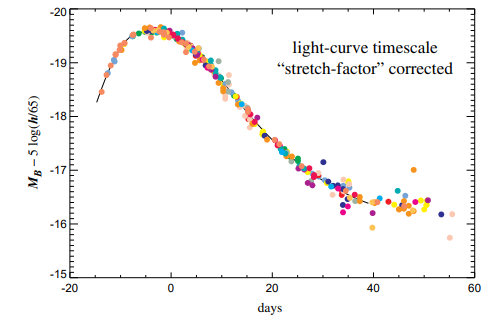}
         \caption{}
         \label{fig:SNIa_lightcurve}
     \end{subfigure}
        \caption{(\textbf{a}) Remnants of G299 Type Ia supernova. {Adapted from}: \cite{g299sup} (\textbf{b}) Typical lightcurve of a Type Ia supernova. {{Adapted from}:} \cite{Signore:2000mg}.}
        \label{fig:SNIa Images}
\end{figure}

Type Ia supernovae emerge as a result of the second mechanism, and more specifically, they emerge in binary star systems when one of the companion stars has a mass lower than the Chandrasekhar limit and thus, ends up as a white dwarf. Once the other companion star reaches its red giant phase, the~white dwarf starts accreting matter from it due to its gravitational field. When it reaches the Chandrasekhar limit, the~degeneracy pressure holding it in equilibrium can no longer keep up with the ever-increasing gravitational pressure and the star shrinks and thus raises its temperature. Eventually, the temperature reaches a point where carbon fusion can take place, which leads to a violent explosion that can be detected by a light curve and has the form shown in Figure~\ref{fig:SNIa Images}b.

The initial fusion begins a runaway thermonuclear process known as carbon detonation. During~this process, large quantities of the radioactive isotope nickel-56 ($^{56}Ni$) are produced. As it is radioactive, it undergoes positron decay to the stable isotope iron-56 ($^{56}Fe$)~\cite{Wright:2017rsu}. This decay chain can be simplified as follows:
\be
^{56}_{28}Ni \rightarrow ^{56}_{27}Co + ^{0}_{1}e^+ +\gamma \rightarrow ^{56}_{23}Fe+2^{0}_{1}e^+ +\gamma
\ee

\begin{figure}[H]
     \begin{subfigure}[b]{6cm}\captionsetup{justification=centering}
         \includegraphics[width=6cm]{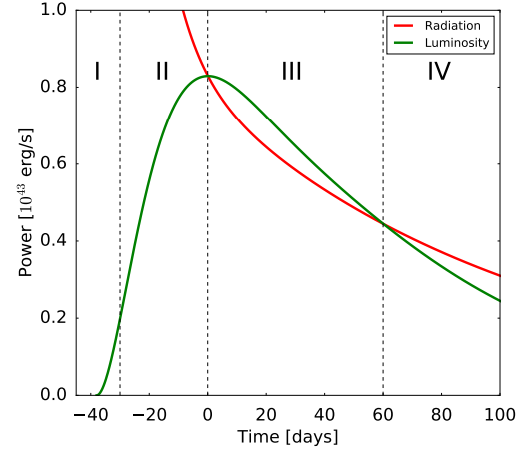}
         \caption{}
         \label{fig: Light_Curve_Phases}
     \end{subfigure}
     \begin{subfigure}[b]{6cm}
     \captionsetup{justification=centering}
         \includegraphics[width = 6cm]{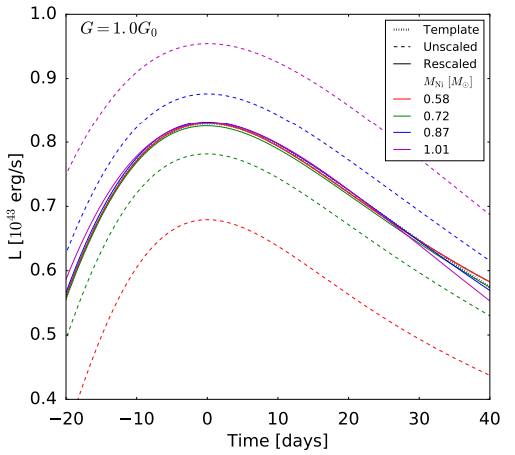}
         \caption{}
         \label{fig:Ni_Abundance}
     \end{subfigure}
        \caption{(\textbf{a}) Typical light curve of Type Ia supernovae along with the instantaneous power produced by the radioactive decay chain of $^{56}Ni$ divided into four phases.  {Adapted from}: \cite{Wright:2017rsu} (\textbf{b}) Typical Type Ia supernova light curves for different amounts of nickel-56 mass produced. { Adapted from}:{} \cite{Wright:2017rsu}.}
        \label{fig:SNIa_Light}
\end{figure}

The radiation produced during this process has the form of short wavelength gamma rays. This radiation does not contribute immediately to the light curve but~only after it has increased its wavelength through interactions with the supernova ejecta. However, not all the gamma rays produced can interact with the ejecta in order to lower their energy, so a percentage of them diffuses through the ejecta. These gamma rays do not contribute to the light curve at a phenomenon called gamma-ray~leakage. 
A qualitative description of the diffusion process can be summarized in four phases, each contributing to the light curve, as shown in  Figure~\ref{fig:SNIa_Light}a.

\underline{{Phase I}
}: At early times, the~outer layers of the ejecta are hot and densely packed and have high opacity to radiation of all wavelengths. However, this instantaneous luminosity observed is only a small fraction of the energy radiated from the decay happening in the center of the ejecta and thus contributes to the light curve as a small initial~brightness.

\underline{{Phase II}}: As the ejecta expand and disperse, its opacity to longer wavelength radiation falls until it becomes completely translucent, and radiation in the UV, optical and infrared range can escape. In~this phase, the light curve rises until it reaches its~peak.

\underline{{Phase III}}: As the ejecta become fully translucent, the~trapped radiation, which has been produced earlier, can escape. This corresponds to a rise of luminosity above the instantaneous power from the radioactive decay until this excess amount of radiation has all escaped. This phase corresponds to the part of the light curve right after the~peak.

\underline{{Phase IV}}:
 The point where the opacity of the ejecta becomes small enough that most of the short wavelength radiation can escape and produce gamma ray leaks, which do not contribute to the light curve. This can be seen in the part of the light curve where the observed luminosity falls again under the instantaneous power from the radioactive~decay.
 
 Furthermore, at~late times, in~the so-called nebular phase, a~significant distribution to the light curve is made by the positrons, which in total can carry about $3.5\%$ of the total decay energy~\cite{Mazzali:2000gk}.
 \paragraph{{Light Curve Features in Standard Gravity}}
\begin{itemize}
    \item The brightness of the light curve at its maximum is proportional to the mass of the synthesized $^{56}Ni$ \cite{Wright:2017rsu,Mazzali:2000gk,Arnett:1982ioj}. This feature can be seen in Figure~\ref{fig:SNIa_Light}b.
    \item The light curve width depends on the optical opacity of the ejecta, the~total ejected mass and the kinetic energy of the explosion~\cite{Mazzali:2000gk}, which is calculated as the difference between the energy produced by nuclear fusion and the gravitational binding energy of the white dwarf progenitor~\cite{Wright:2017rsu}. The~latter contributes only weakly to the light curve width~\cite{Mazzali:2000gk}.
\end{itemize}
\paragraph{{Light Curve Features in Modified Gravity} }
Many studies suggest that the absolute luminosity at peak $L_{peak}$ increases as G decreases. In~\cite{Gaztanaga:2001fh}, the~assumption:
\be
L_{peak}\propto M_{Ni}\propto M_{Ch} \propto G^{-3/2}
\ee
is made, which means that a slow decrease in $G$ results in dimmer distant supernovae than predicted for a standard scenario.
In~\cite{Amendola:1999vu}, the assumption that there is some relation between the peak luminosity and the Chandrasekhar mass is made and the hypothesis:
\be
L_{peak}\propto G^{-\gamma}
\ee
is tested, with~$\gamma>0$ of order unity. These studies have been used to find possible resolutions to the Hubble crisis~\cite{Marra:2021fvf} along with the $f\sigma_8$ tension~\cite{Kazantzidis:2019dvk}.

\begin{figure}[H]
         \includegraphics[width=0.6\textwidth]{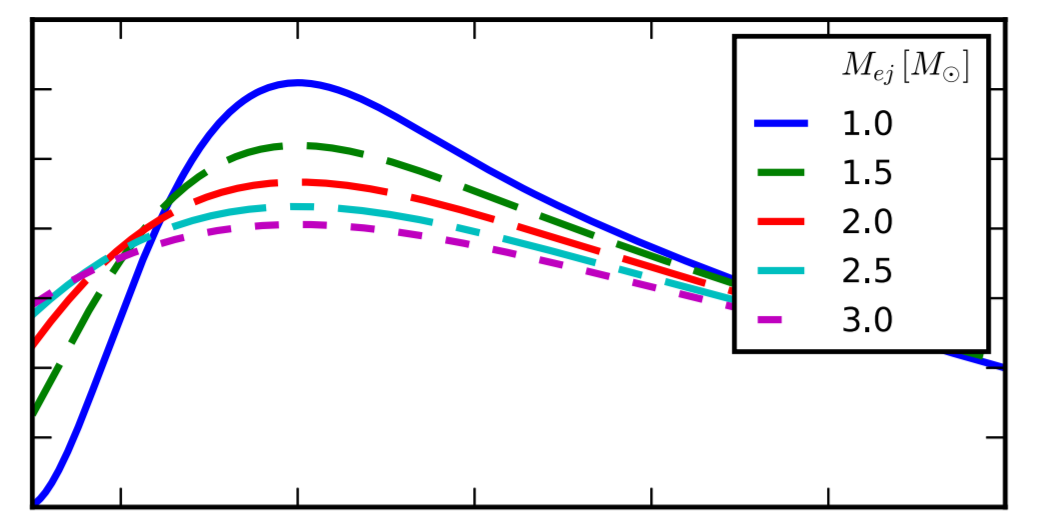}
         \caption{{Type} 
 Ia light curves for different values of Chandrasekhar mass showing that the peak luminosity decreases as the Chandrasekhar mass increases and thus the Gravitational constant decreases. {{Adapted from}:} \cite{Wright:2017rsu}.}
         \label{fig:Grav_Const}
\end{figure}

However, there is a recent study by Wright and Li~\cite{Wright:2017rsu}, in~which a semi-analytical model is used to calculate the light curves of Type Ia supernovae, which suggests that the SNIa peak luminosity decreases with Chandrasekhar mass. More specifically, the relations between $M_{Ni}$ and $G$ and $M_{Ch}$ and $G$ are being tested as to how they affect the peak luminosity. It seems that the dominant effect by varying $G$ in the peak luminosity is through the Chandrasekhar mass $M_{Ch}$ as the light curves produced with different values of nickel-56 mass $M_{Ni}$ closely match the $G = G_0$ light curves due to the relationship between $M_{Ni}$ and $\kappa$. It is shown that $M_{Ch}$ increases as G decreases, which, as~seen in Figure~\ref{fig:Grav_Const}, corresponds to a wider light curve but also to a lower peak luminosity compared to the $G =G_0$ light~curve.
\paragraph{{Observations}}
The absolute luminosity of Type Ia supernovae is almost constant at the peak of brightness, so the distance to it can be determined by measuring its apparent luminosity. Thus, they are a kind of standard candle (by which luminosity can be measured using observations). However, in~reality, things are not so easy. The~intrinsic spread in absolute magnitudes is too large to produce any meaningful cosmological constraints.

On the other hand, at~the end of the 1990s, a high-quality sample of local supernovae $(z<<1)$ helped towards the correlation of the absolute magnitude M and the width of the light curve. Therefore, if~someone measures the apparent magnitude of a Type Ia supernova and the width of its light curve, {they} can predict its absolute magnitude. Thus, the~universal SNIa absolute magnitude is nothing but the corrected magnitude in terms of the light curve width, as seen in Figure \ref{fig:lightcurve_correction}. Therefore, a more appropriate term when referring to them is not standard but  standardizable candles.

 However, there is another correction that we need to apply, since the redshift increases when we observe different parts of the power spectrum (broader, brighter SNe appear slightly bluer than they should be~\cite{Nugent:2002si}). This correction is named K-correction, and we always assume that it has already been included in the estimation of the apparent magnitude that we~use.

Type Ia supernovae are the preferred distance indicators~as:
\begin{figure}[H]
     \begin{subfigure}[b]{6cm}
     \captionsetup{justification=centering}
         \includegraphics[width = 6cm]{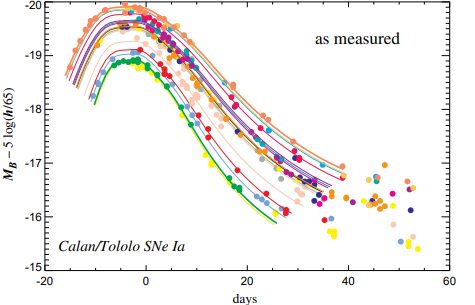}
         \caption{}
         \label{fig:lightcurve_01}
     \end{subfigure}
     \begin{subfigure}[b]{6cm}
     \captionsetup{justification=centering}
         \includegraphics[width=6cm]{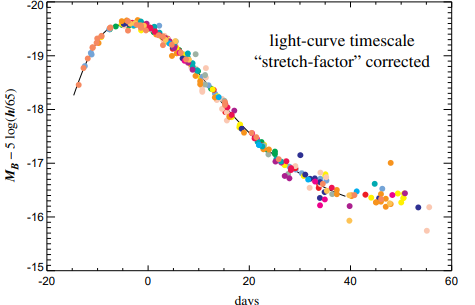}
         \caption{}
         \label{fig:lightcurve_02}
     \end{subfigure}
        \caption{(\textbf{a}) {Light curves} 
 of some local Type Ia supernovae as measured. (\textbf{b}) The light curves of the same supernovae corrected in terms of their width. {{Adapted from}:} \cite{Signore:2000mg}.}
        \label{fig:lightcurve_correction}
\end{figure}
\begin{itemize}
\item They are the most common type of supernova in the Universe.
\item They are extremely luminous, as~at their peak luminosity they can reach an absolute magnitude of about: $M\approx-19$, which is about the absolute magnitude of a \mbox{bright galaxy.}
\item They have a relatively small dispersion of peak absolute magnitude.
\item Their explosion mechanism is fairly uniform and well understood and, according to known physics, has no cosmic evolution.
\item There are a lot of local SNeIa that we can use to test their physics and  calibrate the absolute magnitude for the distant ones.
\end{itemize}

\subsection{Baryonic Acoustic Oscillation Measurements (BAO)}

The high redshift Universe $(z>1100)$ was pretty much homogeneous except from some really small perturbations (all four species were perturbed approximately by the same fractional amount) and consisted mainly of four density components: dark matter, baryons, photons and neutrinos. Photons and baryons were tightly coupled due to Compton scattering. Neutrinos do not interact and move too fast, so gravity cannot stop them. On~the other hand, dark matter becomes attracted and falls into the perturbations' overdensities, due to gravity. The~perturbations of the photon-baryon fluid (they are coupled) have both overdensity and overpressure. This overpressure creates an expanding sound wave moving with the speed of sound $c_s$ of that time. Thus, the perturbation of photons/baryons is carried~outwards.
\begin{figure}[H]
         \includegraphics[width=7cm]{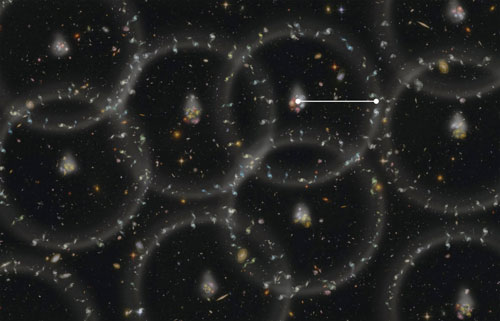}
         \caption{{A cartoon} produced by the BOSS project showing the spheres of baryons around the initial dark matter clumps. {{Adapted from}}: \cite{bao_cart}.}
         \label{fig:BAO_cartoon}
\end{figure}
\paragraph{{Recombination}}
As the universe cools down, there is a point (approximately at redshift $z\approx1320$) that protons and electrons can combine to form hydrogen: $e^- + p^+ \longleftrightarrow H + \gamma$. At~this point, photons do not scatter as efficiently and start to decouple, while the sound speed drops and thus the sound wave begins to slow down.
\paragraph{{Photon Decoupling}}
The same process continues until~the photons decouple completely. Then photons' perturbation begins to smooth out and the sound speed of the baryon perturbation is reduced so much that the pressure wave stops. Thus, we are left with the original dark matter perturbation surrounded by the baryon perturbation in a shell, as shown in Figure \ref{fig:BAO_cartoon}.
\paragraph{{Beyond} Photon Decoupling}
As can be seen in Figure~\ref{fig:BAO Generation Description}e,f, the~two perturbations left attract each other and they start to mix, eventually coming together. Although, the~acoustic peak perturbation is lower than the dark matter one since dark matter mass is dominated by baryons.

\paragraph{{Structure} Formation}
Galaxies form in matter (both dark matter and baryons) overdensities. Most galaxies appear at the original perturbation. However, a $1\%$ enhancement of galaxies appears at the acoustic scale and can be seen in the galaxy correlation~function.

The time that baryons are released from the drag of the photons is known as the drag epoch. We can obtain the redshift $z_d $ of this epoch, which is shortly after recombination, using the fitting formula introduced in~\cite{Eisenstein_1998}:
\be\label{eq:2.13}
z_d = \frac{1291\left(\Omega_mh^2\right)^{0.251}}{1+0.659\left(\Omega_mh^2\right)^{0.828}}\left[ 1+ b_1\left(\Omega_bh^2\right)^{b_2}\right]
\ee
where:
\be
b_1 = 0.313(\Omega_mh^2)^{-0.419}\left[ 1+ 0.607(\Omega_mh^2)^{0.674}\right]
\ee
\be 
b_2 = 0.238(\Omega_mh^2)^{0.223}
\ee

Photon-baryon fluid acoustic waves propagate at the sound speed:
\be
c_s = \frac{c}{\sqrt{3\left(1+R_s\right)}}
\ee
where:
\be
R_s \equiv \frac{3\rho_{b}}{4\rho_{\gamma}} = \frac{3\Omega_b}{4\Omega_{\gamma}}\frac{1}{1+z}
\ee 

It is also easy to prove the equation:
\be
c_s = \frac{c}{\sqrt{3\left(1+31500\cdot\Omega_{b}h^2\left( \frac{T_{cmb}}{2.7}\right)^{-4}a\right)}}
\ee
that we will use in the data analysis~later.\\
Thus:

\be
r_s = \int^{\infty}_{z_d}\frac{c_s(z)}{H(z)} dz
\ee
The sound horizon depends~on:
\begin{itemize}
\item The epoch of recombination, which affects the drag epoch $z_d$.
\item The expansion of the Universe, \emph{H(z)}.
\item The baryon-to-photon ratio, which affects $c_s$.
\end{itemize}
\begin{figure}[H]
     \begin{subfigure}[b]{4cm}
     \captionsetup{justification=centering}
         \includegraphics[width=4cm]{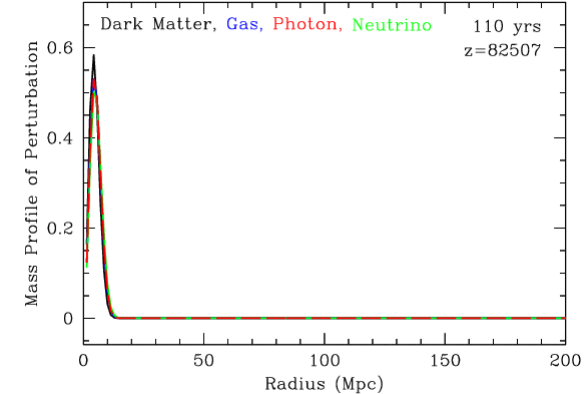}
         \caption{}
         \label{fig:tetete}
         \begin{minipage}{.1cm}
            \vfill
            \end{minipage}
     \end{subfigure}
     \hfill
     \begin{subfigure}[b]{4cm}
     \captionsetup{justification=centering}
         \includegraphics[width=4cm]{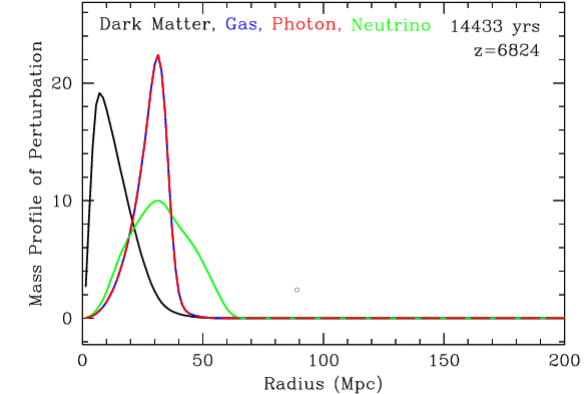}
         \caption{}
         \label{fig:tatatata}
         \begin{minipage}{.1cm}
            \vfill
            \end{minipage}
         \end{subfigure}%
    \hfill
     \begin{subfigure}[b]{4cm}
     \captionsetup{justification=centering}
         \includegraphics[width=4cm]{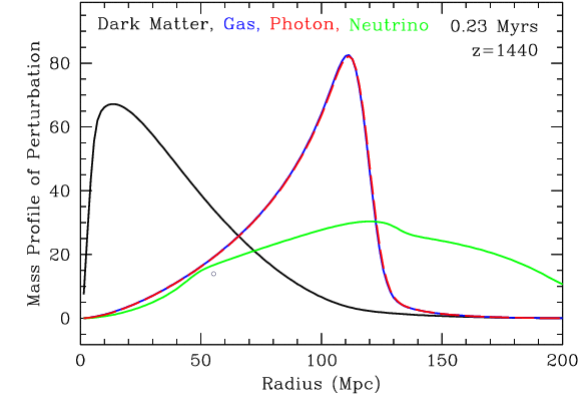}
         \caption{}
         \label{fig:tatatate}
         \begin{minipage}{.1cm}
            \vfill
            \end{minipage}
         \end{subfigure}
        \hfill
     \begin{subfigure}[b]{4cm}
     \captionsetup{justification=centering}
         \centering
         \includegraphics[width=4cm]{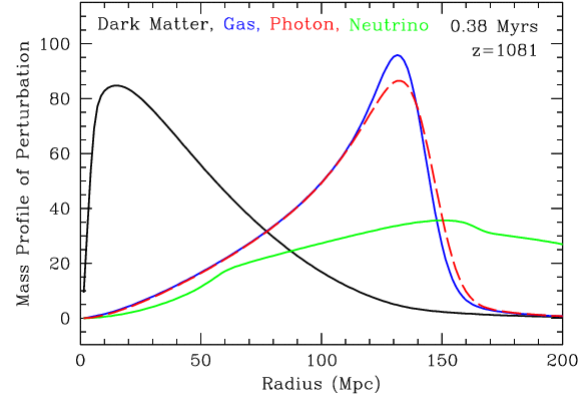}
         \caption{}
         \label{fig:tatatati}
         \begin{minipage}{.1cm}
            \vfill
            \end{minipage}
          \end{subfigure}%
          \hfill
         \begin{subfigure}[b]{4cm}
     \captionsetup{justification=centering}
         \includegraphics[width=4cm]{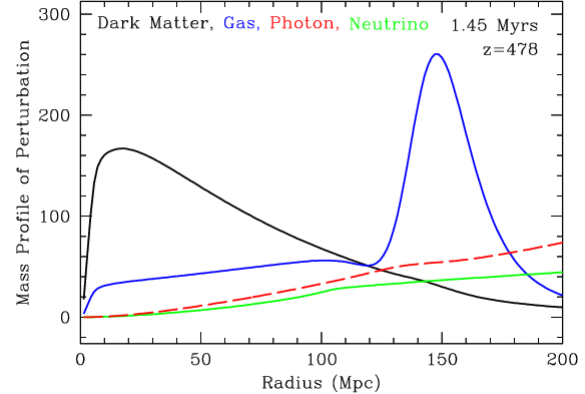}
         \caption{}
         \label{fig:tataatati}
         \begin{minipage}{.1cm}
            \vfill
            \end{minipage}
          \end{subfigure}
          \hfill
         \begin{subfigure}[b]{4cm}
     \captionsetup{justification=centering}
         \includegraphics[width=4cm]{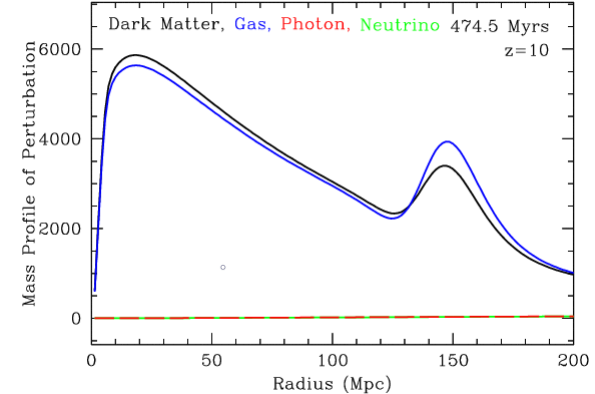}
         \caption{}
         \label{fig:tatataati}
         \begin{minipage}{.1cm}
            \vfill
            \end{minipage}
     \end{subfigure}
        \caption{Perturbation evolution: (\textbf{a}) Early Universe: initial perturbations. (\textbf{b}) Early Universe: neutrinos spread out, acoustic waves form. (\textbf{c}) Early Universe: acoustic waves propagate outwards, dark matter perturbation grows. (\textbf{d}) Recombination: photons do not scatter efficiently, sound speed drops. (\textbf{e}) Photon decoupling: photons spread out, sound wave stalls. (\textbf{f}) Before structure formation: dark matter and baryons attract each other and they mix up, with~dark matter dominating due to much higher mass. {{Adapted from}}: \cite{bao_figs_01}, while the original animation can be found here: \cite{bao_anim}.}
        \label{fig:BAO Generation Description}
        
\end{figure}
\unskip
\subsubsection{BAO~Measurements}
The BAO scale can be found as a peak in the galaxy correlation function or equivalently as damped oscillations in the large-scale structure power spectrum, as~seen in \mbox{Figure~\ref{fig:BAO Images}.}
In spectroscopic surveys, we observe the angular and redshift distributions of galaxies as a power spectrum $P(k_\bot,k_\parallel)$ in the redshift space, where $k_\bot$ and $k_\parallel$ are the wavenumbers perpendicular and parallel to the direction of light, respectively. Two possible measurements that we can perform are in the line-of-sight dimension or in the transverse direction measuring the ratios:
\be 
\delta z_s = \frac{r_s(z_\star)H(z)}{c}
\ee
\be
\theta_s = \frac{r_s(z_\star)}{(1+z)d_A(z)}
\ee
respectively, with~$d_A$ being the angular diameter distance and $r_s(z_\star)$ the sound horizon at the decoupling epoch.
Through these observables, we can define the distances:
\be
D_H(z) = \frac{c}{H(z)}
\ee
\be
D_M(z) = (1+z)d_A(z)
\ee

The first helps us measure the expansion rate H(z), while the second is just the comoving angular diameter distance. Here, we can see from~Equation~(\ref{eq:1.46}) that the comoving angular diameter distance $D_M$ corresponds to the metric distance $d_m$, which we defined in Equation~(\ref{eq:1.39}) 

We can also use the spherically averaged spectrum to obtain a combined distance scale ratio:
\be
\left[\theta_s^2(z)\delta z_s(z)\right]^{1/3}\equiv\frac{r_s(z_\star)}{\left[\left(1+z\right)^2{d_A}^2(z)\frac{c}{H(z)}\right]^{1/3}}
\ee
which relates with the distance:
\be
D_V \equiv\left[ \left(1+z\right)^2{d_A}^2(z)\frac{c}{H(z)}\right]^{1/3}= \left[cz\frac{{D_M}^2(z)}{H(z)} \right]^{1/3}
\ee

However, as~we saw earlier, $r_s$ can change due to deviations of the cosmological parameters. This means that BAO measurements do not really constrain the values mentioned above, but~they constrain the~values:
\begin{figure}[H]
     \begin{subfigure}[b]{6.3cm}
     \captionsetup{justification=centering}
         \includegraphics[width=6.3cm]{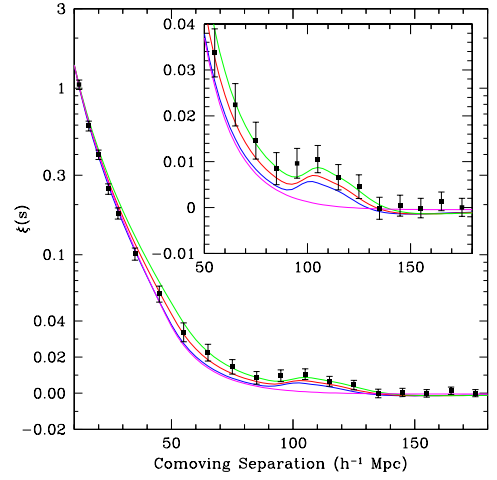}
         \caption{}
         \label{fig:BAO_01}
     \end{subfigure}
     \begin{subfigure}[b]{6cm}
     \captionsetup{justification=centering}
         \includegraphics[width=6cm]{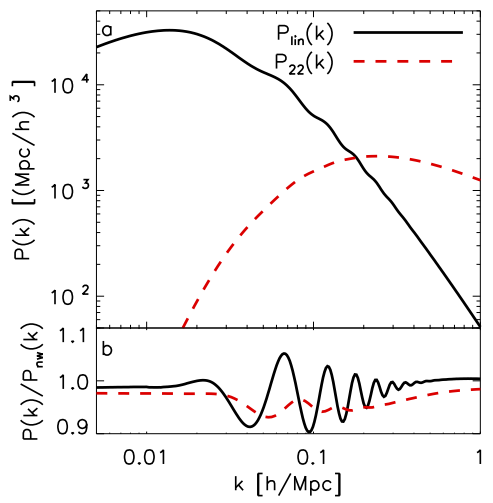}
         \caption{}
         \label{fig:BAO_02}
     \end{subfigure}
        \caption{(\textbf{a}) The large-scale redshift-space correlation function of the SDSS LRG sample. {Adapted from}: \cite{Eisenstein_2005}. (\textbf{b}) Damped oscillations in the large-scale structure power spectrum. {{Adapted from}}: \cite{bao_figs_02}.}
        \label{fig:BAO Images}
\end{figure}

\be
D_M \times \frac{r_s^{fid}}{r_s}
\ee
\be
D_V \times \frac{r_s^{fid}}{r_s}
\ee
\be 
D_H \times \frac{r_s^{fid}}{r_s}
\ee
 where $r_s^{fid}$ is the sound horizon in the context of the fiducial cosmology assumed in the construction of the large-scale structure correlation~function.

\subsection{CMB~Measurement}
\begin{figure}[H]
     \begin{subfigure}[b]{7cm}
     \captionsetup{justification=centering}
         \includegraphics[width=7cm]{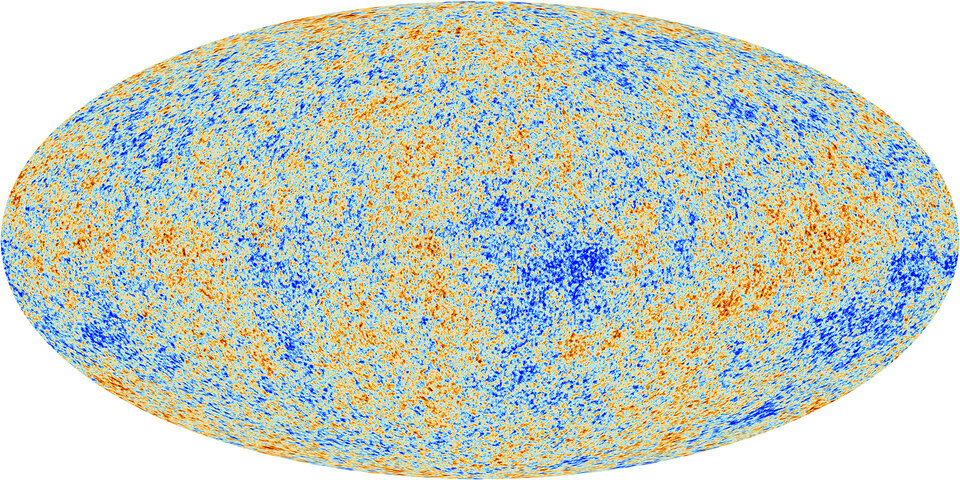}
         \caption{}
         \label{fig:CMB_01}
     \end{subfigure}
     \hfill
     \begin{subfigure}[b]{5cm}
     \captionsetup{justification=centering}
         \includegraphics[width=5cm]{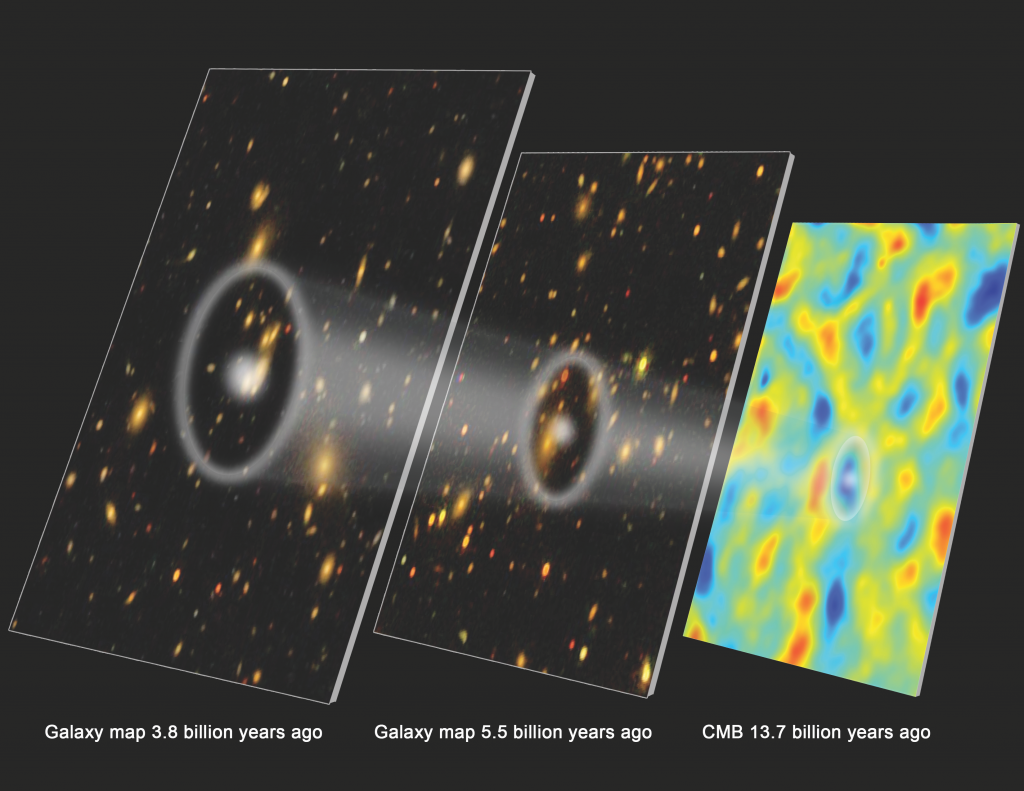}
         \caption{}
         \label{fig:CMB_02}
     \end{subfigure}
        \caption{(\textbf{a}) {The} cosmic microwave background (CMB) as seen by the Planck satellite. {Adapted from}: \cite{cmb_001}. (\textbf{b}) BAO evolution since the CMB. {{Adapted from}:} \cite{bao_anim}.}
        \label{fig:CMB Images}
\end{figure}

The CMB, shown in Figure \ref{fig:CMB_01}, can be treated as a BAO measurement at $z = z_\star = 1090$, as shown in Figure \ref{fig:CMB_02}, measuring the angular scale of the sound horizon at a high redshift~\cite{Aubourg:2014yra}. Thus, similar to the BAO measurements, we can define the characteristic angle of the location of the peaks:
\be
\theta_A = \frac{r_s(z_\star)}{D_M(z_\star)}
\ee
where $r_s(z)$ is the comoving sound horizon at redshift z and $z_\star$ is the redshift to the photon-decoupling surface.

The multipole $l$ (for the full analysis, see~\cite{Amendola:2015ksp}) corresponding to the angle $\theta_A$ is:
\be
l_A = \frac{\pi}{\theta_A} = \pi\frac{D_M(z_\star)}{r_s(z_\star)}
\ee

\begin{figure}[H]
         \includegraphics[width=12cm]{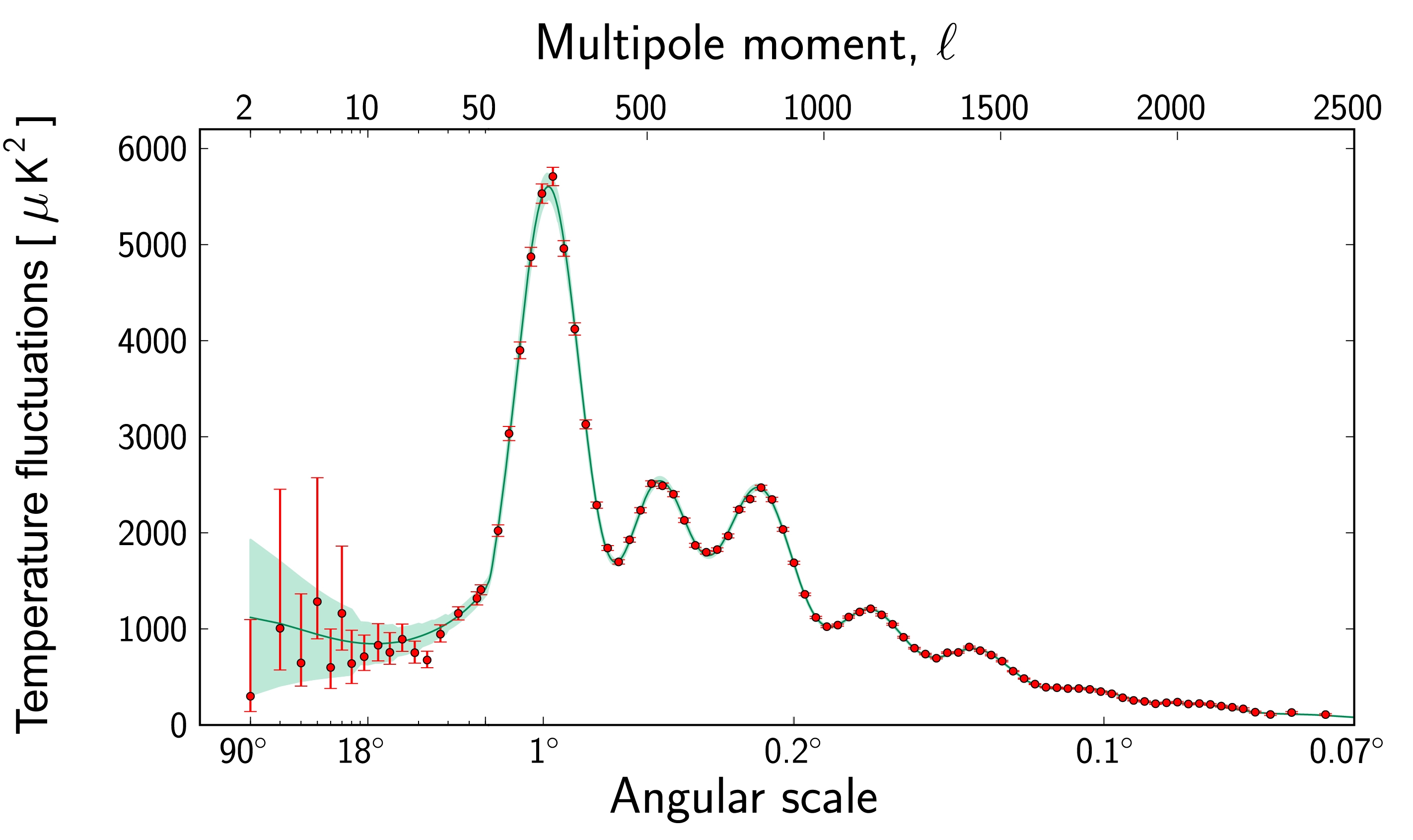}
         \caption{{The} CMB power spectrum versus the multipole moment l and the angular size $\theta$. The~curve shows the theoretical prediction of the power spectrum, while the red points represent the Planck data as of March 2013. {{Adapted from}:} \cite{cmb_002}.}
         \label{fig:CMB_multipole}
\end{figure}

Now, the~comoving angular diameter distance at that epoch $D_M(z_\star)$ can be \mbox{expressed as:}
\be\label{eq:2.31}
D_M(z_\star) = \frac{c}{H_0}\frac{1}{\sqrt{\Omega_m}}R
\ee
where R is the CMB shift~parameter.
\paragraph{{Shift} Parameter R}
From References~\cite{Alestas:2020mvb,Efstathiou:1998xx,Elgaroy:2007bv}, we know that the CMB power spectrum, which can be seen in Figure \ref{fig:CMB_multipole}, will be almost identical if these parameters are~fixed:
\begin{itemize}
\item The physical density parameters for matter $\omega_m$, baryons $\omega_b$, radiation $\omega_r$ and curvature $\omega_k$.
\item The primordial fluctuation spectrum.
\item The flat-universe comoving angular diameter distance to the recombination surface $D_M(z_\star)$.
\end{itemize}

Furthermore, the~product: $\sqrt{\omega_m}\cdot D_M(z_\star)$ does not depend on $H_0$ since:
\be
\sqrt{\omega_m}\cdot D_M(z_\star) = \sqrt{\Omega_mh^2}\int^{z_\star}_0\frac{dz}{H_0E(z)} = \frac{\sqrt{\Omega_m}}{100}\int^{z_\star}_0\frac{dz}{E(z)}
\ee
since: $H_0 \equiv 100\cdot h\cdot kms^{-1}Mpc^{-1}$.
This combination is the well-known shift parameter:
\be
R\equiv\sqrt{\omega_m}D_M(z_\star) = \sqrt{\Omega_m h^2}D_M(z_\star)
\ee

In general, the~shift parameter is defined as~\cite{Efstathiou:1998xx,Elgaroy:2007bv}:
\be
R = \left( \frac{\omega_m}{\omega_k}\right)^{1/2}S(\omega_k^{1/2}y)
\ee
where: $S(x)$ is defined in Equation~(\ref{eq:1.3})  and y is:
\be
y = \int^1_{a_\star}\frac{da}{a\dot{a}} = \int^1_{a_\star}\frac{da}{a^2H(a)}
\ee

In a flat universe:
\be
S(\omega_k^{1/2}y) = \omega_k^{1/2}y 
\ee

Thus:
\be
R = \sqrt{\frac{\omega_m}{\omega_k}}\sqrt{\omega_k}y = \sqrt{\omega_m}\int^1_{a_\star}\frac{da}{a^2H(a)} = \sqrt{\omega_m}\int^{z_\star}_0\frac{dz}{H(z)}
\ee

However, it is also common in the bibliography~\cite{Zhai_2019,Amendola:2015ksp} to define the shift parameter as:
\be
R\equiv\sqrt{\Omega_m H_0^2}D_M(z_\star)
\ee
in natural units (NU), or~as:
\be
R\equiv\frac{\sqrt{\Omega_m H_0^2}D_M(z_\star)}{c}
\ee
in~{SI} units.

In this analysis, the~definition used for the shift parameter is the latter~one.
\paragraph{{Photon} Decoupling Epoch}
In order to study the CMB, we should be able to find the value of the  redshift to the photon-decoupling surface $z_\star$. In~spite of the complexity of such a procedure, there is a fitting formula derived by Hu and Sugiyama~\cite{Hu:1995en,Zhai_2019}: 

\be
z_\star = 1048\left[1+0.00124\left(\Omega_bh^2 \right)^{-0.738} \right]\left[ 1+g_1\left(\Omega_mh^2 \right)^{g_2}\right]
\ee
where:
\be
g_1 = \frac{0.0783\left(\Omega_bh^2\right)^{-0.238}}{1+39.5\left(\Omega_bh^2\right)^{0.763}}
\ee
\be
g_2 = \frac{0.560}{1+21.1\left(\Omega_bh^2\right)^{1.81}}
\ee

This fitting formula is designed to be valid at a percent level, assuming that the parameters $\Omega_b$, $\Omega_m$ are inside the ranges:
\be
0.0025\lesssim\Omega_b\lesssim 0.25
\ee
\be
0.025\lesssim \Omega_m\lesssim 0.64
\ee

\section{Data~Analysis}
\label{sec3}
\subsection{Bayes~Theorem}

Bayes Theorem states that:
\be
p(\theta|x) = \frac{p(\theta,x)}{p(x)} = \frac{p(x|\theta)p(\theta)}{p(x)}
\ee
where:
\begin{itemize}
\item $p(\theta|x)$: The posterior probability distribution $p(\theta|x)$ for the parameters $\theta$ and data $x$. It is the probability that the parameters will obtain certain values after completing the experiment and making some assumptions~\cite{Heavens:2009nx}.
\item $p(x|\theta)$: It is called likelihood, and we also refer to it as $\mathcal{L}(x;\theta)$
\item $p(\theta)$: The prior probability distribution $p(\theta)$ for the parameters $\theta$. It expresses what we know about the parameters before performing the experiment, including the results of previous experiments or theory. For~example, we know that the age of the Universe must be positive.\\ In the absence of any previous information, it is common to adopt the principle of indifference and assume that all values of the parameters are equally likely and take $p(\theta) = constant$. As~a bound, someone can either use some finite bounds or use infinite bounds and work with an unnormalized prior. This prior is called a flat prior.
\item $p(x)$: The evidence.
\end{itemize}
\subsection{\texorpdfstring{Maximum~Likelihood}{}}

Consider data consisting of M independent measurements $X_{i,obs}$ with known standard deviation $\sigma_i$. Consider also a theoretical model prediction $X_{th}(p)$ to be tested by the data, with~p representing the parameters used in the model. The~$\chi^2$ function of the model as a function of its parameters is defined as~\cite{Farooq:2013syn}:
 \be\label{eq:3.1}
 \chi^2(p) = \sum_{i=1}^M\frac{\left[ X_{th}(p)-X_{i,obs}\right]^2}{\sigma_i^2}
 \ee
 
 In reality, $\chi^2$ quantifies the discrepancy between the predictions of the theoretical model and the observations at a particular value of the parameter p. Thus, a~small value of $\chi^2$ indicates a good fit. The~values $p_0$ that minimize the $\chi^2(p)$ function are called best-fit~parameters.
 \subsubsection{Non-Independent~Measurements}
 \paragraph{{Covariance}}
 If X and Y are two random variables, their covariance, denoted as $cov(X,Y)$, is defined by~\cite{2003theory}:
 \be
 cov(X,Y) = \left<(X-\left<X\right>)(Y-\left<Y\right>) \right>
 \ee
 where $\left< X\right>$ and $\left< Y\right>$ are the mean values of $X$ and $Y$, respectively.
 
 In general, for~discrete random variables:
 \be
 cov(X,Y) = \sigma_{ij} = \sum_{ij}(x_i-\left<x_i\right>)(y_j-\left<y_j\right>)h(x_i,y_j)
 \ee
 with $h(x_i,y_j)$ referring to the joint distribution of \emph{X} and~\emph{Y}.
 
 For continuous random variables:
 \be
 cov(X,Y) = \int dx \int dy (x-\left<X\right>)(y-\left<Y\right>)f(x,y)
 \ee
 where, again, $f(x,y)$ is the joint distribution of X and~Y.
 
It is evident that:
 \be
 cov(X,Y) = \left<(X-\left<X\right>)(X-\left<X\right>) \right> = \left<(X-\left<X\right>)\right> ^2 = Var(X) = \sigma^2_X
 \ee
 \paragraph{{Correlation}}

 The correlation of X and Y is denoted as $\rho(X,Y)$ and is defined by:
 \be
 \rho(X,Y) = \frac{cov(X,Y)}{\sigma_X \sigma_Y}
 \ee
while for discrete variables, we have:
\be
\rho_{ij} = \frac{\sigma_{ij}}{\sigma_i\sigma_j}
\ee
 \paragraph{{Covariance Matrix}}
 Having that in mind, we can construct the covariance matrix for the measurement vector $X_i$ as:
 \be
V_{ij} = cov\left[X_i,X_j\right] = \begin{bmatrix}
\sigma^2_1 & \rho_{12}\sigma_1\sigma_2 & \dots & \rho_{1N}\sigma_1\sigma_N\\
\rho_{21}\sigma_1\sigma_2 & \sigma^2_2 & \dots & \rho_{2N}\sigma_2\sigma_N  \\
\vdots & \vdots &  \ddots & \vdots \\
\rho_{N1}\sigma_N\sigma_1 & \rho_{N2}\sigma_N\sigma_2 & \dots & \sigma^2_N
\end{bmatrix}
\ee

As we can see, the~covariance matrix is always symmetric and square.
\paragraph{\texorpdfstring{$\chi^2$ {Function}}{}}
When we do not have  independent data, we cannot define the $\chi^2$ function as before since we should incorporate  the covariance matrix in its definition. Thus, $\chi^2(p)$ takes the form:
\be
 \chi^2(p) = \left[X_{th}(p)-X_{i,obs}\right]^T\cdot V^{-1}\left[X_{th}(p)-X_{i,obs}\right]
\ee
\subsubsection{Chi-by-Eye}
There is a useful ``chi-by-eye'' rule, which states that a fit is good and believable when the minimum $\chi^2$ is roughly equal to the number of data minus the number of parameters, and it is increasingly true for a large number of data~\cite{Verde:2009tu}.
\subsection{Likelihood~Function}
We define the corresponding likelihood function $\mathcal{L}$ as:
\be 
\mathcal{L}(p) = exp\left[-\frac{1}{2}\chi^2(p) \right]
\ee

The likelihood function maximizes as $\chi^2(p)$ minimizes. The~parameters that result in a higher value of the likelihood function are more likely to be the true~parameters.

If the model has two parameters, we can visualize the likelihood function as a surface sitting above the parameter space while, in~general, for~a model with N parameters, the likelihood function takes the shape of a hypersurface spanned by the parameter vector \mbox{$p$ \cite{MYUNG200390}.}
\subsubsection{Fisher~Matrix}
\textls[-35]{Sometimes we want to estimate the errors of the parameters from the likelihood function.}

At first, we assume a flat prior so we can identify the posterior through the~likelihood.

Then, we can expand the log-likelihood $ln\mathcal{L}$ as a Taylor series around its maximum~\cite{Verde:2009tu}:
$$
\ln\mathcal{L}(p) = \ln\mathcal{L}(p_0)+\sum_{j = 1}^N \frac{\partial\ln\mathcal{L}(p)}{\partial p_j}\bigg|_{p_0} \left(p_j - p_{j,0}\right)+$$
\be
 + \frac{1}{2!}\sum_{i = 1}^N\sum_{j = 1}^N \frac{\partial^2\ln\mathcal{L}(p)}{\partial p_i\partial p_j}\bigg|_{p_0}\left(p_i - p_{i,0}\right)\left(p_j - p_{j,0}\right)+\dots
\ee

Thus, we expand around the maximum log-likelihood $\ln\mathcal{L}(p_0)$:
\be
\frac{\partial\ln\mathcal{L}(p)}{\partial p_j}\bigg|_{p_0} = 0
\ee
for every $p_j$.

Furthermore, we stop the expansion to the quadratic term, which implies that we say that the likelihood surface is a multivariate Gaussian. Thus:
\be\label{eq:3.7}
\ln\mathcal{L}(p) \approx \ln\mathcal{L}(p_0) + \frac{1}{2!}\sum_{ij}\left(p_i - p_{i,0}\right)\frac{\partial^2\ln\mathcal{L}(p)}{\partial p_i\partial p_j}\bigg|_{p_0}\left(p_j - p_{j,0}\right)
\ee
\paragraph{{Hessian Matrix}}

The Hessian Matrix is defined as:
\be
\mathcal{H}_{ij} = -\frac{\partial^2\ln\mathcal{L}(p)}{\partial p_i\partial p_j}
\ee

It encloses information about the errors of the parameters and their covariance. If~the Hessian is not diagonal, the parameter estimates are correlated, which means that they have a similar effect on the data, and it can be difficult to discern them using the data. However, it is not certain that the parameters themselves will be correlated.
\paragraph{{Conditional Error}}
If all parameters are fixed except, e.g., $p_i$, we can estimate its error as:
\be
\sigma_{p_i} = \frac{1}{\sqrt{\mathcal{H}_{ii}}}
\ee

This is called a conditional error. It is the minimum error bar attainable on $p_i$, when the other parameters are known~\cite{Heavens:2009nx}, but~it is uninteresting and almost never used.
\paragraph{{Fisher Matrix}}
In 1935, Fisher answered the question of how accurately  someone can measure model parameters from a given dataset. The~Fisher information matrix is defined as:
\be
F_{ij} = - \left< \frac{\partial^2\ln\mathcal{L}(p)}{\partial p_i\partial p_j}\right>
\ee
or as:
\be
F_{ij} = \left<\mathcal{H}\right>
\ee

In practice, we choose a fiducial model and compute the Fisher matrix using~\cite{Verde_stat}. Substituting:
\be
\Delta\ln\mathcal{L} = \ln\mathcal{L}(p) - \ln\mathcal{L}(p_0)
\ee
in the Equation~(\ref{eq:3.7}), for~a one parameter case, we obtain:
\be\label{eq:3.20}
\Delta\ln\mathcal{L} = \frac{1}{2} F_{ii} \left(p_i - p_{i,0}\right)^2
\ee

Here, by identifying that $2\Delta\ln\mathcal{L}$ is really $\Delta \chi^2$, we see that the $1\sigma$ displacement for the parameter $p_i$, when $2\Delta\ln\mathcal{L} = 1$, is:
\be
\sigma_{p_i} = \frac{1}{\sqrt{F_{ii}}}
\ee
which is analogous to the conditional error.
In general,
\be
\sigma^2_{ij} \geq \left(F^{-1}\right)_{ij}
\ee
\be
\sigma_{p_i} \geq \left(F^{-1}\right)^{1/2}_{ii}
\ee
where $F^{-1}$ refers to the inverse of the Fisher information matrix, and $\sigma_{p_i}$ is the expected marginal error~\cite{Heavens:2009nx,Verde:2009tu}.

The Fisher matrix approach always gives you an optimistic estimate of the errors and \textls[-35]{this is obvious from the inequalities above,  which are forms of the  Cramér--Rao inequality~\cite{Heavens:2009nx,Verde:2009tu}.}
The inequality becomes an equality only if the likelihood is~Gaussian.
\subsection{Marginalization}
Most of the time, even though we use many parameters in a model, we may not have interest in all of them or we may have included some nuisance parameters, i.e.,~parameters  whose values we know with limited~accuracy.

Thus, we want to report the confidence level of the cosmological parameters that are interesting to us regardless of the value of the uninteresting/nuisance ones. This happens through a process called marginalization, where we marginalize over them~\cite{Verde:2009tu}. For~example, in~cosmological models with dark energy, a nuisance parameter is the Hubble constant $H_0$, or~in cosmological models with curvature, we want to know the value of $\Omega_k$ regardless of the values of $\Omega_m$ or $\Omega_\Lambda$. To~perform the marginalization process, we need, at~first, to~estimate a prior distribution $\pi(\nu)$. A~reasonable choice is to use a Gaussian probability density function with a mean value of $\nu_0$ (the most likely value) and variance $\sigma_\nu$.

In this way, we can build a posterior likelihood function that will depend only on the interesting parameters p and not on the parameter $\nu$:
\be
\mathcal{L}(p) = \int \mathcal{L}(p,\nu)\pi(\nu)d\nu 
\ee

While if~we consider the reasonable choice of $\pi(\nu)$ (Gaussian), we obtain:
\be
\mathcal{L}(p) = \frac{1}{\sqrt{2\pi\sigma_\nu^2}}\int\mathcal{L}(p,\nu)exp\left[ -\frac{(\nu-\nu_0)^2}{2\sigma_\nu^2}\right] d\nu
\ee

\textls[-35]{Thus, in~the same way as before, by minimizing $\chi^2(p)$, we can find the best-fit parameters
~\cite{Farooq:2013syn}.}
\begin{figure}[H]
         \includegraphics[width=10cm]{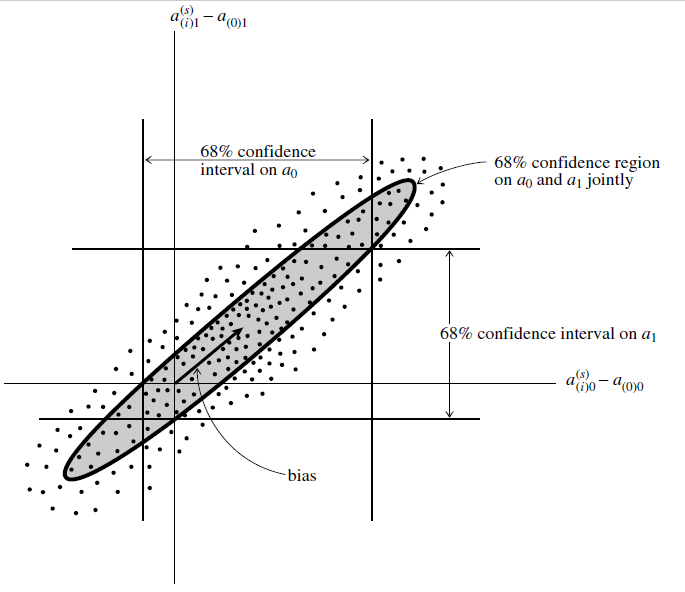}
         \caption{{Probability} distribution for the case M = 2 (two parameters $a_0$ and $a_1$). Furthermore, there are three different confidence regions all at the same confidence level. The~first region is defined by the vertical lines and represents a $68\%$ confidence interval for the variable $a_0$ without regard to the value of $a_1$, while the second one is defined by the horizontal lines and represents a $68\%$ confidence interval for the variable $a_1$ without regard to the value of $a_0$. The~third one is the ellipse, which shows a $68\%$ confidence interval for $a_0$ and $a_1$, jointly. {{Adapted from}}: \cite{10.5555/1403886}.}
         \label{fig:Conf_lvl}
\end{figure}
\unskip
\subsection{Confidence~Limits}
It is common practice not to present every detail of the probability distribution of errors in parameter estimation but to summarize it in the form of confidence limits. The~full probability distribution is a function defined in the M-dimensional space of parameters p (M = number of parameters). 

A confidence region is a region in  M-dimensional space that contains a certain percentage of the total probability distribution. The~ideal is to find a small region that contains a large percentage of the total probability~distribution. 

When we perform an analysis, we are free to pick both the confidence level and the shape of the confidence region. The~only requirement is that the region we pick contains the stated percentage of probability. The~most commonly used percentages are: $68.27\%,95.45\%,99.73\%$, which correspond to standard deviations $1\sigma, 2\sigma, 3\sigma$ from the most likely value, while the convention when we want to choose a shape for a confidence region is: for one dimension, we use a line segment centered on the measured value, and for two or higher dimensions, it is most common to use ellipses or~ellipsoids.

The whole point of the confidence level is to inspire confidence; for example, when we say that we have a confidence region with a confidence limit of $99\%$, there is a $99\%$ chance that the true parameter falls within this region around the measured~value. An example is shown in Figure \ref{fig:Conf_lvl}.
\begin{figure}[H]
         \includegraphics[width=12cm]{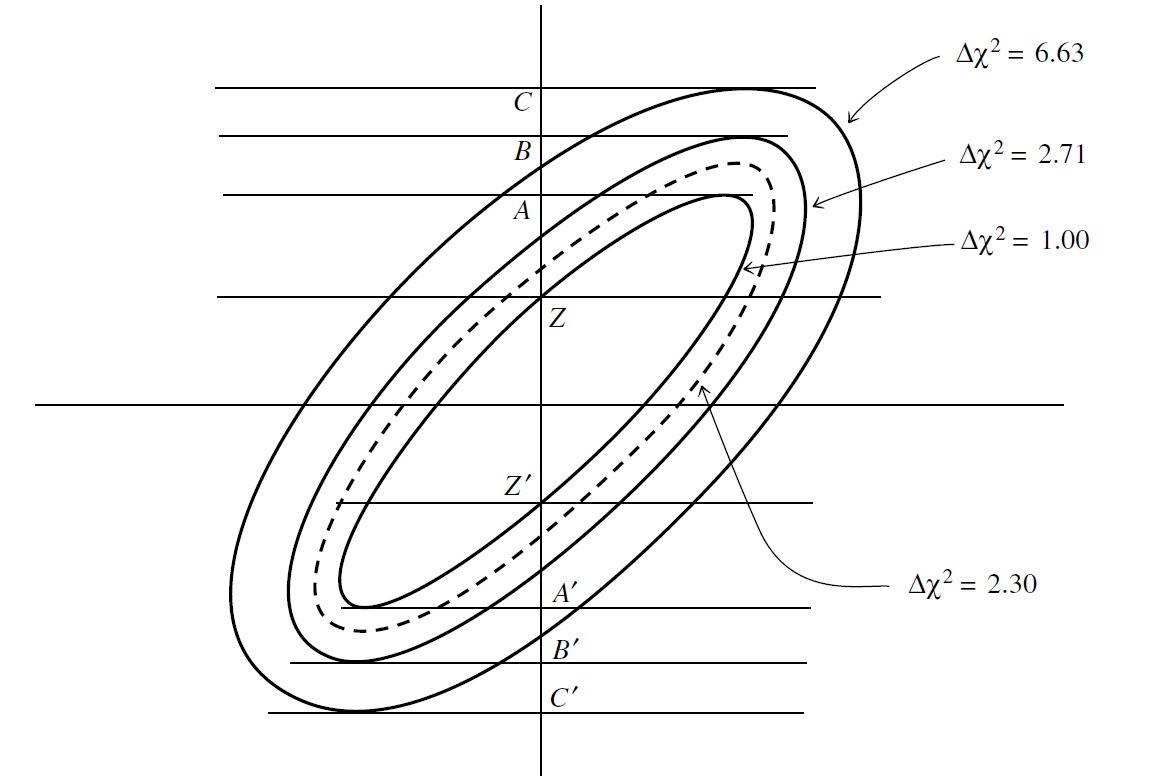}
         \caption{{Confidence} 
 regions derived using a constant $\Delta\chi^2$ region as a boundary. Here, the~shape of the confidence region is chosen to be an ellipse. It is important to note that the intervals AA', BB', CC' are the ones that contain the percentage of the normally distributed data that correspond to the respective $\Delta\chi^2$. {Adapted from}:  \cite{10.5555/1403886}.}
         \label{fig:Conf_reg_01}
\end{figure}

\subsubsection{\texorpdfstring{Constant $\chi^2$ Boundaries as Confidence~Limits}{}}
In order to obtain the minimum value of $\chi^2$ for the observed data set, we use the value $p = p_0$ for the parameters. If~the vector of the parameters' values $p$ is perturbed away from $p_0$, $\chi^2$ increases.
The region in which $\chi^2$ does not increase more than a fixed amount $\Delta\chi^2$ defines some M-dimensional confidence region around $p_0$. If~$\Delta\chi^2$ is a large number,  the confidence region will be large, while smaller values for $\Delta\chi^2$ correspond to smaller~regions.

Although we are free to use whatever value we want for $\Delta\chi^2$, there are some special values that correspond to the most commonly used confidence limits: $68.27\%$, $95.45\%$, $99.73\%$. These values do not only depend on the preferred confidence limits but also on the number of parameters M (or equivalently, the degrees of freedom of the model) as shown in Table \ref{tab: Delta-chi_square_01}. The scale of the likelihood contours as the value of $\Delta\chi^2$ changes is shown in Figure \ref {fig:Conf_reg_01}.
\begin{specialtable}[H]
\caption{{Table containing} $\Delta\chi^2$ values as a function of confidence level $p$ and the number of parameters M. The~process used to derive these values can be seen in Appendix \ref{subsection:Derivation_01}.}
\label{tab: Delta-chi_square_01}

\setlength{\cellWidtha}{\columnwidth/6-2\tabcolsep+0.0in}
\setlength{\cellWidthb}{\columnwidth/6-2\tabcolsep+0.0in}
\setlength{\cellWidthc}{\columnwidth/6-2\tabcolsep+0.0in}
\setlength{\cellWidthd}{\columnwidth/6-2\tabcolsep+0.0in}
\setlength{\cellWidthe}{\columnwidth/6-2\tabcolsep+0.0in}
\setlength{\cellWidthf}{\columnwidth/6-2\tabcolsep+0.0in}
\scalebox{1}[1]{\begin{tabularx}{\columnwidth}{>{\PreserveBackslash\centering}m{\cellWidtha}>{\PreserveBackslash\centering}m{\cellWidthb}>{\PreserveBackslash\centering}m{\cellWidthc}>{\PreserveBackslash\centering}m{\cellWidthd}>{\PreserveBackslash\centering}m{\cellWidthe}>{\PreserveBackslash\centering}m{\cellWidthf}}

 \toprule
 
 \multirow{2}{4em}{\centering \boldmath{$\sigma$} }& \multirow{2}{5em}{\centering \textbf{Probability}}&\multicolumn{4}{c}{\textbf{M}}\\

&& \textbf{1}&\textbf{2}&\textbf{3}&\textbf{4} \\
\midrule

 1& $68.27\%$ & 1.00 & 2.30 & 3.53 & 4.72\\

 2& $95.45\%$ & 4.00 & 6.18 & 8.02 & 9.72\\
 3 & $99.73\%$ & 9.00 & 11.8 & 14.2 & 16.3\\
\bottomrule
\end{tabularx}}
\end{specialtable}

\vspace{-12pt}
\subsubsection{Errors}
When we use a model, it is really important to know not only the favored values obtained by the data but also their respective errors. In~order to simplify the procedure of finding the errors and have a good approximation of them, we can use a linear model of \mbox{the form:}
\be
y(x) = a + bx
\ee
with errors in both axes: $\sigma_{x_i}, \sigma_{y_i}$

At first, we need to calculate the $\chi^2$ function for this model. To~do, that we need to know the weighted sum of variances, which in this case is the variance of the linear combination: $y_i - a - bx_i $ of the random variables $x_i,y_i$ \cite{10.5555/1403886}.
\be
Var(y_i - a - bx_i) = Var(y_i)+b^2Var(x_i) = \sigma^2_{y_i} + b^2\sigma^2_{x_i} \equiv \frac{1}{w_i}
\ee

Thus:
\begin{figure}[H]
         \includegraphics[width=12cm]{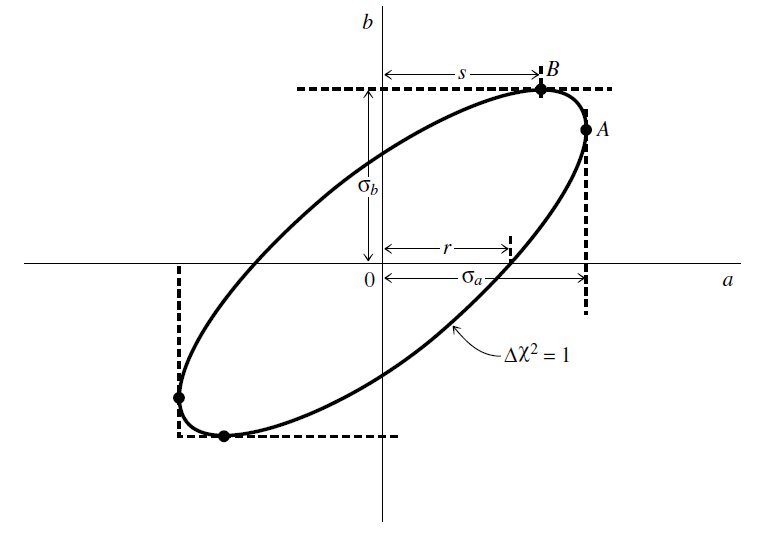}
         \caption{{Standard} errors for the parameters a and b found as the projections of the ellipse, which is the confidence region boundary with $\Delta \chi^2 = 1$, on~the a and b axes. {{Adapted from}}: \cite{10.5555/1403886}.}
         \label{fig:Errors_01}
\end{figure}
\be
\chi^2(a,b) = \sum_{i = 0}^{N-1}w_i(y_i - a - bx_i)^2 = \sum_{i = 0}^{N-1}\frac{(y_i - a - bx_i)^2}{\sigma^2_{y_i} + b^2\sigma^2_{x_i}}
\ee

The goal is to minimize $\chi^2(a,b)$ with respect to a and b, so we need to solve:
\be\label{eq:3.10}
\frac{\partial \chi^2}{\partial a} = 0
\ee
\be\label{eq:3.11}
\frac{\partial \chi^2}{\partial b} = 0
\ee
in order to find the best-fit values for the parameters a and~b.

As we saw earlier, a confidence region boundary where $\chi^2$ takes a value greater than its minimum: $\chi^2 + \Delta\chi^2$ with $\Delta\chi^2 = 1$, defines a $1\sigma$ region. Thus, by taking the projections onto the a and b axes, we obtain the standard errors for the parameters a and b, respectively, as shown in Figure \ref{fig:Errors_01}.\\ In the linear case, we can obtain these projections by Taylor expanding $\chi^2$ as follows:
\end{paracol}
$$\chi^2(a+\Delta a,b+\Delta b) = \chi^2(a,b) + \frac{\partial \chi^2}{\partial a} (a+\Delta a- a) + \frac{\partial \chi^2}{\partial b} (b+\Delta b - b)+
\frac{1}{2!}\frac{\partial^2 \chi^2}{\partial a^2}(a+\Delta a- a)^2 +$$
\begin{paracol}{2}
\switchcolumn

\be
+\frac{1}{2!}\frac{\partial^2 \chi^2}{\partial b^2}(b+\Delta b - b)^2 + \frac{\partial^2 \chi^2}{\partial a \partial b}(a+\Delta a- a) (b+\Delta b - b) +...
\ee
which due to Equations~(\ref{eq:3.10})  and~(\ref{eq:3.11}), which are true at the minimum of $\chi^2$, and~the fact that: $\Delta \chi^2 \equiv \chi^2(a+\Delta a,b+\Delta b) - \chi^2(a,b)$ gives:

\be
\Delta \chi^2 \approx \frac{1}{2}\left[\frac{\partial^2 \chi^2}{\partial a^2}(\Delta a)^2 + \frac{\partial^2 \chi^2}{\partial b^2}(\Delta b)^2 \right] + \frac{\partial^2 \chi^2}{\partial a \partial b}\Delta a \Delta b
\ee

Our goal is to solve the equation $\Delta \chi^2 = 1$ numerically for all the parameters of our model. As a result of this, in~combination with Equation~(\ref{eq:3.20}),  we find the components of the Fisher matrix, and from the diagonal components of its inverse, we obtain the marginal errors for the~parameters.

\subsection{\texorpdfstring{$\chi^2$ for CMB~Data}{}}
In this analysis, we use the central values for $R$, v$l_A$ and $\Omega_bh^2$ from Planck 2018 for a flat universe. The~data can be found in~\cite{Zhai_2019} and are expressed in terms of a data vector and a covariance matrix.
The data vector is: 
\be
v = \begin{pmatrix}
1.74963  \\
301.80845  \\
0.02237   
\end{pmatrix}
\ee
and the corresponding covariance matrix is:
\be
C_v = 10^8\times \begin{pmatrix}
1598.9554 & 17112.007 & -36.311179  \\
17112.007 &  811208.45 & -494.79813 \\
-36.311179 & -494.79813 & 2.1242182
\end{pmatrix}
\ee

The only remaining task is to find the inverse matrix of $C_v$ and define the $\chi^2_{CMB}$.

It is useful to define a new vector:
\be
vec = v_{th}-v_{obs} = \begin{pmatrix}
R - 1.74963  \\
l_A - 301.80845  \\
\Omega_bh^2 - 0.02237   
\end{pmatrix}
\ee

Thus, the~$\chi^2$ function takes the form:
\be\label{eq:3.36}
\chi^2_{CMB} = vec^T\cdot C^{-1}_{CMB}\cdot vec
\ee
\subsection{\texorpdfstring{$\chi^2$ for BAO~Data}{}}
\subsubsection{6dFGS and~WiggleZ}
Both surveys give a measurement for the ratio:
\be
d_z \equiv \frac{r_s(z_d)}{D_V(z)} 
\ee

From 6dFGS we obtain: $d_z (z = 0.106) = 0.336 \pm 0.015$, while from WiggleZ, we obtain three correlated measurements:
$$d_z (z = 0.44) = 0.073$$
$$d_z (z = 0.6) = 0.0726$$
$$d_z (z = 0.73) = 0.0592$$
with the inverse of the covariance matrix given as:
\be
C^{-1}_{WiggleZ} = \begin{pmatrix}
1040.3 & -807.5 & 336.8  \\
-807.5 &  3720.3 & -1551.9 \\
336.8 & -1551.9 & 2914.9
\end{pmatrix}
\ee

Having four measurements, we produce the data vectors:
\be
v = \begin{pmatrix}
z  \\
d_z  \\
\sigma_{d_z}  
\end{pmatrix} = \begin{pmatrix}
0.106  \\
0.336  \\
0.015   
\end{pmatrix} , \begin{pmatrix}
0.44  \\
0.073  \\
0.031
\end{pmatrix},\begin{pmatrix}
0.6  \\
0.0726  \\
0.0164   
\end{pmatrix},\begin{pmatrix}
0.73  \\
0.0592  \\
0.0185   
\end{pmatrix}
\ee
and the total inverse covariance matrix:
\be
C^{-1} =\begin{pmatrix}
\left(\frac{1}{0.015}\right)^2& 0 & 0 & 0\\
0 & 1040.3 & -807.5 & 336.8  \\
0 & -807.5 &  3720.3 & -1551.9 \\
0 & 336.8 & -1551.9 & 2914.9
\end{pmatrix}
\ee

The vector used to define the $\chi^2$ function is:
\be
vec = d_z^{th}-d_z(z)
\ee
and the $\chi^2$ function for these measurements is:
\be
\chi^2_{6dFGs,Wiggle} = vec^{T}\cdot C^{-1}\cdot vec
\ee
\subsubsection{SDSS}

From this survey, we obtain a measurement of:
\be
\frac{D_V}{r_d}\times r_d^{fid}
\ee
where:
\be
r_d = r_s(z_d)
\ee
and its value in the context of the fiducial cosmology is:
\be
r_d^{fid} = 149.28 Mpc
\ee

Thus, we obtain the data vectors:
\be
v = \begin{pmatrix}
z  \\
\frac{D_V}{r_d} \\
\sigma  
\end{pmatrix} = \begin{pmatrix}
0.15  \\
4.465666824  \\
0.1681350461   
\end{pmatrix} , \begin{pmatrix}
0.32  \\
8.4673  \\
0.167
\end{pmatrix},\begin{pmatrix}
0.57  \\
13.7728  \\
0.134   
\end{pmatrix}
\ee

The $\chi^2$ function: $\chi^2_{SDSS}$ is obtained by using Equation~(\ref{eq:3.1})  for the parameter $\frac{D_V}{r_d}$, which is essentially the ratio: $\frac{1}{d_z}$. Thus:
\be
\chi^2_{SDSS} = \sum_{i=1}^3\frac{\left[ \frac{1}{d_z(z_i)}-\left(\frac{D_V}{r_d}\right)_i\right]^2}{\sigma_i^2}
\ee
\subsubsection{\texorpdfstring{Ly-$\alpha$}{}}
From this survey, we obtain two measurements. The~first measurement corresponds to: $\frac{D_M}{r_d}(z = 2.34) = 37.41 \pm 1.86$ and the second to: $\frac{D_H}{r_d}(z = 2.34) = 8.86 \pm 0.29$. The~data vectors for these measurements are:
\be 
v_1 = \begin{pmatrix}
z\\
\frac{D_M}{r_d}  \\
\sigma
\end{pmatrix} = \begin{pmatrix}
2.34\\
37.41  \\
1.86
\end{pmatrix}
and \hspace{0.25cm} v_2 =\begin{pmatrix}
z \\
\frac{D_H}{r_d}  \\
\sigma
\end{pmatrix}=\begin{pmatrix}
2.34 \\
8.86  \\
0.29
\end{pmatrix}
\ee

In this analysis, $d_A$ is used instead of $D_M$ so the new data vector for the first measurement is: 
\be
v_1 = \begin{pmatrix}
z\\
\frac{d_A}{r_d}  \\
\sigma_{d_A}
\end{pmatrix}= \begin{pmatrix}
z\\
\frac{D_M}{(1+z)\cdot r_d}  \\
\frac{\sigma}{1+z}
\end{pmatrix} = \begin{pmatrix}
2.34\\
11.20 \\
0.56
\end{pmatrix}
\ee

The corresponding inverse covariance matrix for $u_1$ and $u_2$ is:
\be
C^{-1} =\begin{pmatrix}
\left(\frac{1}{0.56}\right)^2 & 0  \\
0 & \left(\frac{1}{0.29}\right)^2\\
\end{pmatrix}
\ee

The vector used to define the $\chi^2$ function is:
\be
vec_{Ly-\alpha} = \begin{pmatrix}
\frac{d_A}{r_d} - \left(\frac{d_A}{r_d}\right)_{th}\\
\frac{D_H}{r_d} - \left(\frac{D_H}{r_d}\right)_{th} \\
\end{pmatrix}
\ee
and the $\chi^2$ function for these measurements is:
\be
\chi^2_{Ly-\alpha} = vec_{Ly-\alpha}^{T}\cdot C^{-1}\cdot vec_{Ly-\alpha}
\ee

Finally, the~total $\chi^2$ function for the BAO data is the sum of the individual $\chi^2$ functions for the data from the four surveys. Thus:
\be\label{eq:3.27}
\chi^2_{BAO} = \chi^2_{6dFGs,Wiggle} + \chi^2_{SDSS} + \chi^2_{Ly-\alpha} 
\ee

\subsection{\texorpdfstring{$\chi^2$ for SNIa~Data}{}}
For the SNIa data, we use the Pantheon dataset. A~representation of a subset of the Pantheon data called the binned Pantheon dataset is shown in Figure~\ref{fig:BinPanth}.
\begin{figure}[H]
         \includegraphics[width=12cm]{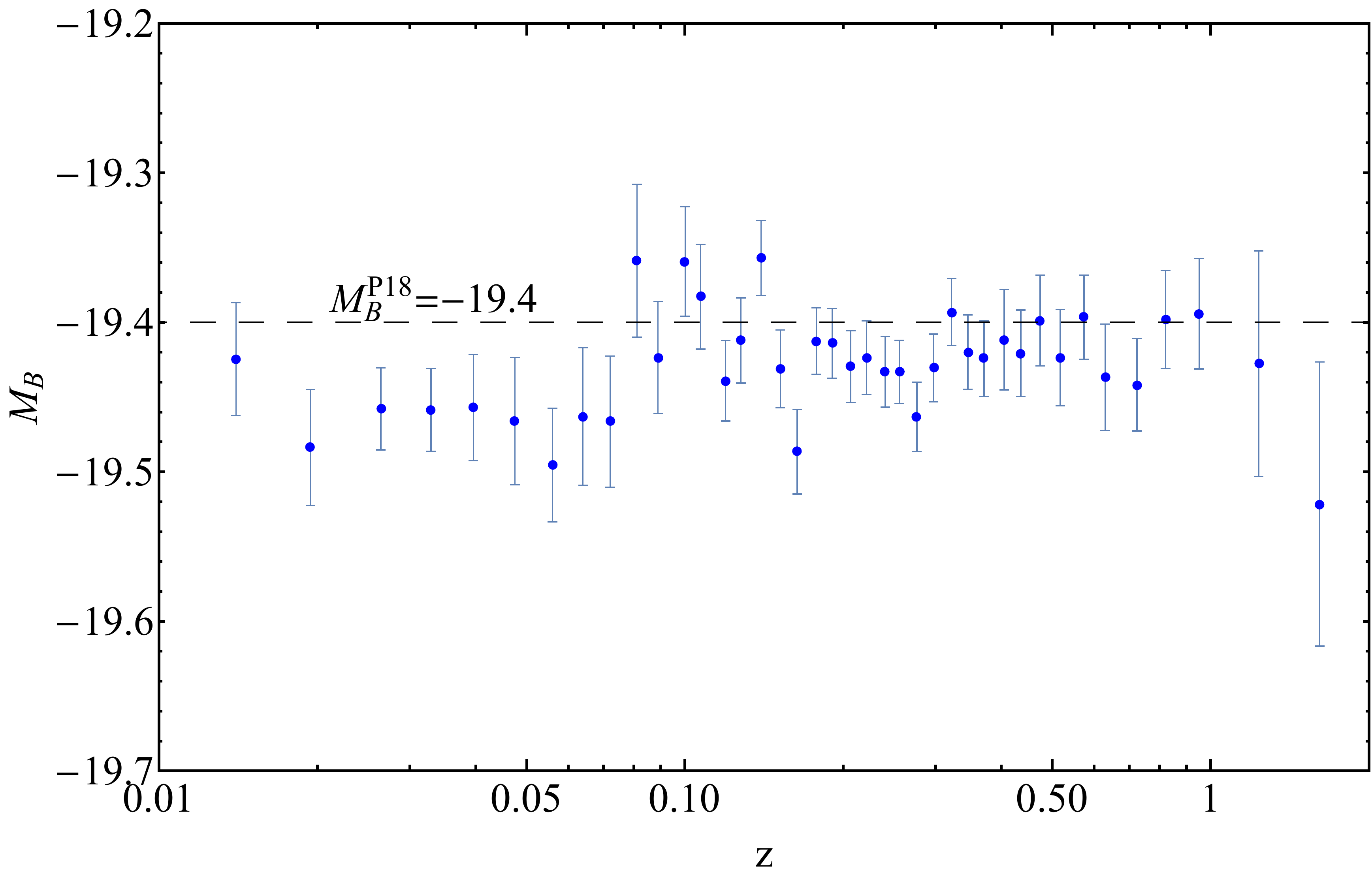}
         \caption{Plot of the inferred SNIa absolute magnitudes $M_{B,i} = m_{B,i} - \mu(z_i)$ of the binned Pantheon data~\cite{Scolnic:2017caz} under the assumption of a Planck/$\Lambda CDM$ luminosity distance. Furthermore, there is a straight line at the value: $M^{p18}_B = -19.4$, which is the value of the SNIa absolute magnitude calibrated via BAO and CMB measurements found in: \cite{Marra:2021fvf}.}
         \label{fig:BinPanth}
\end{figure}
The data vector used for the definition of $\chi^2$ is:
\be
V_i^{Pantheon} = m^{obs}(z_i) - m_i^{th}
\ee
where the apparent magnitude $m_{th}$ is obtained using Equation~(\ref{eq:1.87})  or equivalently as:
\be\label{eq:3.55}
m_{th} = M + 5\log_{10}\left[D_L(z)\right] + 5\log_{10}\left(\frac{c/H_0}{1Mpc}\right) + 25
\ee
where:
\be
D_L(z) \equiv \frac{H_0d_L(z)}{c}
\ee
is the Hubble-free luminosity distance~\cite{Kazantzidis:2020tko}.
However, the~parameters $M$ and $H_0$ are degenerate. Therefore, it is common either to marginalize them as nuisance parameters~\cite{SDSS:2014iwm,Scolnic:2017caz,Conley_2010} or to use a new parameter $\mathcal{M}$, which is a degenerate combination of the two:
$$\mathcal{M} \equiv M + 5\log_{10}\left[\frac{c/H_0}{1Mpc}\right] + 25 = M + 5\log_{10}\left[ \frac{299792.458km s^{-1}}{100h km s^{-1} Mpc^{-1}1Mpc}\right] =$$
\be
= M + 5 \cdot 3.47682 -5\log_{10}(h) +25 = M -5\log_{10}(h) + 42.38
\ee
to make the analysis of a cosmological model using only SNIa data. Then, the $m_{th}$ is defined in terms of $\mathcal{M}$ as:
\be
m_{th} = 5\log_{10}\left[ D_L(z)\right] + \mathcal{M}
\ee

Thus, the~appropriate $\chi^2$ function is:
\be\label{eq:3.59}
\chi^2_{Pantheon} = V^i_{Pantheon}\left[C_{ij}^{total}\right]^{-1}V^j_{Pantheon}
\ee
where the total covariance matrix $C_{ij}^{total}$ is obtained as:
\be
C_{ij}^{total} = D_{ij} + C_{systematic}
\ee
with $D_{ij}$ being the diagonal matrix:
\be
D_{ij} =\begin{pmatrix}
\sigma^2_{m_{obs,1}} & 0 & 0 & \dots\\
0 & \sigma^2_{m_{obs,2}} & 0 & \dots  \\
\vdots & \vdots &  \ddots & \vdots \\
0 & 0 & \dots & \sigma^2_{m_{obs,N}}
\end{pmatrix}
\ee
and $C_{systematic}$ is a non-diagonal matrix associated with the systematic uncertainties that emerge from the bias corrections method~\cite{Scolnic:2017caz,Kazantzidis:2020tko} and can also be  found in the same sources with the~dataset.
\subsection{\texorpdfstring{Total $\chi^2$ function}{}}
As in the BAO analysis Equation~(\ref{eq:3.27}),  the total $\chi^2$ function is obtained as the sum of all the individual $\chi^2$ functions used in the~analysis.

Thus:
\be
\chi^2_{total} = \chi^2_{CMB} + \chi^2_{BAO} + \chi^2_{Pantheon}
\ee
which is the function that we want to minimize in order to find the best-fit parameters for a given cosmological model that we want to~test.

Useful information about data analysis can be found in Appendix \ref{section:datanal}.
\section{Dark Energy~Models}\label{sec4}
The discovery of cosmic acceleration~\cite{Riess:1998cb,Perlmutter:1998np} has lead to a wide range of approaches for its description and its modeling. The~main approaches include either the introduction of a fluid  in the context of general relativity  with a parametrized equation-of-state $w(z)$ in the energy momentum tensor or the introduction of new scalar fields in the context of GR or modified gravity. {In the present analysis, we assume a flat universe. This assumption is valid since the Planck collaboration~\cite{Aghanim:2018eyx} derived a~value for the curvature density using the joint constraint from the CMB and BAO measurements: $\Omega_k = 0.001\pm 0.002$, which is consistent with a flat universe.}

\subsection{\texorpdfstring{Spatially-Flat $\Lambda$CDM~Model}{}}

The simplest form of dark energy is the cosmological constant $\Lambda$ \cite{Carroll:2000fy}. It corresponds to a time-independent energy density:
\be
\rho_\Lambda = \frac{\Lambda}{8\pi G}
\ee
obtained from an ideal fluid with equation of state $ w= -1$ and non-relativistic matter, i.e.,~cold dark matter (CDM). This corresponds to the $\Lambda CDM$ model. Assuming a flat universe in the presence of matter and a cosmological constant, Equation~(\ref{eq:1.26}) gives:
\be
H^2(z) = \frac{8\pi G}{3}\rho_m + \frac{\Lambda}{3}
\ee
or in terms of the dimensionless density parameters:
\be
H^2(z) = H_0^2\left[ \Omega_m(1+z)^3 + \Omega_\Lambda\right]
\ee
with:
\be
\Omega_m + \Omega_\Lambda = 1
\ee

This is the minimal standard model of cosmology, since the predicted Hubble parameter $H(z)$ has {only two free parameters} that can be constrained through observations.
Even though $\Lambda CDM$ is simple and is not yet excluded by the observations, it faces some challenges both theoretically and~observationally.
\subsubsection{{Theoretical~Challenges}}

 The two most notable theoretical challenges are the fine tuning problem~\cite{Weinberg:1988cp,Sahni:2002kh} and the coincidence problem~\cite{Princeton_01,Velten:2014nra}.
\paragraph{{Fine-Tuning Problem}}
Since  we now observe the cosmic acceleration, we require that the cosmological constant $\Lambda$ is of the same order of magnitude as the square of the present Hubble parameter $H_0$.
\be
\Lambda \approx H_0^2 = \left( 100h\frac{km}{s Mpc}\right)^2 = \left( 100h\cdot 3.24\times10^{-20}\frac{Mpc}{Mpc}\cdot6.58\times10^{-16}eV\right)^2
\ee

Using the unit conversion for km to Mpc and frequency $s^{-1}$ to the corresponding energy in eV, we obtain:
\be
\Lambda \approx \left( 2.132 h \times 10^{-42} GeV\right)^2
\ee

This value corresponds to an energy density:
\be
\frac{\Lambda}{3} = \rho_\Lambda \frac{8\pi G}{3}\Rightarrow \rho_\Lambda = \frac{\Lambda}{8\pi G} \approx \frac{1}{8\pi G}\left( 2.132 h \times 10^{-42} GeV\right)^2
\ee
which by substituting $h \approx 0.7$ and $G\propto m_{pl}^{-2}$ gives:
\end{paracol}
$$
\rho_\Lambda \approx \frac{1}{8\pi}\left( 2.132\cdot 0.7 \times 10^{-42} GeV\right)^2 \cdot m_{pl}^2 = \frac{2.227\times10^{-84}GeV^2\cdot1.4884\times 10^{38}GeV^2}{8\pi}$$
\begin{paracol}{2}
\switchcolumn

\be
\approx0.132\times 10^{-46}GeV^4\approx10^{-47} GeV^4
\ee

We can assume that this energy density comes from the vacuum energy of empty space, which is mathematically equivalent to the cosmological constant~\cite{Frieman:2008sn,Carroll:2000fy}. We can obtain this vacuum energy by summing over all the zero-point energies of some field with mass $m$, momentum $k$ and frequency $\omega$ up to a cut-off scale $k_{max} >> m$ \cite{Amendola:2015ksp}.
\be
\rho_{vac} = \int_0^{k_{max}} \frac{d^3k}{(2\pi)^3}\frac{\sqrt{k^2 + m^2}}{2} = \int_0^{k_{max}} \frac{4\pi k^2dk}{(2\pi)^3}\frac{1}{2}\sqrt{k^2 + m^2}
\ee

Since the integral is dominated by the large k modes $(k>>m)$:
\be
\rho_{vac} \approx \int_0^{k_{max}}\frac{4\pi k^2dk}{8\pi^3}\frac{1}{2}k = \int_0^{k_{max}}\frac{\pi k^3dk}{4\pi^3} = \frac{k_{max}^4}{16\pi^2}
\ee

A reasonable cut-off scale is the Planck scale $(m_{pl})$ up to which general relativity is believed to hold. Thus:
\be
\rho_{vac} \approx \frac{m_{pl}^4}{16\pi^2}\approx\frac{\left(1.22\times10^{19}GeV\right)^4}{16\pi^2}\approx0.014\times 10^{76} GeV^4\approx 10^{74} GeV^4 
\ee

We can see that:
\be
\frac{\rho_{vac}}{\rho_\Lambda}\approx\frac{10^{74}GeV^4}{10^{-47}GeV^4}\approx 10^{121}
\ee
which means that the predicted value is 121 orders of magnitude larger than the observed~one.

Another cut-off scale that can be used is the one when supersymmetry (if it exists) breaks at around 1 $TeV$ \cite{Farooq:2013syn}. Then:
\be
\rho_{vac} \approx \frac{\left(10^3GeV\right)^4}{16\pi^2}\approx 6.33\times10^9 GeV^4
\ee
which is a value 56 orders of magnitude larger than the observed one.
This discrepancy between the much larger initial values for the vacuum energy is also called the ``smallness'' problem~\cite{Weinberg:1988cp} and is also  referred to as the cosmological constant \mbox{problem/puzzle~\cite{Carroll:2000fy,Farooq:2013syn}.}
\paragraph{{Coincidence Problem}}
The lack of explanation for why dark energy has the same order of magnitude as non-relativistic matter density at the present epoch, which is obvious in Figure~\ref{fig:Energy densities evolution}, is called the coincidence problem~\cite{Farooq:2013syn}. This is quite bizarre since $\rho_m$ scales as $a^{-3}$, while $\rho_\Lambda$ scales as $a^0$ and:
\be
\frac{\rho_\Lambda}{\rho_m}\propto a^3
\ee
which indicates that $\Lambda$ was negligible in the past and will dominate the future. If~the cosmological constant is considered to be an initial condition in the early Universe, it seems really unlikely that $\Lambda$ should have a value comparable to matter at the present cosmological epoch while galaxies and other large-scale structures have formed~\cite{Farooq:2013syn}. A~really interesting fact is that if the cosmological constant had a couple orders of magnitude higher energy density, large-scale structure would not have formed, while if it was some orders of magnitude lower, it would not have been detectable. A~possible solution to this problem is the anthropic principle.
\paragraph{{Anthropic Principle}}
 The general idea of this principle is that physical theories should take into consideration the existence of life on Earth~\cite{Amendola:2015ksp}.
 
Carter was the one that used the expression ``Anthropic Principle'' and proposed two variants for it~\cite{1974IAUS...63..291C}. The~first is the weak anthropic principle, which states the spacetime position of life in the universe is privileged to the extent of being compatible with our existence as observers, while the second one, the~strong anthropic principle, states that the universe and the fundamental physical constants must be such as to admit the creation of observers within it at some~point.
\subsubsection{Observational~Challenges}
In general, these observational challenges occur when different observations favor different values for the same parameter, and the most notable ones are the $H_0$ and growth tension. A~more detailed analysis for these challenges along with other tensions and curiosities concerning the $\Lambda CDM$ model can be found in: \cite{Perivolaropoulos:2021jda}.
\paragraph{\texorpdfstring{$H_0$ {Tension}}{}}
As discussed in the Introduction, using a $\Lambda CDM$ background cosmology and data from the CMB and BAO measurements, a value of $H_0 = 67.4 \pm 0.5\,\rm km s^{-1}\, Mpc^{-1}$ is obtained~\cite{Aghanim:2018eyx} for the Hubble constant. On~the other hand, using local measurements coming from SnIa data leads to a value of: $H_0 = 74.03 \pm 1.42\,\rm km s^{-1} \,Mpc^{-1}$ \cite{Riess:2019cxk}. In~general, this discrepancy can range from $4.4\sigma$ to over $5\sigma$ depending on which local data are used~\cite{Wong:2019kwg,Camarena:2019moy}. More detailed presentations of the subject along with reviews of the solutions that have been proposed can be found in \cite{DiValentino:2021izs,DiValentino:2020zio}.
\paragraph{{Growth Tension}}
The amplitude of the primordial power spectrum, measured through the parameter $\sigma_8$, which is the linear amplitude of matter fluctuations on scales of $8\,\rm h^{-1}Mpc$, and~the matter density parameter $\Omega_{0m}$ are the two parameters that affect the growth rate and magnitude of linear cosmological perturbations~\cite{Kazantzidis:2020tko}. Dynamical probes for the cosmological perturbations' growth rate, which measure it directly, such as weak lensing~\cite{Kohlinger:2017sxk,Joudaki:2017zdt,Abbott:2017wau} and redshift space distortion~\cite{Kazantzidis:2019dvk,Macaulay:2013swa,Tsujikawa:2015mga}, indicate that the observed growth rate is weaker than the theoretical prediction obtained in the context of a $\Lambda CDM$ background using the observed background expansion rate measured through Type Ia supernovae, BAO and CMB data, which are geometric probes. Dynamical probes favor smaller values for both $\Omega_{0m}$ and $\sigma_8$ than the geometric ones, and this discrepancy varies from around 2--3 $\sigma$ \cite{Johnson:2015aaa,Kazantzidis:2018rnb,Nesseris:2017vor,Nunes:2021ipq} to even more than $5\sigma$ \cite{Skara:2019usd} depending on the model parametrization and the dataset used. However, if~the constraints from the CMB on the $\Lambda CDM$ background are not taken into account, leaving only the constraints from the SNIa data, the~tension decreases to a level below $2\sigma$ \cite{LHuillier:2017ani}. The tension also decreases when marginalized confidence contours are used~\cite{Quelle:2019vam} and can be resolved completely within the Minimal theory of massive gravity scenario \cite{deAraujo:2021cnd}. A brief analysis of the growth tension along with some proposed solutions can be found in \cite{DiValentino:2020vvd}.

\subsubsection{Fitting the \lcdm parameters: Maximum~Likelihood}
\paragraph{BAO and CMB Data}
In the Mathematica code used, we construct the $\chi^2$ function for the model: $\chi^2(\Omega_{0m},h)$ where we use the value: $\Omega_{b}h^2 = 0.02236$ from~\cite{Aghanim:2018eyx}, with~which we perform two separate analyses: one solely with BAO data and one with both the BAO and the CMB data.
At first, we use only the BAO data:
\be
\chi^2(\Omega_{0m},h) = \chi^2_{BAO}
\ee
with $\chi^2_{BAO}$ defined in Equation~(\ref{eq:3.27}).

Using both the BAO and CMB data, we obtain:
\be
\chi^2(\Omega_{0m},h) = \chi^2_{BAO}+\chi^2_{CMB}
\ee
with $\chi^2_{CMB}$ defined in Equation~(\ref{eq:3.36}).
\paragraph{{SNIa Data}}
In this analysis, we use the Pantheon dataset and try to constrain the parameter $\Omega_{0,m}$ and the degenerate combination $\mathcal{M}$. Thus, we construct the $\chi^2 $ function as:
\be
\chi^2 (\Omega_{0,m},\mathcal{M}) = \chi^2_{Pantheon}
\ee
with $\chi^2_{Pantheon}$ defined in Equation~(\ref{eq:3.59}).

\paragraph{{Combined Data}}
At last, we combine all the data, and the new $\chi^2$ function is defined as:
\be
\chi^2(\Omega_{0m},\mathcal{M},h) = \chi^2_{Pantheon} + \chi^2_{BAO} +\chi^2_{CMB}
\ee
since BAO and CMB measurements explicitly constrain the Hubble constant. 
Implementing the maximum likelihood method~\cite{Verde:2009tu,10.5555/1403886,Arjona:2018jhh}{, i.e.,~minimizing the $\chi^2$ function over all free parameters}, we obtain the best-fit values for each data combination, which can be seen in  Table~\ref{tab: results_01}.
\begin{specialtable}[H]
\caption{\textls[-25]{Table containing the best-fit parameters for the $\Lambda CDM$ model for different data~combinations.}}
\label{tab: results_01}
\setlength{\cellWidtha}{\columnwidth/4-2\tabcolsep-0.9in}
\setlength{\cellWidthb}{\columnwidth/4-2\tabcolsep+0.3in}
\setlength{\cellWidthc}{\columnwidth/4-2\tabcolsep+0.3in}
\setlength{\cellWidthd}{\columnwidth/4-2\tabcolsep+0.3in}
\scalebox{1}[1]{\begin{tabularx}{\columnwidth}{>{\PreserveBackslash\centering}m{\cellWidtha}>{\PreserveBackslash\centering}m{\cellWidthb}>{\PreserveBackslash\centering}m{\cellWidthc}>{\PreserveBackslash\centering}m{\cellWidthd}}

 \toprule
 
 \multirow{2}{2em}{\centering  }&\multicolumn{3}{c}{\boldmath{$\Lambda CDM$}}\\

& \textbf{BAO + CMB} & \textbf{SNIa} & \textbf{Combined} \\
\midrule

 $\Omega_{0m}$& $0.3178 \pm 0.0059$ & $0.299 \pm 0.022$ & $0.3169 \pm 0.0057$  \\
 $\mathcal{M}$ & $-$ & $23.809 \pm 0.011 $ & $23.817 \pm 0.049 $  \\
 $h$ & $0.6718 \pm 0.0039 $ & $-$ & $0.6724 \pm 0.0038$  \\
 $\chi^2$ & 6.3927 & 1025.63 & 1032.7 \\
\bottomrule
\end{tabularx}}
\end{specialtable}
\paragraph{{Results}}
The constraints obtained using the Pantheon dataset  are in agreement with corresponding previous studies~\cite{Scolnic:2017caz,Kazantzidis:2020tko,Zhao:2019azy}.

From Figure~\ref{fig:LCDM_01}, we can see that the BAO measurements cannot significantly constrain  the parameters $h$ and $\Omega_m$, while the addition of the CMB measurements dramatically improves the constraints. As~expected, the preferred value of $h$ is consistent with (\ref{P18}) and inconsistent with (\ref{hr20}), which demonstrates the $H_0$ tension. Since the best-fit value of $\cal M$ remains practically unaffected by the addition of BAO and CMB data, we conclude that the $M$-tension remains present.

Figure~\ref{fig:LCDM_02}  shows the constraints obtained from the Pantheon dataset. When the measurements from the geometric probes (BAO and CMB) are added, the~constraints on  $\Omega_m$ are dramatically improved, while the constraints on $\cal M$ are practically~unaffected.
\begin{figure}[H]
\includegraphics[scale=0.6]{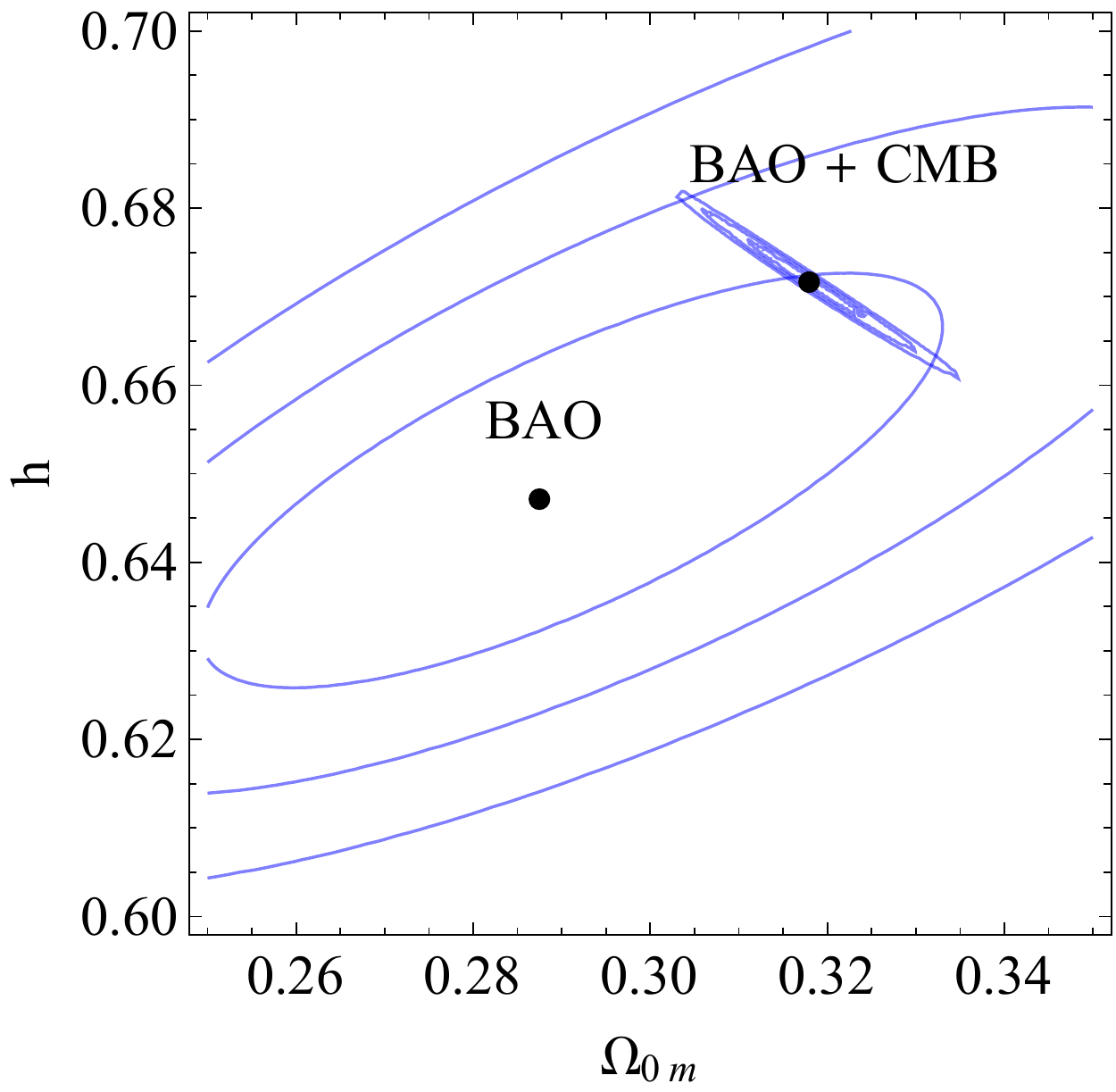}
\caption{The $1\sigma-3\sigma$ confidence contours in the parametric space $(h-\Omega_{0m})$ for the BAO data and the BAO + CMB data combination. Notice the dramatic improvement of the constraints when the CMB data points are~included.}
\label{fig:LCDM_01}
\end{figure}
\vspace{-6pt}
\begin{figure}[H]
\includegraphics[width = 10cm]{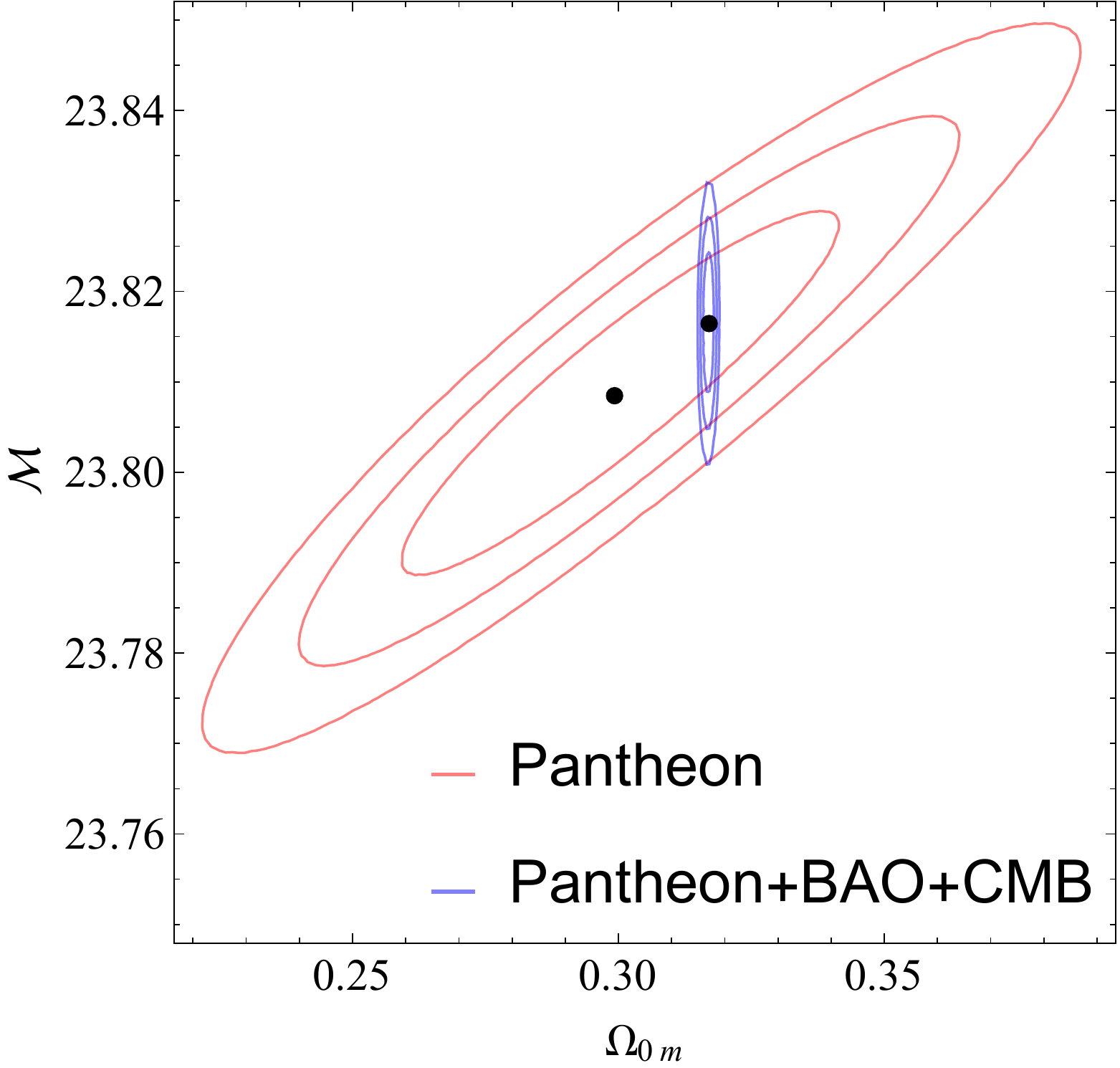}
\caption{The $1\sigma - 3\sigma$ confidence contours in the parametric space $(\mathcal{M}-\Omega_{0m})$.The red contour corresponds to the full Pantheon dataset while the blue contour corresponds to the Pantheon+CMB+BAO data combination. As~expected, when the CMB+BAO data are included the constraints on $\Omega_{0m}$ improve~dramatically. }
\label{fig:LCDM_02}
\end{figure}
\subsection{Spatially-Flat wCDM~Model}

This generic parametrization of the dark energy equation-of-state parameter is of \mbox{the form}
\be
w_0 = \frac{p_{DE}}{\rho_{DE}}< -\frac{1}{3}
\ee
which is assumed to be an arbitrary constant. When $w_0 = -1$, the $WCDM$ model reduces to the $\Lambda CDM$ model. Assuming a flat universe in presence of matter and a spatially- homogeneous fluid with $w<-1/3$, Equation~(\ref{eq:1.26})  gives:
\be
H^2(z) = \frac{8\pi G}{3}\left( \rho_m + \rho_{DE} \right)
\ee
or in terms of the dimensionless density parameters:
\be
H^2(z) = H^2_0\left[ \Omega_m(1+z)^3 + \Omega_{DE}(1+z)^{3(1+w_0)}\right]
\ee
with:
\be
\Omega_m + \Omega_{DE} = 1
\ee

In order to impose constraints on the model parameters, we  use the combined data  from the Pantheon dataset and BAO and CMB measurements. However, instead of $\mathcal{M}$, we consider  $M$ and $h$ separately, and thus, the $m_{th}$ used in the Pantheon dataset analysis is the one defined in Equation~(\ref{eq:3.55}). Thus, the~constructed $\chi^2$ function is \mbox{defined as:}
\be
\chi^2(\Omega_{0m},\mathcal{M},w_0,h) = \chi^2_{Pantheon} + \chi^2_{BAO} +\chi^2_{CMB}
\ee

Implementing the maximum likelihood method, we obtain the best-fit values, which can be seen in Table~\ref{tab: results_02}. {In order to produce the contour plots for the $wCDM$ model, we vary all four parameters in the $\chi^2$ function and then we show a two-dimensional projection of the ellipsoid produced in the four-dimensional parameter space.}

\begin{specialtable}[H]
\caption{Table containing the best-fit values for the $\Lambda CDM$, $wCDM$ and $CPL$ models using both the CMB and BAO measurements and data from Type Ia~supernovae.}
\label{tab: results_02}
\setlength{\cellWidtha}{\columnwidth/4-2\tabcolsep-0.9in}
\setlength{\cellWidthb}{\columnwidth/4-2\tabcolsep+0.3in}
\setlength{\cellWidthc}{\columnwidth/4-2\tabcolsep+0.3in}
\setlength{\cellWidthd}{\columnwidth/4-2\tabcolsep+0.3in}
\scalebox{1}[1]{\begin{tabularx}{\columnwidth}{>{\PreserveBackslash\centering}m{\cellWidtha}>{\PreserveBackslash\centering}m{\cellWidthb}>{\PreserveBackslash\centering}m{\cellWidthc}>{\PreserveBackslash\centering}m{\cellWidthd}}
 \toprule
 
 \multirow{2}{2em}{\centering  }&\multicolumn{1}{c}{\boldmath{$\Lambda CDM$}}&\multirow{2}{4em}{\centering \textbf{wCDM}}&\multirow{2}{4em}{\centering \textbf{CPL}}\\

& \textbf{Combined}&  & \\
\midrule

 $\Omega_{0m}$ & $0.3169 \pm 0.0057$  & $0.315 \pm 0.008$&$0.315 \pm 0.013$\\

 $w_0$& $-1$  & $-1.01 \pm 0.03$&$-1.07 \pm 0.15$\\
 $w_a$ &$-$& $-$ & $0.24 \pm 0.47$ \\
 $\mathcal{M}$  & $ 23.812 \pm 0.006$  & $-$&$-$\\
 $M$  & $-$  & $ -19.42 \pm 0.02$&$ -19.43 \pm 0.02$\\
 $h$  & $0.6724 \pm 0.0038$  & $0.675 \pm 0.008$& $0.674 \pm 0.011$\\
 $\chi^2$ & 1032.7  & 1032.6& 1031.97\\
\bottomrule
\end{tabularx}}
\end{specialtable}
\paragraph{{Results}}
When we compare the values favored by the data for the $wCDM$ model with the respective values for the $\Lambda CDM$ model, we can see that it favors smaller values for the equation-of-state $w_0$ and the matter density $\Omega_{0,m}$ along with higher values for the dimensionless Hubble parameter $h$, while giving, at~the same time, a~slightly better fit that can be seen from the lower value of $\chi^2$. In~addition, the favored range of the SnIa absolute magnitude $M$ remains inconsistent with the value range indicated by local Cepheid calibrators (\mbox{Equation \eqref{R20M}}), i.e.,~the $M$-tension remains along with the Hubble tension, as shown in Figure~\ref{fig: wcdm_res_001}.
\begin{figure}[H]
     \begin{subfigure}[b]{4.35cm}
     \captionsetup{justification=centering}
         \includegraphics[width=4.35cm]{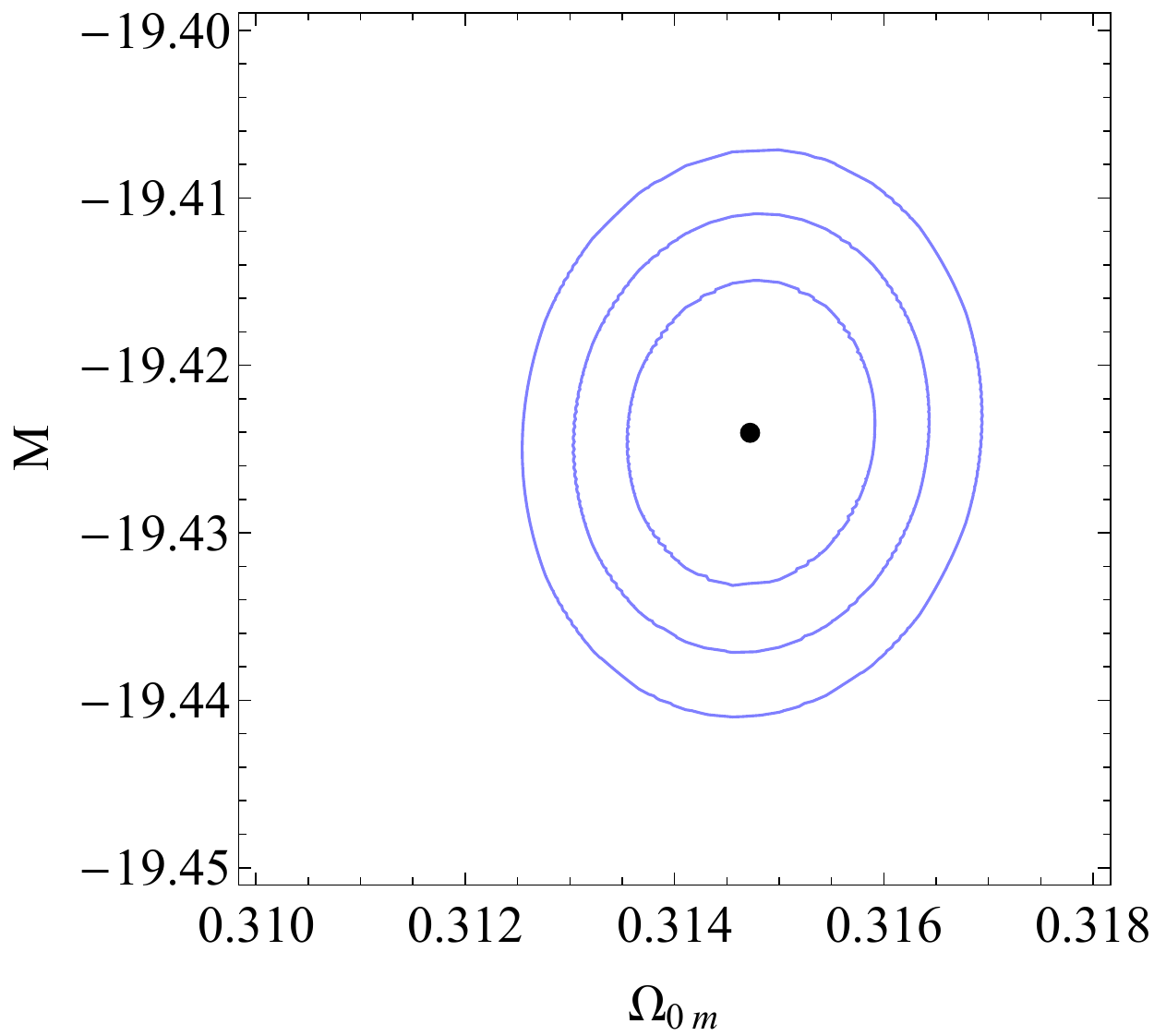}
         \caption{}
         \label{fig:wcdm_001}
     \end{subfigure}
     \hfill
     \begin{subfigure}[b]{8cm}
     \captionsetup{justification=centering}
         \includegraphics[width=8cm]{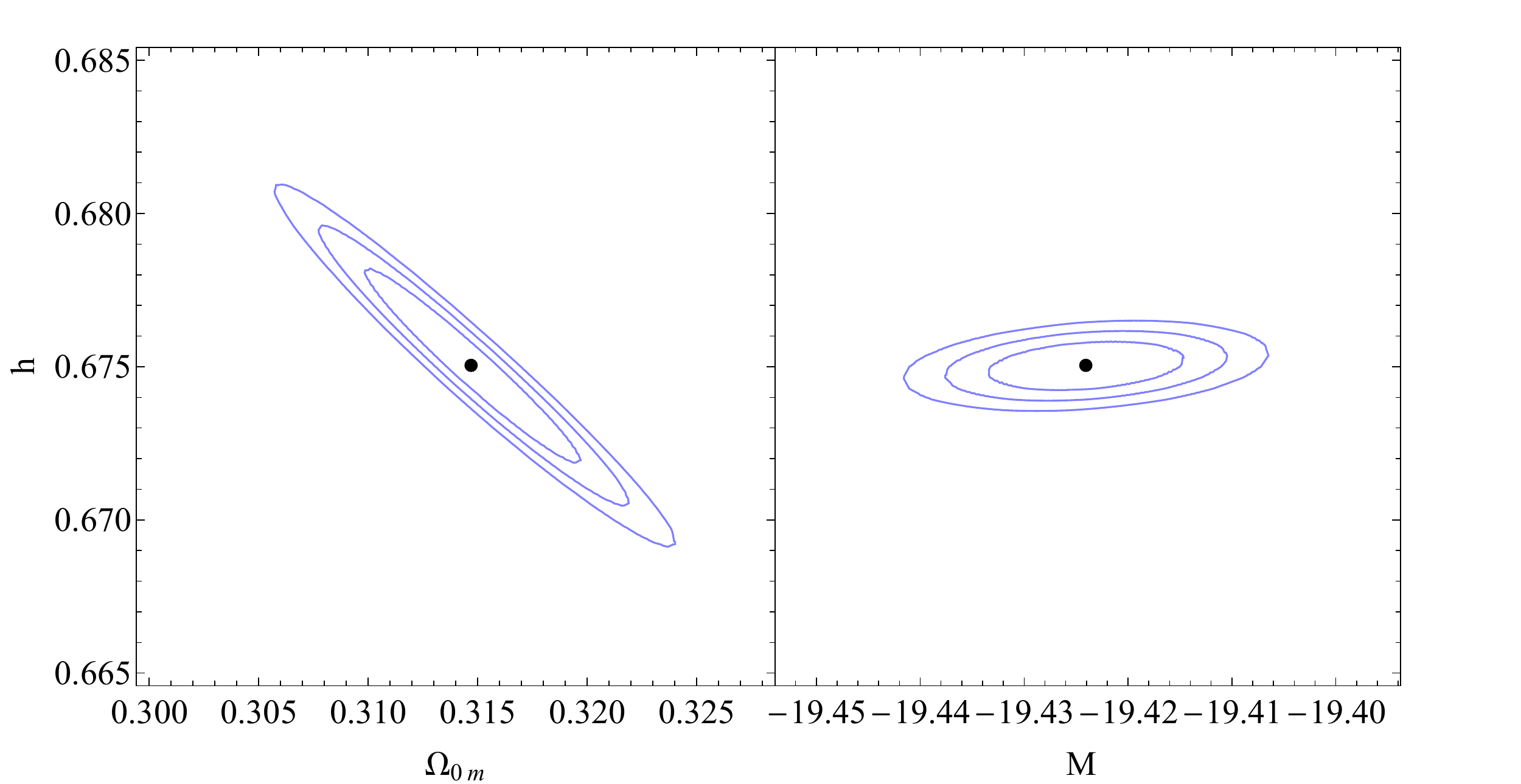}
         \caption{}
         \label{fig:wcdm_00002}
     \end{subfigure}
     \begin{subfigure}[b]{13.7cm}
     \captionsetup{justification=centering}
         \includegraphics[width=13.7cm]{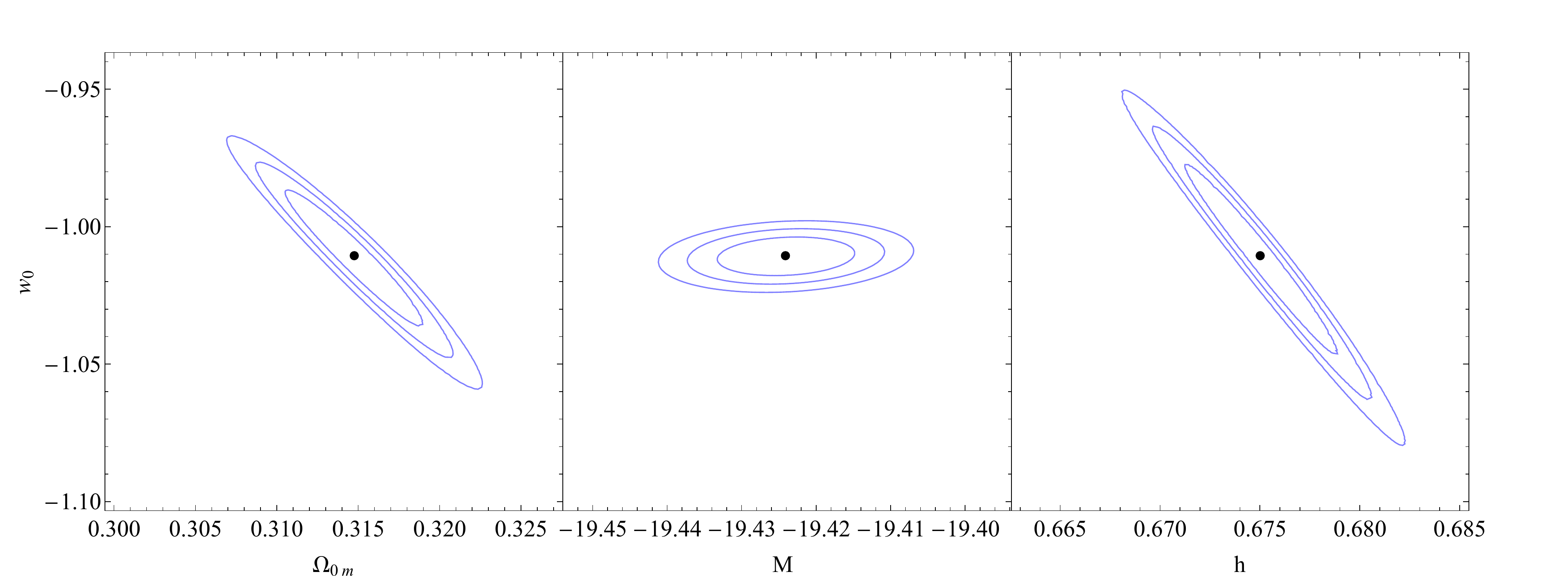}
         \caption{}
         \label{fig:wcdm_004}
     \end{subfigure}
        \caption{Confidence contours for the wCDM model: (\textbf{a}) The $1\sigma - 3\sigma$ confidence contour in the parametric space $(M-\Omega_{0m})$. (\textbf{b}) The $1\sigma - 3\sigma$ confidence contours in the parametric spaces $(h-\Omega_{0m})$ and $(h-M)$. (\textbf{c}) The $1\sigma - 3\sigma$ confidence contours in the parametric spaces $(w_0-\Omega_{0m})$, $(w_0-M)$ and $(w_0-h)$.}
        \label{fig: wcdm_res_001}
\end{figure}
\subsection{Chevallier--Polarski--Linder (CPL) Parametrization}
A commonly used ansatz
, proposed by Chevalier-Polarski~\cite{Chevallier:2000qy} and Linder~\cite{Linder:2002et}, allows for dynamical dark energy and is based on an expansion of $w(a)$ around the present value of the scale factor $a=1$:
\be
w(a) = w_0 + w_a (1-a) 
\ee

This can be derived as the Taylor series expansion of $w(a)$ around $a = 1$ including only linear~terms:

\vspace{-12pt} 
\be
w(a) = w(a = 1) + (a-1)w'(a = 1)+\frac{1}{2}(a-1)^2w''(a = 1)+\mathcal{O}\left[(a-1)^3\right]
\ee
where $w'$ denotes the derivative in terms of the scale factor $a$.

In terms of the redshift $z$, the~equation-of-state $w(z)$ takes the form:
\be
w(z) = w_0+w_a\left( 1-\frac{1}{1+z}\right) = w_0+w_a\frac{z}{1+z}
\ee

Its present value  $w(a_0)$ is: 
\be
w(a_0) = w_0 + w_a(1-a_0) = w_0
\ee
and the value of its slope:

\be
w'(a) = \frac{dw(a)}{da} = -w_a
\ee

Again, implementing the maximum likelihood method, we obtain the best-fit values, which can be seen in  Table~\ref{tab: results_02}. {In order to produce the contour plots for the $CPL$ model, we vary all five parameters in the $\chi^2$ function, and then we show a two-dimensional projection of the ellipsoid produced in the five-dimensional parameter space.}
\begin{figure}[H]
     \begin{subfigure}[b]{4.3cm}
     \captionsetup{justification=centering}
         \includegraphics[width=4.3cm]{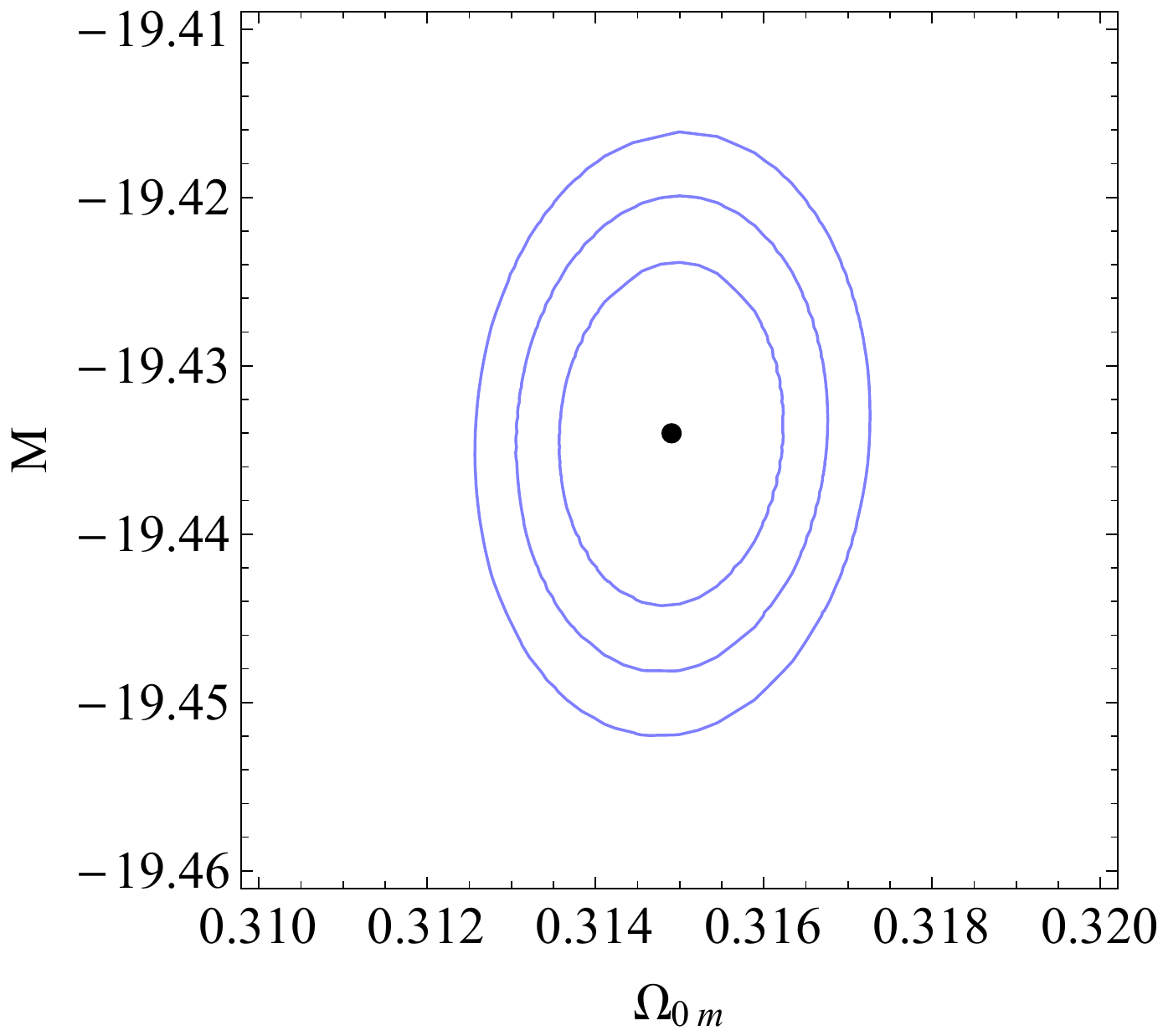}
         \caption{}
         \label{fig:cpl_001}
     \end{subfigure}
     \hfill
     \begin{subfigure}[b]{8cm}
     \captionsetup{justification=centering}
         \includegraphics[width=8cm]{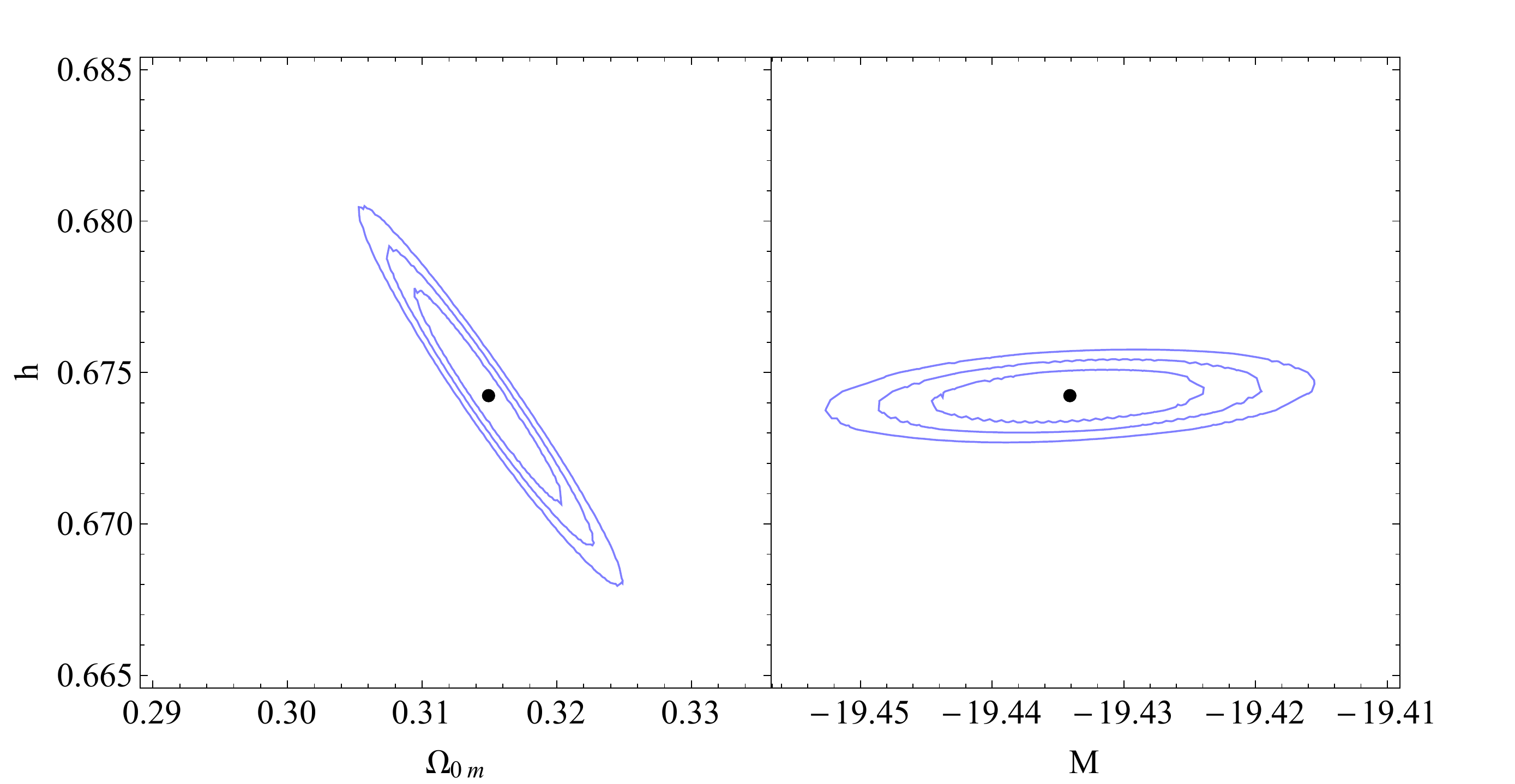}
         \caption{}
         \label{fig:cpl_002}
     \end{subfigure}
     \begin{subfigure}[b]{12.7cm}
     \captionsetup{justification=centering}
         \includegraphics[width=12.7cm]{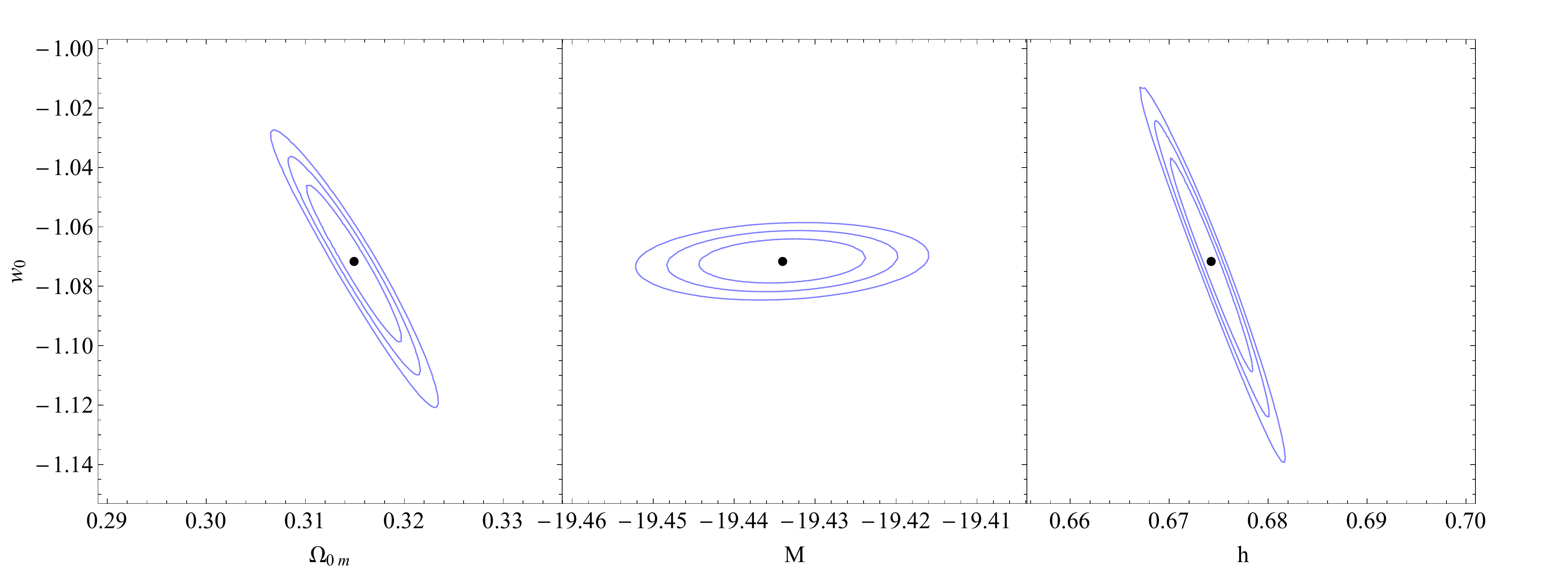}
         \caption{}
         \label{fig:cpl_003}
     \end{subfigure}
        \caption{{Confidence} contours for the CPL model: (\textbf{a}) The $1\sigma - 3\sigma$ confidence contour in the parametric space $(M-\Omega_{0m})$. (\textbf{b}) The $1\sigma - 3\sigma$ confidence contours in the parametric spaces $(h-\Omega_{0m})$ and $(h-M)$. (\textbf{c}) The $1\sigma - 3\sigma$ confidence contours in the parametric spaces $(w_{0}-\Omega_{0m})$, $(w_{0}-M)$ and $(w_{0}-h)$.}
        \label{fig: cpl_res_001}
\end{figure}
\begin{figure}[H]
\includegraphics[width=14cm]{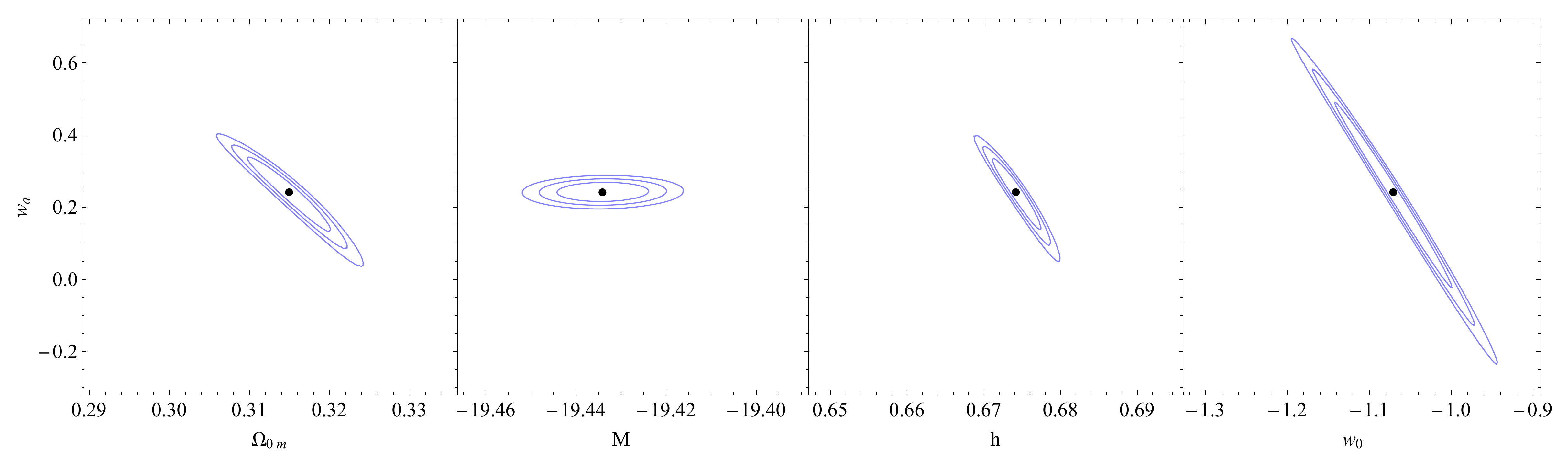}
\caption{{The} $1\sigma - 3\sigma$ confidence contours in the parametric spaces: $(w_a-\Omega_{0m})$, $(w_a-M)$, $(w_a-h)$ and $(w_a-w_0)$ for the CPL~model.}
\label{fig:cpl_kappa}
\end{figure}

\paragraph{{Results}}

When we compare the values favored by the data for the $CPL$ model with the respective values for the $\Lambda CDM$ and $wCDM$ models, we can see that $CPL$  favors even smaller values for the equation of state $w_0$ at the present time. Furthermore, it gives a value for the matter density $\Omega_{0,m}$ around that predicted from the $wCDM$ analysis with  for the dimensionless Hubble parameter $h$, indicating also the presence of the Hubble tension. The~$M$ tension also remains  since the preferred value of $M$ is significantly lower than the corresponding Cepheid value, as seen in Figure \ref{fig: cpl_res_001}. Furthermore, the~positive values preferred for $w_a$, as can be seen in Figure \ref{fig:cpl_kappa}, indicate a crossing of the phantom divide line at $w=-1$. Finally, despite the additional parameters involved, the~quality of fit is similar to that of \lcdm and $wCDM$ since $\chi^2$ is similar in all three parametrizations, as shown in Table~\ref{tab: results_02}.

\subsection{Adding the Local $H_0$ Determination}
Taking into account the locally determined the value of the Hubble constant (\ref{hr20})) (hereafter abbreviated as "the Riess data point") corresponds to adding a term in $\chi^2$ as~follows
\be
\chi^2_{Riess}(h) = \frac{(h-h_{Riess})^2}{\sigma_{Riess}^2} = \frac{(h-0.7403)^2}{0.0142^2}
\ee
to observe how the best-fit parameters are affected.
Thus, the new $\chi^2$ function is:
\be
\chi^2_{new} = \chi^2_{CMB} + \chi^2_{BAO} + \chi^2_{Pantheon} + \chi^2_{Riess}
\ee

The new contours produced with the use of the new $\chi^2$ function for every model can be seen in Figures~\ref{fig:Riess_lcdm}  and \ref{fig: wcdm_riess_001}, while we obtain the best-fit values, which can be seen in  Table~\ref{tab: results_03}, by~implementing the maximum likelihood method.

 \begin{figure}[H]
     \begin{subfigure}[b]{6.55cm}
     \captionsetup{justification=centering}
         \includegraphics[width=6.5cm]{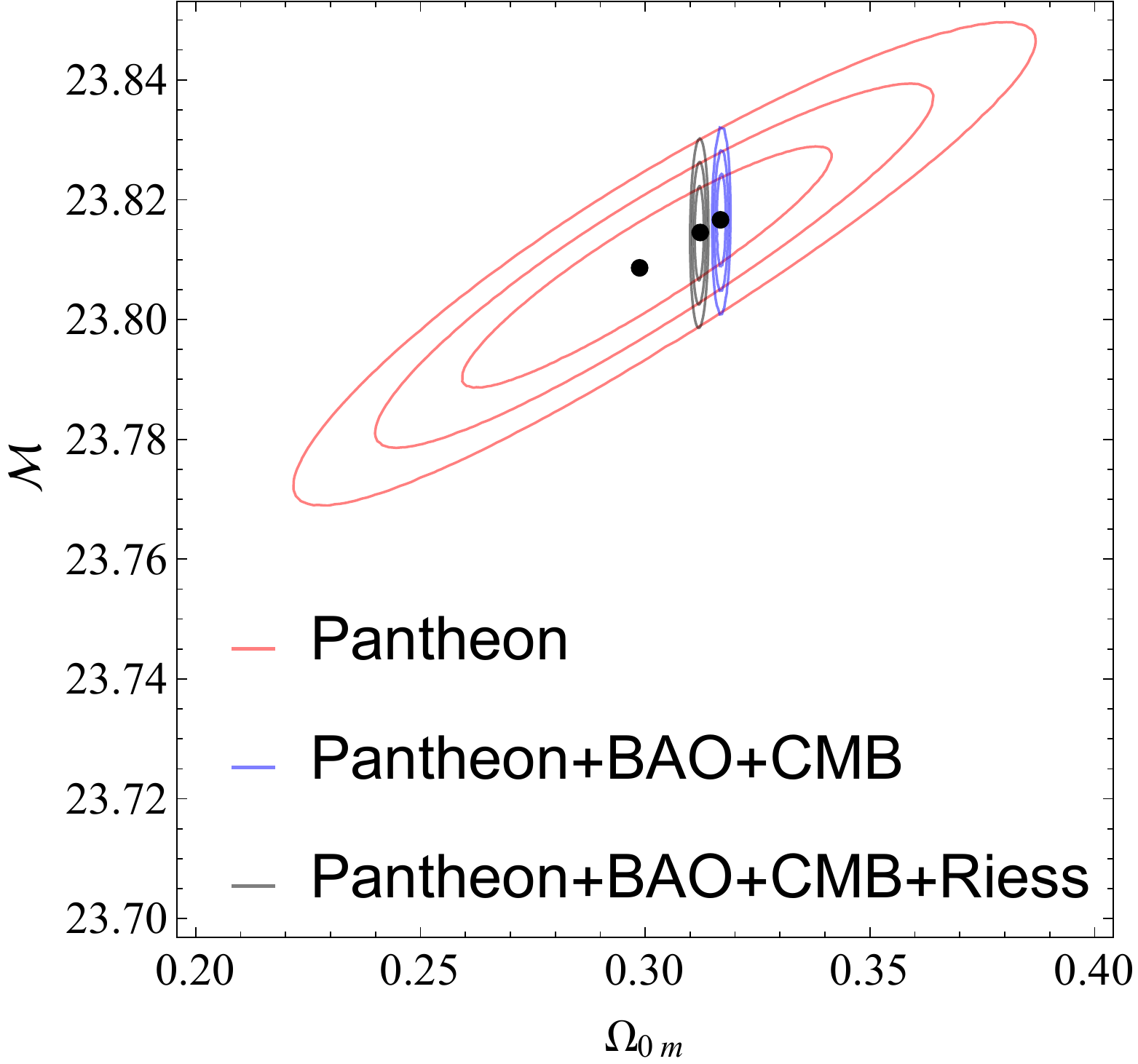}
         \caption{}
         \label{fig:lcdm_riess_01}
     \end{subfigure}
     \begin{subfigure}[b]{6.55cm}
     \captionsetup{justification=centering}
         \includegraphics[width=6.5cm]{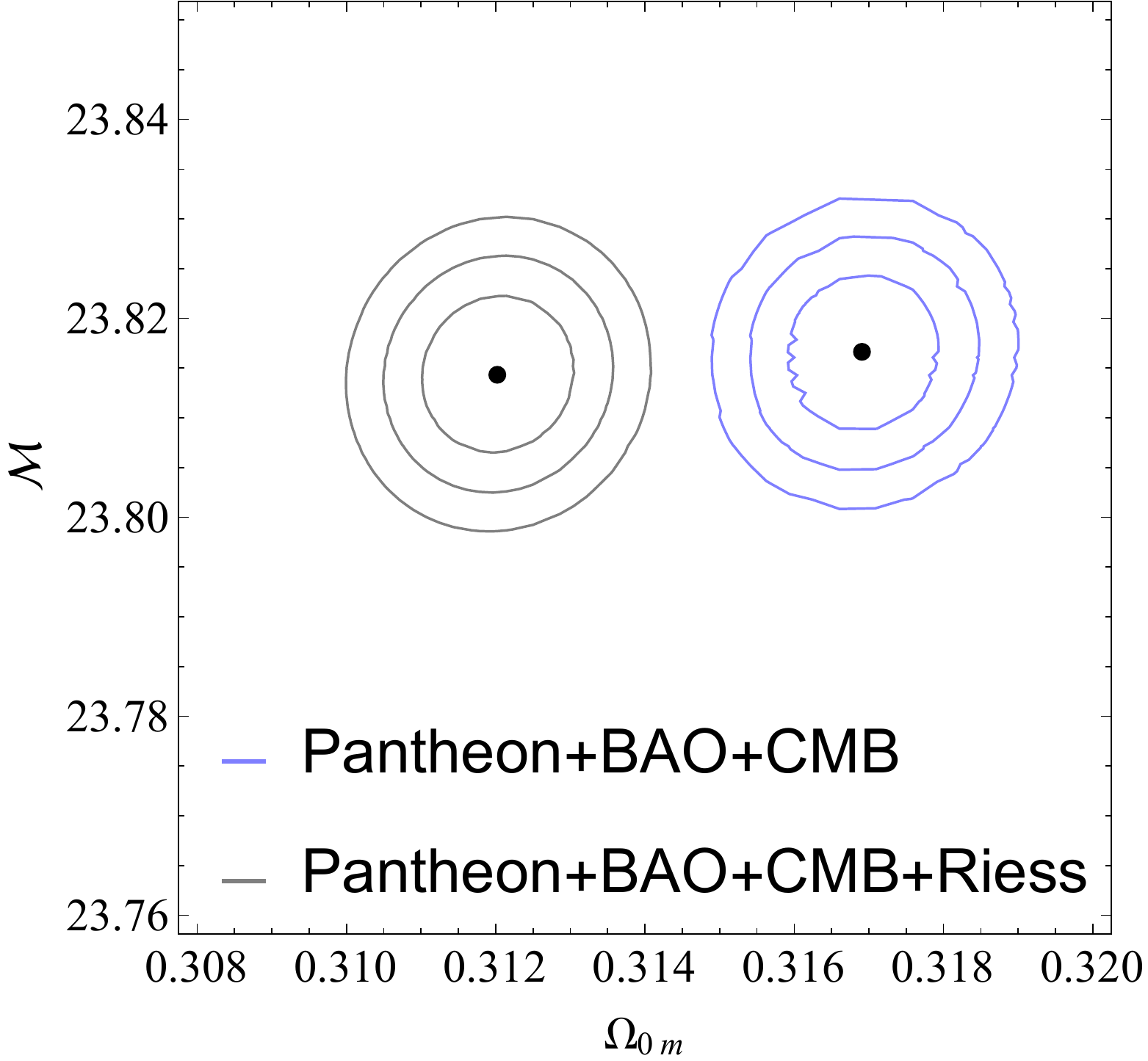}
         \caption{}
         \label{fig:lcdm_riess_02}
     \end{subfigure}
        \caption{Confidence contours for the $\Lambda CDM$ model with the addition of the Riess point: (\textbf{a}) The $1\sigma - 3\sigma$ confidence contour in the parametric space $(\mathcal{M}-\Omega_{0m})$ for SNIa data, SNIa+BAO+CMB and SNIa+BAO+CMB+Riess data. (\textbf{b}) The $1\sigma - 3\sigma$ confidence contour in the parametric space $(\mathcal{M}-\Omega_{0m})$ for the SNIa+BAO+CMB and SNIa+BAO+CMB+Riess~data.}
        \label{fig:Riess_lcdm}
\end{figure}
\vspace{-6pt}
\begin{specialtable}[H]
\caption{Table containing the best-fit values for the $\Lambda CDM$, $wCDM$ and $CPL$ models using both the CMB and BAO measurements, data from Type Ia supernovae and the Riess~point.}
\label{tab: results_03}
\setlength{\cellWidtha}{\columnwidth/4-2\tabcolsep-0.9in}
\setlength{\cellWidthb}{\columnwidth/4-2\tabcolsep+0.3in}
\setlength{\cellWidthc}{\columnwidth/4-2\tabcolsep+0.3in}
\setlength{\cellWidthd}{\columnwidth/4-2\tabcolsep+0.3in}
\scalebox{1}[1]{\begin{tabularx}{\columnwidth}{>{\PreserveBackslash\centering}m{\cellWidtha}>{\PreserveBackslash\centering}m{\cellWidthb}>{\PreserveBackslash\centering}m{\cellWidthc}>{\PreserveBackslash\centering}m{\cellWidthd}}

 \toprule
 &\boldmath{$\Lambda CDM$}&{ \textbf{{wCDM}}}&{ \textbf{CPL}}\\
\midrule

 $\Omega_{0m}$ & $0.3120 \pm 0.0055$  & $0.299 \pm 0.007$&$0.298 \pm 0.013$\\
 $w_0$& $-1$  & $-1.07 \pm 0.03$&$-1.12 \pm 0.16$\\
 $w_a$ &$-$& $-$ & $0.19 \pm 0.56$ \\
 $\mathcal{M}$  & $ 23.814 \pm 0.049$  & $-$&$-$\\
 $M$  & $-$  & $ -19.39 \pm 0.02$&$ -19.39 \pm 0.02$\\
 $h$  & $0.6757 \pm 0.0037$  & $0.693 \pm 0.008$& $0.693 \pm 0.011$\\
 $\chi^2$ & 1054.48  & 1047.86& 1047.59\\
\bottomrule
\end{tabularx}}
\end{specialtable}
\vspace{-6pt}
 \begin{figure}[H]
     \begin{subfigure}[b]{4.25cm}
     \captionsetup{justification=centering}
         \includegraphics[width=4.25cm]{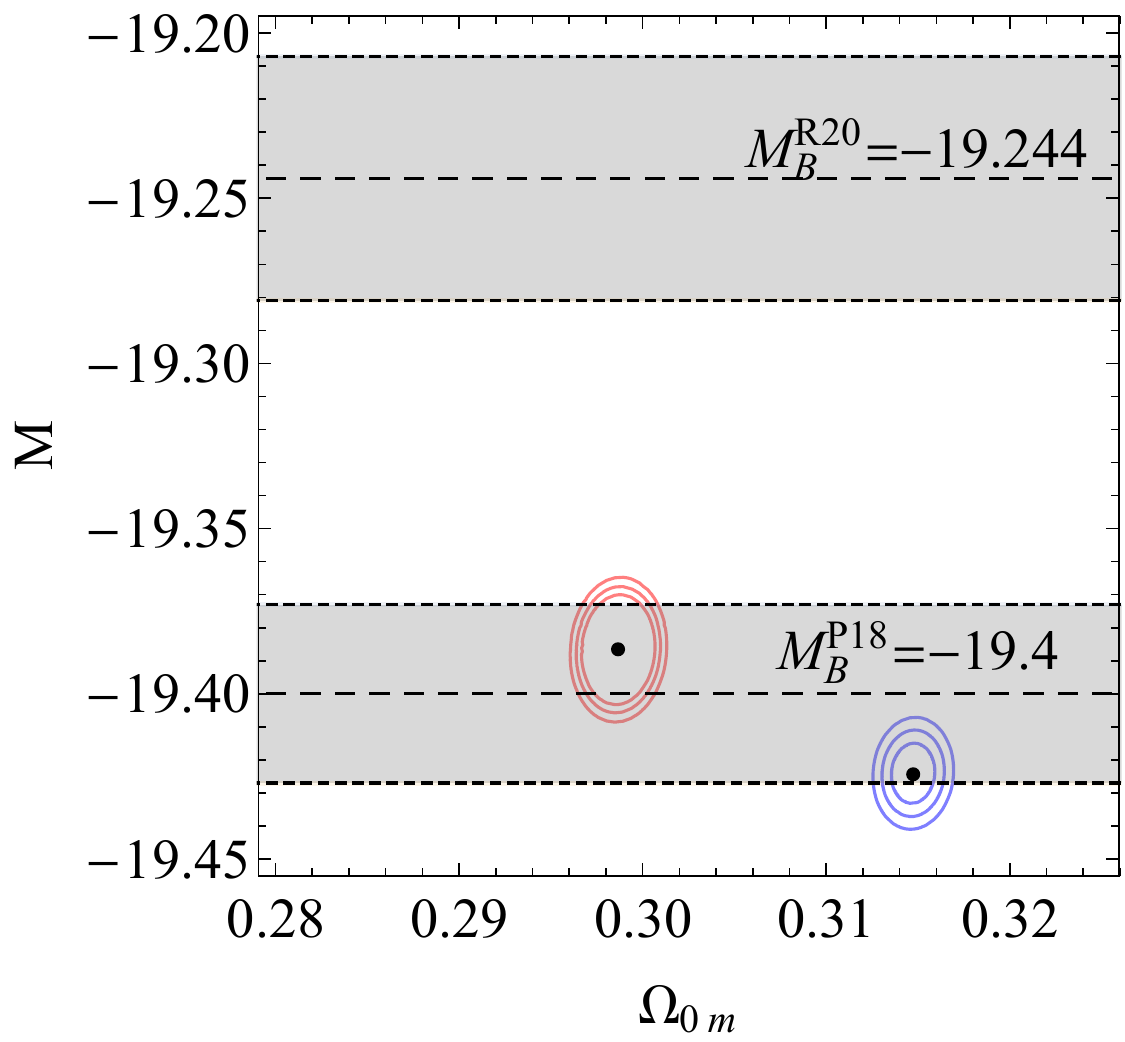}
         \caption{}
         \label{fig:wcdm_riess_001}
     \end{subfigure}
     \hfill
     \begin{subfigure}[b]{8cm}
     \captionsetup{justification=centering}
         \includegraphics[width=8cm]{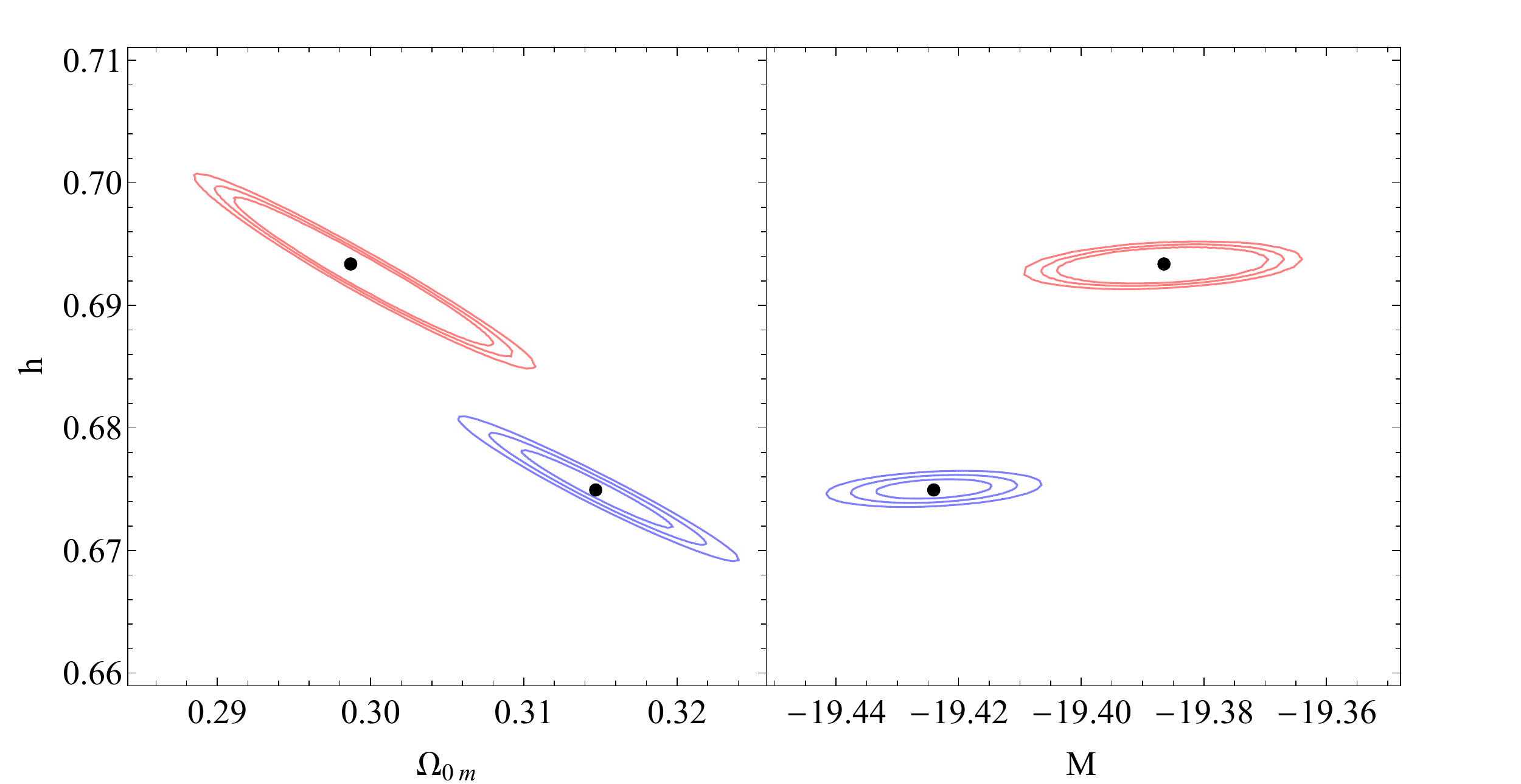}
         \caption{}
         \label{fig:wcdm_riess_00002}
     \end{subfigure}
     \begin{subfigure}[b]{12.8cm}
     \captionsetup{justification=centering}
         \includegraphics[width=12.8cm]{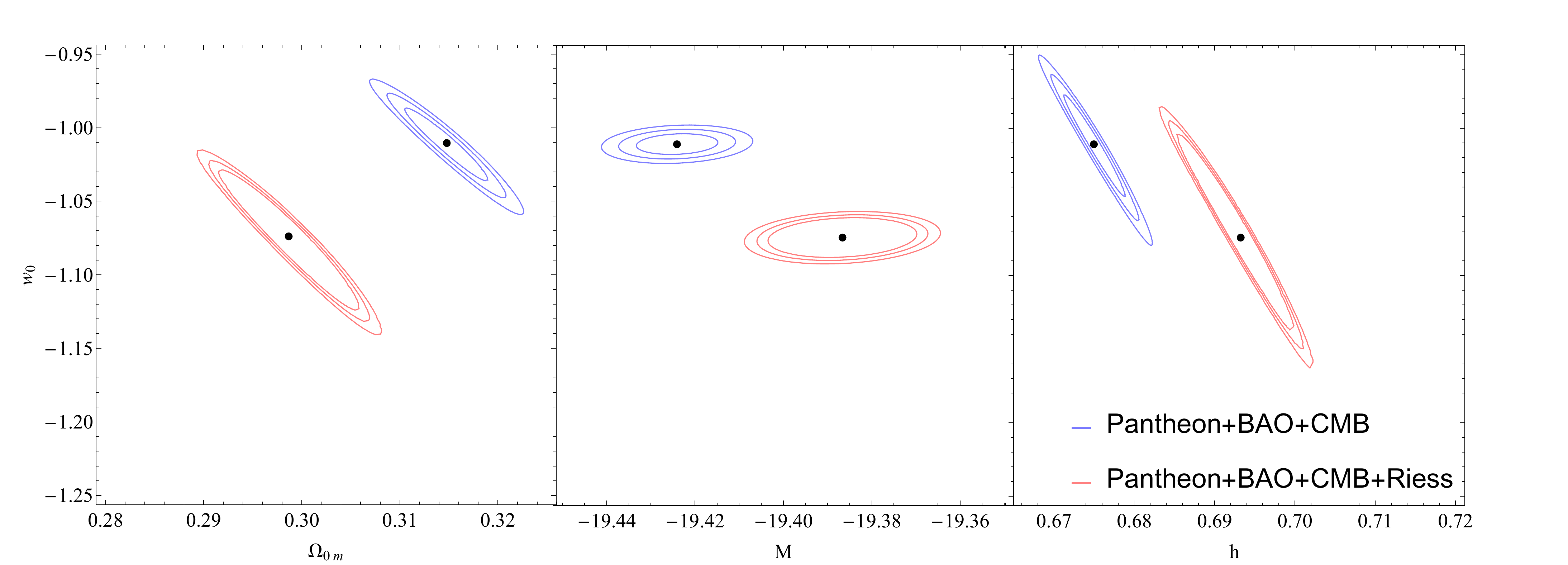}
         \caption{}
         \label{fig:wcdm_riess_004}
     \end{subfigure}
        \caption{Confidence contours for the wCDM model with the addition of the Riess point: \mbox{(\textbf{a}) The} $1\sigma - 3\sigma$ confidence contour in the parametric space $(M-\Omega_{0m})$. (\textbf{b}) The $1\sigma - 3\sigma$ confidence contours in the parametric spaces $(h-\Omega_{0m})$ and $(h-M)$. (\textbf{c}) The $1\sigma - 3\sigma$ confidence contours in the parametric spaces $(w_0-\Omega_{0m})$, $(w_0-M)$ and $(w_0-h)$.}
        \label{fig: wcdm_riess_001}
\end{figure}
\vspace{-6pt}

\subsubsection{Results}

 \paragraph{\texorpdfstring{$\Lambda${CDM}}{}}  
 In Figure~\ref{fig:Riess_lcdm} and in Table~\ref{tab: results_03}, we can see that adding the Riess point leads to slightly  smaller values for both $\mathcal{M}$ and matter density $\Omega_{0,m}$. The~quality of fit also worsens significantly since $\chi^2$ increases by about 22 units compared to the case when the Riess point is not~included.

 \paragraph{{wCDM}}
 In Figure~\ref{fig: wcdm_riess_001}a  and in Table~\ref{tab: results_03} we can see that the addition of the Riess point indicates that the best-fit absolute remains in tension with the value indicated by local Cepheid calibrators $M^{R20}$. In~addition, the~new data combination prefers a somewhat higher value for the Hubble constant and a lower value for {the} matter density. However, the~Hubble and $M$ tensions remain despite the introduction of the Riess point.  Lastly, $wCDM$  in this case favors a lower value for the equation of state $w$, thus moving further away from the cosmological constant value $w=-1$. The~quality of fit also worsens significantly since $\chi^2$ increases by about 15  units compared to the case when the Riess point is not included. This worsening is not as bad as in the case of \lcdm but indicates that the Riess point is clearly not consistent with the rest of the cosmological data even in the context of $wCDM$.
 \paragraph{{CPL}}
From Figure~\ref{fig: cpl_riess_001}a,  it is clear that the addition of the Riess point has similar effects in the $CPL$ as in $wCDM$ model; somewhat higher values for the Hubble constant, lower values for the matter density,  slightly higher SNIa absolute magnitude, and lower value for the equation of state at the present epoch $w_0$. Lastly, the~parameter $w_a$ remains almost unaffected as the change in its best-fit value is really small. Clearly, however, the Hubble and $M$ tensions remain for this $H(z)$ deformation~model. 

The quality of fit is very similar as in the case of $wCDM$ despite the {extra} parameter. It is also  significantly worse compared to the case when the Riess point is not included. This worsening is not as bad as in the case of \lcdm but indicates that the Riess point is clearly not consistent with the rest of the cosmological data also in the context of $CPL$.
\begin{figure}[H]
     \begin{subfigure}[b]{4.25cm}
     \captionsetup{justification=centering}
         \includegraphics[width=4.25cm]{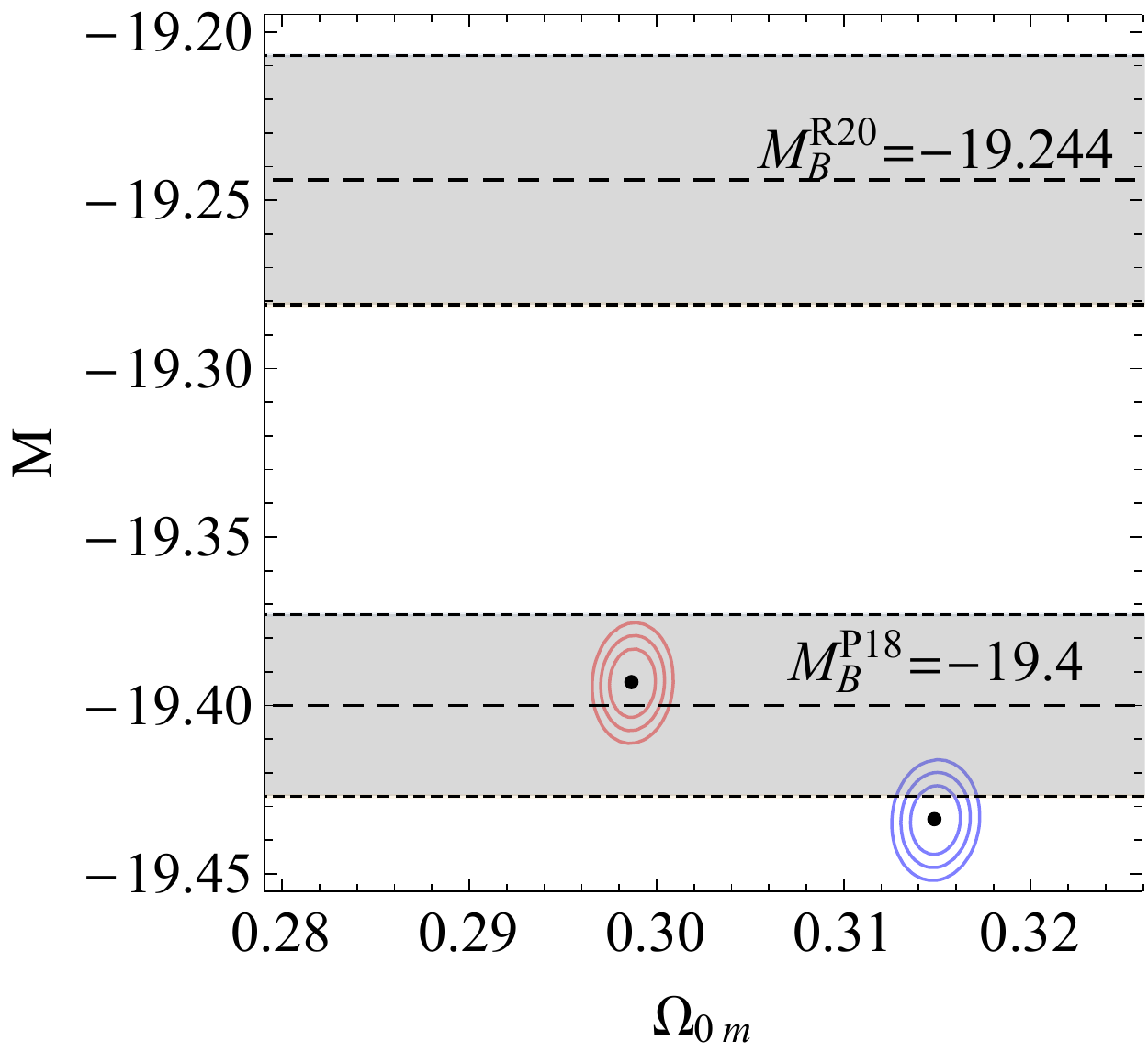}
         \centering
         \caption{}
         \label{fig:cpl_riess_001}
     \end{subfigure}
     \hfill
     \begin{subfigure}[b]{8cm}
     \captionsetup{justification=centering}
         \includegraphics[width=8cm]{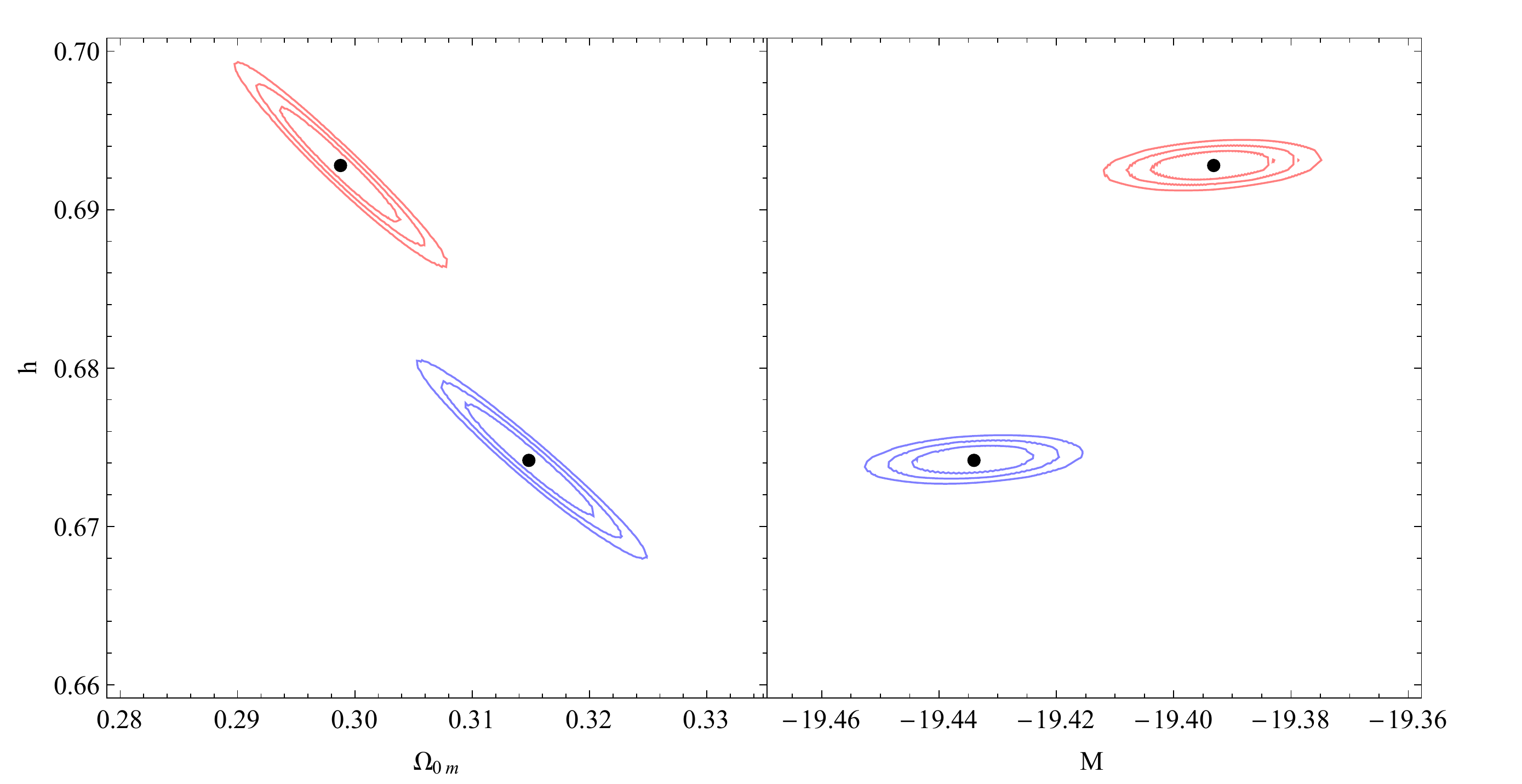}
         \caption{}
         \label{fig:cpl_riess_00002}
     \end{subfigure}
     \begin{subfigure}[b]{13.7cm}
     \captionsetup{justification=centering}
         \includegraphics[width=13.7cm]{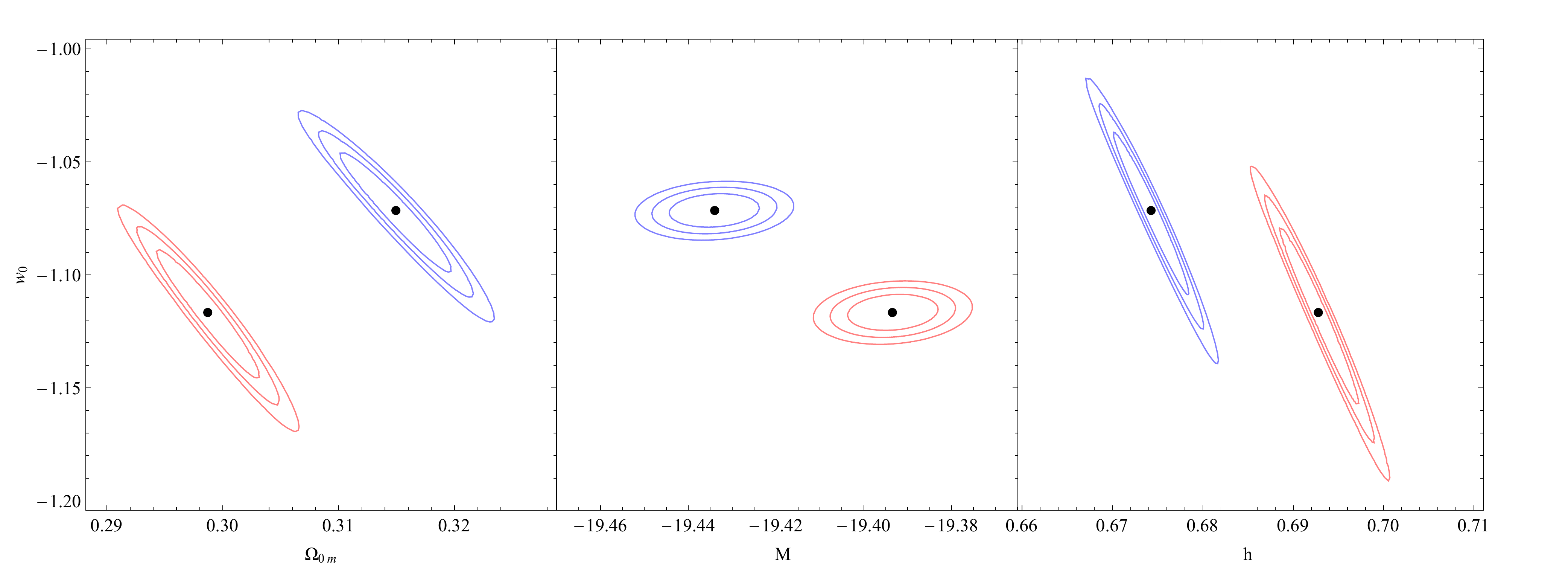}
         \caption{}
         \label{fig:cpl_riess_004}
     \end{subfigure}
     \begin{subfigure}[b]{13.5cm}
     \captionsetup{justification=centering}
         \includegraphics[width=13.5cm]{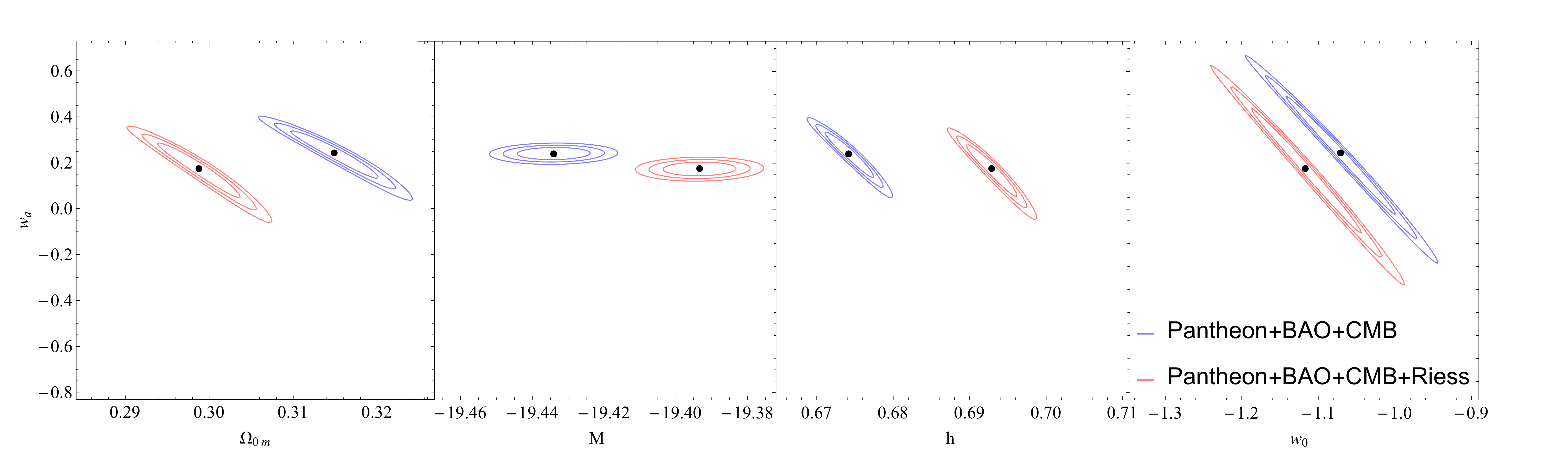}
         \caption{}
         \label{fig:cplriess}
     \end{subfigure}
        \caption{Confidence contours for the CPL model with the addition of the Riess point: (\textbf{a}) The $1\sigma - 3\sigma$ confidence contour in the parametric space $(M-\Omega_{0m})$. (\textbf{b}) The $1\sigma - 3\sigma$ confidence contours in the parametric spaces $(h-\Omega_{0m})$ and $(h-M)$. (\textbf{c}) The $1\sigma - 3\sigma$ confidence contours in the parametric spaces $(w_{0}-\Omega_{0m})$, $(w_{0}-M)$ and $(w_{0}-h)$. (\textbf{d}) The $1\sigma - 3\sigma$ confidence contours in the parametric spaces $(w_{a}-\Omega_{0m})$, $(w_{a}-M)$,$(w_{a}-h)$ and ($w_a-w_0$).}
        \label{fig: cpl_riess_001}
\end{figure}

The decrease of the preferred value of $\Omega_{0m}$ in all cases when the Riess point is included is anticipated due to the property of the CMB anisotropy spectrum to favor a fixed value of $\Omega_{0m}h^2$. Thus, we anticipate that the introduction of the Riess point, which tends to raise $h^2$, would also tend to lower the favored value of $\Omega_{0m}$ in order to maintain the product $\Omega_{0m}h^2$ as approximately constant and consistent with the observed CMB anisotropy~spectrum.

\section{Reconstruction of Dark~Energy}\label{sec5}
\subsection{Scalar Field Dark Energy Models (Quintessence)}
In the context of quintessence~\cite{Caldwell:1997ii}, we consider a self-interacting canonical scalar field $\phi$ minimally coupled to gravity playing the role of dark energy~\cite{Linder:2007wa}.

This scalar field is described by the Lagrangian density:
\be
\mathcal{L} = \frac{1}{2}\partial_\mu\phi\partial^\mu\phi - V(\phi)
\ee
where $V(\phi)$ is the potential energy density of the field $\phi$.

The stress-energy tensor can be obtained as~\cite{Amendola:2015ksp}:
\be
T_{\mu\nu} = \frac{-2}{\sqrt{-g}}\frac{\delta(\sqrt{-g}\mathcal{L})}{\delta g^{\mu\nu}}
\ee
which reduces to:
\be
T_{\mu\nu} = \frac{-2}{\sqrt{-g}}\frac{\partial\sqrt{-g}}{\partial g^{\mu\nu}}\mathcal{L} + \frac{-2}{\sqrt{-g}}\frac{\partial\mathcal{L}}{\partial g^{\mu\nu}}\sqrt{-g}
\ee

At this point, it is important to calculate the terms:
\be
\frac{\partial\sqrt{-g}}{\partial g^{\mu\nu}} = -\frac{1}{2\sqrt{-g}}\frac{\partial g}{\partial g^{\mu\nu}} = -\frac{1}{2\sqrt{-g}}\frac{g g^{\alpha\beta}\partial g_{\alpha\beta}}{\partial g^{\mu\nu}} = \frac{1}{2}\sqrt{-g}g_{\mu\nu}
\ee
where we used:
\be
dg=gg^{ab}dg_{ab}
\ee 

The detailed proof can be found at Appendix \ref{section:proof1}.

Furthermore:
\be
\frac{\partial\mathcal{L}}{\partial g^{\mu\nu}} = \frac{\partial}{\partial g^{\mu\nu}}\left[ \frac{1}{2} g^{\alpha\beta}\partial_\alpha\phi\partial_\beta\phi - V(\phi)\right] = \frac{1}{2} \delta^\alpha_\mu\delta^\beta_\nu\partial_\alpha\phi\partial_\beta\phi = \frac{1}{2} \partial_\mu\phi\partial_\nu\phi
\ee

Thus, the~stress-energy tensor takes the form:
\be
T_{\mu\nu} =- g_{\mu\nu}\mathcal{L} -2\frac{\partial\mathcal{L}}{\partial g^{\mu\nu}}
\ee
which reduces to:
\be
T_{\mu\nu} = \partial_\mu\phi\partial_\nu\phi-g_{\mu\nu}\mathcal{L}
\ee
or by substituting $\mathcal{L}$, we obtain:
\be
T_{\mu\nu} = \partial_\mu\phi\partial_\nu\phi-g_{\mu\nu}\left(\frac{1}{2}\partial_\mu\phi\partial^\mu\phi - V(\phi)\right)
\ee

Assuming that the scalar field is close to spatially uniform on cosmological scales, we can neglect its spatial derivatives $\partial_i \phi$ compared to its time derivatives $\dot{\phi}$. Thus:
\be
T_{0i} = T_{i0} = 0
\ee
\be
T_{ij} = 0\hspace{0.5cm}, i\neq j
\ee
which means that $T_{\mu\nu}$ is~diagonal.

We can obtain the energy density and the pressure of the field as:
\be\label{eq:4.34}
\rho_\phi = T^0_0 = g^{0\mu}T_{\mu0} = g^{00}T_{00} = \frac{\dot{\phi}^2}{2}+V(\phi)
\ee
\be\label{eq:4.35}
p_\phi = T^i_i = g^{i\mu}T_{\mu i} = g^{ii}T_{ii} = \frac{\dot{\phi}^2}{2}-V(\phi)
\ee

Thus, the~equation-of-state parameter is:
\be
w(\phi) = \frac{\frac{\dot{\phi}^2}{2}-V(\phi)}{\frac{\dot{\phi}^2}{2}+V(\phi)} = \frac{\dot{\phi}^2-2V(\phi)}{\dot{\phi}^2+2V(\phi)}
\ee

In general, for~$\dot{\phi}>> V(\phi)$, i.e.,~for a kinetic energy dominated field:
\be
w(\phi) = \frac{\dot{\phi}^2-2V(\phi)}{\dot{\phi}^2+2V(\phi)}\approx \frac{\dot{\phi}}{\dot{\phi}}\approx 1
\ee
while for $\dot{\phi}<< V(\phi)$, i.e.,~for a potential energy dominated field:
\be\label{eq:150}
w(\phi) = \frac{\dot{\phi}^2-2V(\phi)}{\dot{\phi}^2+2V(\phi)}\approx \frac{-2V(\phi)}{2V(\phi)}\approx -1
\ee
while we can effectively obtain  the cosmological constant $w = -1$ for $\dot{\phi} = 0$.
Quintessence can play the role of dark energy if:
\be
w(\phi) < -\frac{1}{3}\Rightarrow \frac{\dot{\phi}^2-2V(\phi)}{\dot{\phi}^2+2V(\phi)}< -\frac{1}{3}\Rightarrow 3\dot{\phi}-6V(\phi)< -\dot{\phi}-2V(\phi)\Rightarrow \dot{\phi}<V(\phi)
\ee

However, this is not enough since dark energy domination today requires $w=-1$ not only now but for an extended period of time (roughly between $z\approx 1$ and now). Thus, it is a requirement that the condition $\dot{\phi}<V(\phi)$ holds for a while. This can happen if the time derivative of this condition is also fulfilled:

\be
\left|\frac{d}{dt}\dot{\phi}^2\right| < \left|\frac{d}{dt} V(\phi)\right|\Rightarrow\left|2\dot{\phi}\ddot{\phi}\right|< \left|\frac{dV(\phi)}{d\phi}\dot{\phi}\right|\Rightarrow\left|\ddot{\phi}\right|<\left|\frac{V'(\phi)}{2}\right|<\left|V'(\phi)\right|
\ee
where $V'(\phi) = dV(\phi)/d\phi$.

 In summary, a scalar field can play the role of dark energy if:
\be
\dot{\phi}^2<V(\phi)
\ee
\be
\left|\ddot{\phi} \right|<\left|V'(\phi) \right|
\ee

These are the  slow-roll conditions~\cite{Boehm_dark}.
The time evolution of the scalar field is determined by the Klein--Gordon equation:

\be
\ddot{\phi} + 3\left( \frac{\dot{a}}{a}\right)\dot{\phi}+\frac{dV}{d\phi} = 0
\ee
which can be obtained by the variation of the action:
\be
S_\phi =\int \sqrt{-g}\mathcal{L}_\phi\left(\phi,\partial_a\phi \right)d^4x
\ee

The detailed proof can be found in Appendix \ref{section:klein}.

Assuming a flat universe in the presence of matter and the quintessence Equation~(\ref{eq:1.26})  gives:

\be
H^2 = \frac{8\pi G}{3}\left( \rho_m + \rho_\phi\right) = \frac{8\pi G}{3}\left( \rho_m + \frac{1}{2}\dot{\phi}^2+V(\phi)\right)
\ee
or in terms of the dimensionless density parameters:
\be
H^2(z) = H_0^2\left[ \Omega_{0m}(1+z)^3+\Omega_{0\phi}\exp{\left( \int_0^z\frac{3[1+w(z')]}{1+z'}\right)}\right]
\ee
where we used Equation~(\ref{eq:1.25})  for the dynamical energy density $\rho_\phi$. Furthermore, 

\be
\Omega_{0m} + \Omega_{0\phi} = 1
\ee

This model has $\Omega_{0m}$ as a parameter, along with the number of parameters that are used to parametrize the equation of state $w(z)$.

There are two ways to approach the effect that different dark energy models have in the cosmological expansion: either calculate the equation of state for some specified theory and then its effects on the cosmological expansion or start from the observations of the cosmological expansion and then reconstruct the scalar field physics responsible for the effects~observed.

The latter approach is made difficult due to some issues~\cite{Linder:2007wa}:

\begin{itemize}
  \item Noisiness of measurements of the expansion.
  \item Translation from the measured quantity to $\rho_{DE}$ and $w$ through one or two derivatives.
  \item Range of the scale factor or equivalently redshift coverage: $z = \frac{1}{a}-1$
\end{itemize}

However, since we already have the {constraints} for the model parameters, this is the right approach, and others have used it before~\cite{Mortonson:2009qq,Huterer:1998qv,Copeland:2006wr,Nakamura:1998mt,Rajvanshi:2019wmw,Guo:2005ata,Pantazis:2016nky,Scherrer:2015tra,Bonilla:2020wbn}.

\subsection{Reconstruction~Equations}

At first we have the parametrization of the equation of state \emph{w}:
\be
w(z) = w_0 +w_a\frac{z}{1+z}
\ee

From this equation, we see that in reality we can assume that $wCDM$ and $\Lambda CDM$ are special cases of the $CPL$ model with $w_a = 0$ and $w_0 = \text{const}$ and $w_0 = -1$, respectively. Thus, we can reconstruct their fields, too.

Furthermore, we can obtain the energy density using Equation~(\ref{eq:1.25}):

\be
\rho_{DE}(z) = \rho_{DE,0}\exp\left[\int_0^z3\frac{1+w_0+w_a\frac{z'}{1+z'}}{1+z'}dz' \right] = \rho_{DE,0}(1+z)^{3(1+w_0+w_a)}e^{-3w_a\frac{z}{1+z}}
\ee

Then, the~Hubble parameter $H(z)$ is:
\be
H^2(z) = H_0^2\left[ \Omega_{0,m}(1+z)^3 + (1-\Omega_{0,m})(1+z)^{3(1+w_0+w_a)}e^{-3w_a\frac{z}{1+z}}\right]
\ee

From Equations~(\ref{eq:4.34})  and~(\ref{eq:4.35}), we can obtain the potential in terms of the redshift:
\be
V(z) =\frac{1}{2}\left(\rho_{DE}-p_{DE}\right) =  \frac{1}{2}\left[1-w(z)\right]\rho_{DE}(z) = \frac{1}{2}\left( 1- w_0-w_a\frac{z}{1+z}\right)\rho_{DE}(z)
\ee
and the kinetic~term:
\vspace{-6pt}
\be
\frac{1}{2}\Dot{\phi}^2(z) = \frac{1}{2}\left(\rho_{DE}+p_{DE}\right) = \frac{1}{2}\left[ 1+w(z)\right]\rho_{DE}(z) = \frac{1}{2}\left( 1+ w_0+w_a\frac{z}{1+z}\right)\rho_{DE}(z)
\ee

In this way, we obtain the field $\phi$ in terms of the redshift as:
$$\Dot{\phi} = \big|\left[1+w(z)\right]\rho_{DE}(z)\big|^{1/2} \Rightarrow \phi = \int_{t_e}^{t_0}\big|\left[1+w(z)\right]\rho_{DE}(z)\big|^{1/2}dt =$$
\be
 = \int_{a_e}^{a_0}\big|\left[1+w(z)\right]\rho_{DE}(z)\big|^{1/2}\frac{da}{\Dot{a}} = \int_{0}^{z}\big|\left[1+w(z')\right]\rho_{DE}(z')\big|^{1/2}\frac{dz'}{(1+z')H(z')}
\ee
which by substituting the equation of state $w$ and Hubble parameter $H(z)$ for the CPL model gives:
\begin{figure}[H]
     \begin{subfigure}[b]{6.2cm}
     \captionsetup{justification=centering}
         \includegraphics[width=6.2cm]{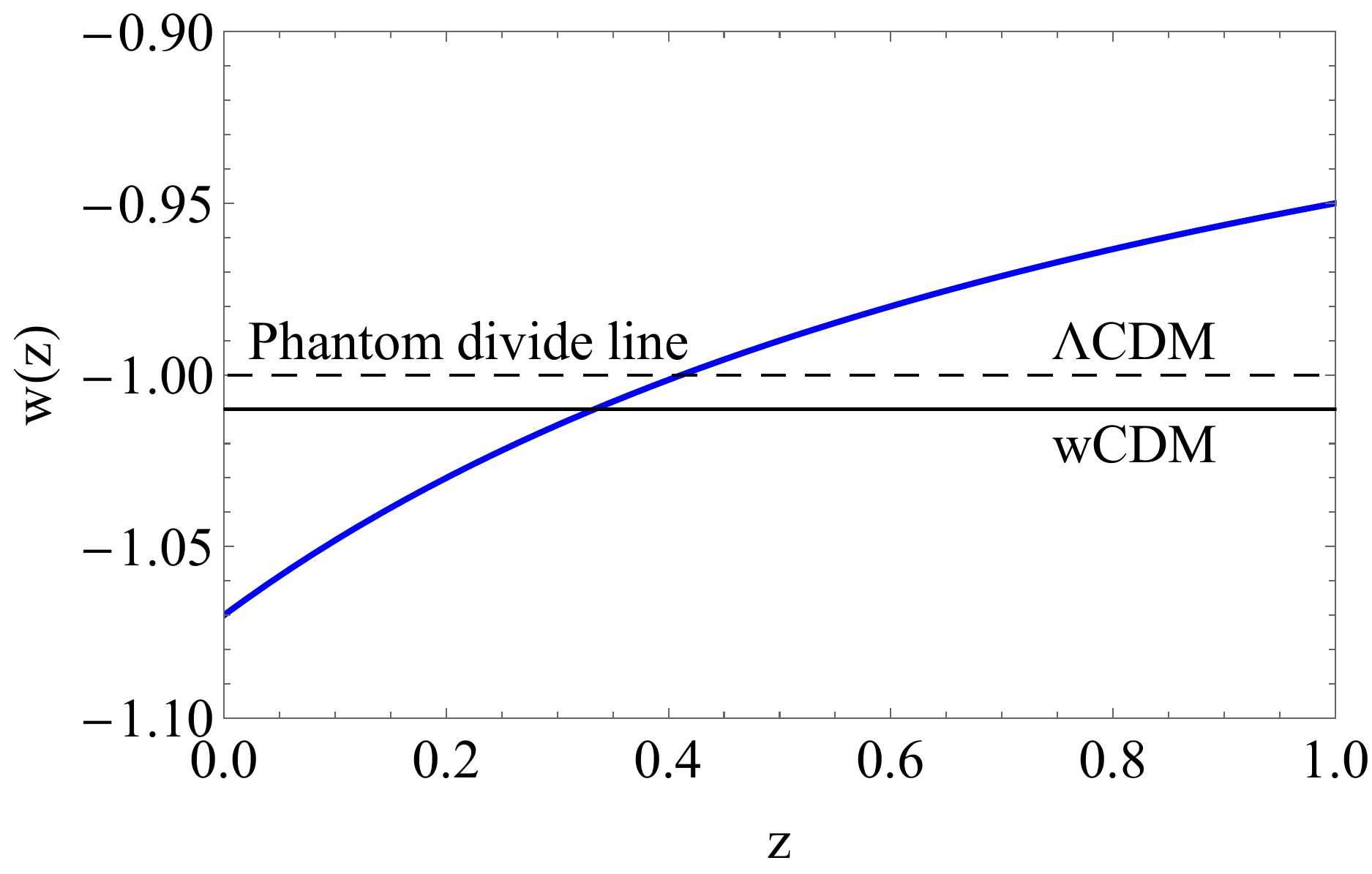}
         \caption{}
         \label{fig:rec1}
         \begin{minipage}{.1cm}
            \vfill
            \end{minipage}
     \end{subfigure}
     \hfill
     \begin{subfigure}[b]{6cm}
     \captionsetup{justification=centering}
         \includegraphics[width=6cm]{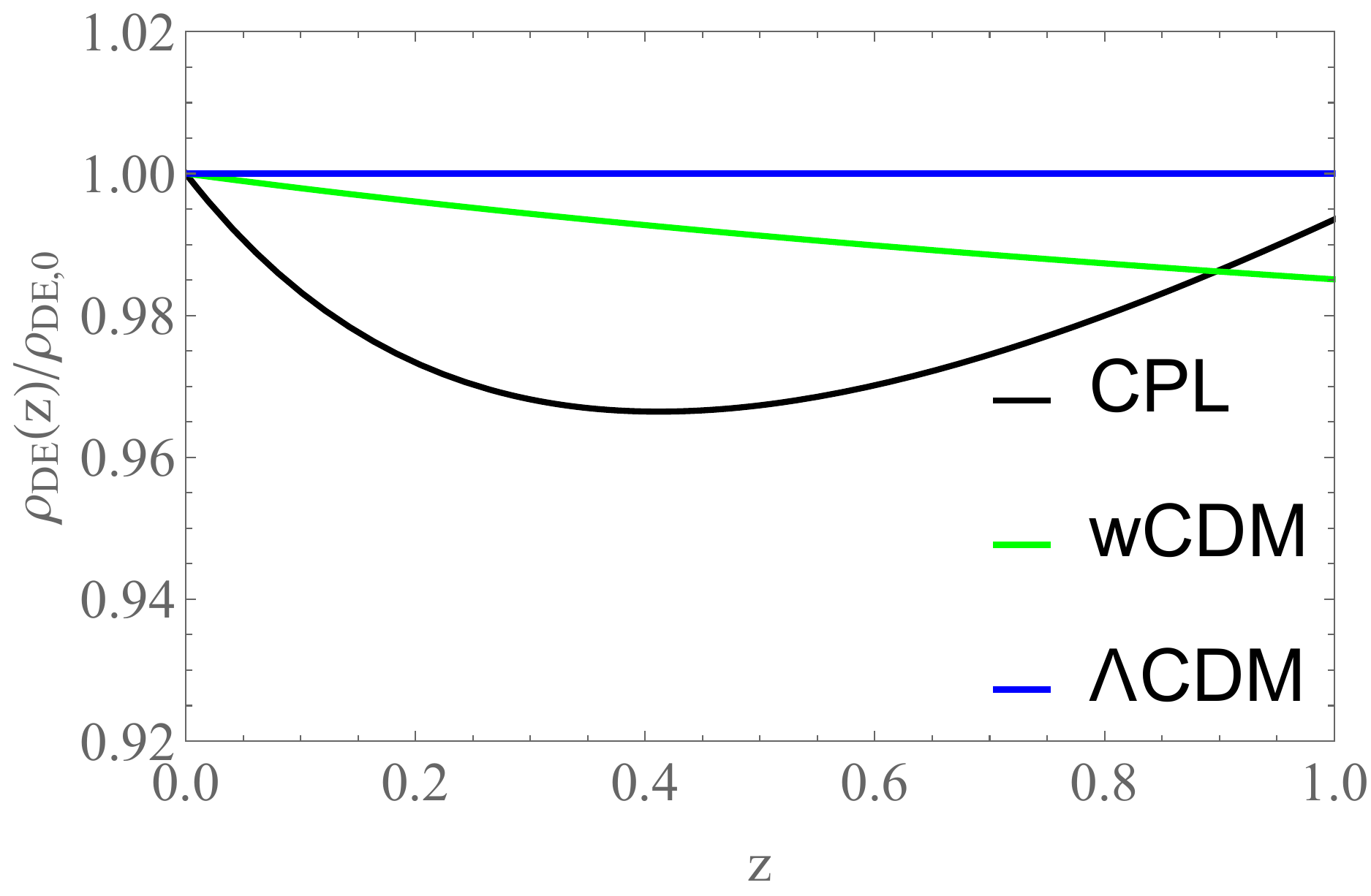}
         \caption{}
         \label{fig:rec2}
         \begin{minipage}{.1cm}
            \vfill
            \end{minipage}
     \end{subfigure}
     \begin{subfigure}[b]{10cm}
     \captionsetup{justification=centering}
         \includegraphics[width=10cm]{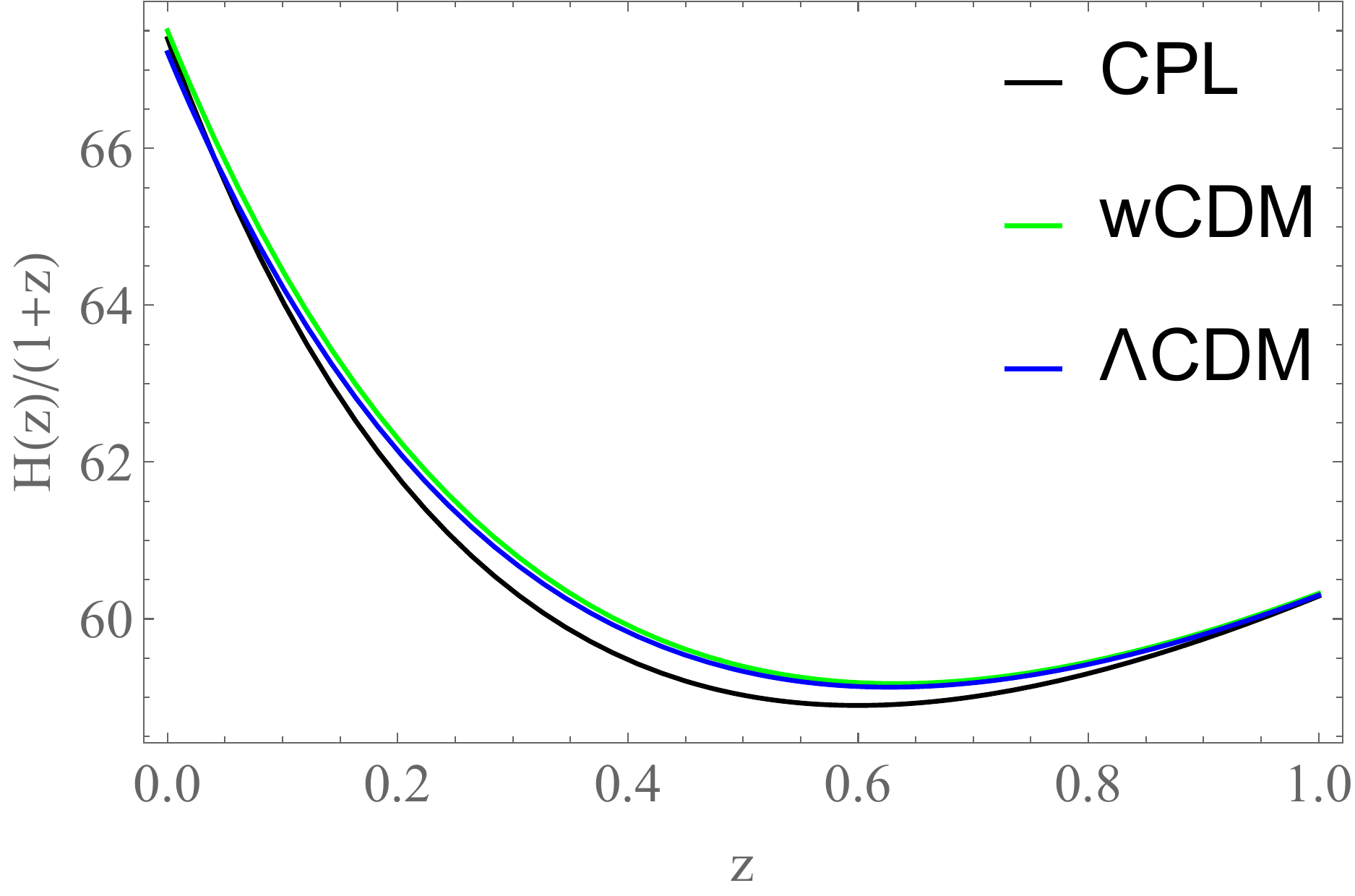}
         \caption{}
         \label{fig:rec3}
     \end{subfigure}
        \caption{Plots for the field reconstruction: (\textbf{a}) Plot of the equation of state $w$ in terms of the redshift z for the three models considered. (\textbf{b}) Plot of the energy density $\rho_{DE}$ in terms of the redshift z for the three models considered. (\textbf{c})  Plot of the fraction $\frac{H(z)}{1+z}$ in terms of the redshift z for the three models~considered.}
        \label{fig: recon_001}
\end{figure}
\be
\phi(z) = \rho_{DE,0}\int_0^z\frac{\big|1+w_0+w_a\frac{z'}{1+z'}\big|^{1/2}}{(1+z')\left[1-\Omega_{0,m}+\Omega_{0,m}(1+z')^{-3(w_0+w_a)}e^{3w_a\frac{z'}{1+z'}}\right]}dz'
\ee

Furthermore, we can reconstruct the potential in terms of the field $V(\phi)$ by constructing points with the values of $V(z)$ and $\phi(z)$ calculated for many different values of the redshift and then plotting~them.

Thus, the~only remaining objective is to substitute the best-fit values for the model parameters obtained through maximum likelihood~estimation.

\subsection{Results}
We can see from Figure~\ref{fig: recon_001}a  that the best-fit value for the wCDM model gives a phantom field, well explained here~\cite{Dabrowski:2014wia}, i.e.,~a field, with $w<-1$ which should have a negative kinetic~term.

Furthermore, we can see that the data favor a value for the equation of state at the present epoch $w_0$ that makes the CPL model a phantom field and a value for $w_a$ that makes it raise that value with redshift. Thus, at~some point (at around $z\approx 0.4$) it crosses the phantom divide line ($w = -1$), i.e., the line that separates the physics that obeys the null energy condition ($\rho+p \geq 0$) from the physics that violates it~\cite{Linder:2007wa}. This feature is responsible for the weird behavior in the plots of the fields $\phi$ and $V(\phi)$ in Figure \ref{fig: recon_002}.

{It is expected that both models will break down trying to explain this phantom regime since, as~seen in
Equation~(\ref{eq:150}), $w$ is allowed to take values higher than $-1$.}
In general, though, ghost solutions are problematic because they lead to instabilities~\cite{Sbisa:2014pzo,Wolf:2019hzy}.
\begin{figure}[H]
     \begin{subfigure}[b]{6.1cm}
     \captionsetup{justification=centering}
         \includegraphics[width=6.1cm]{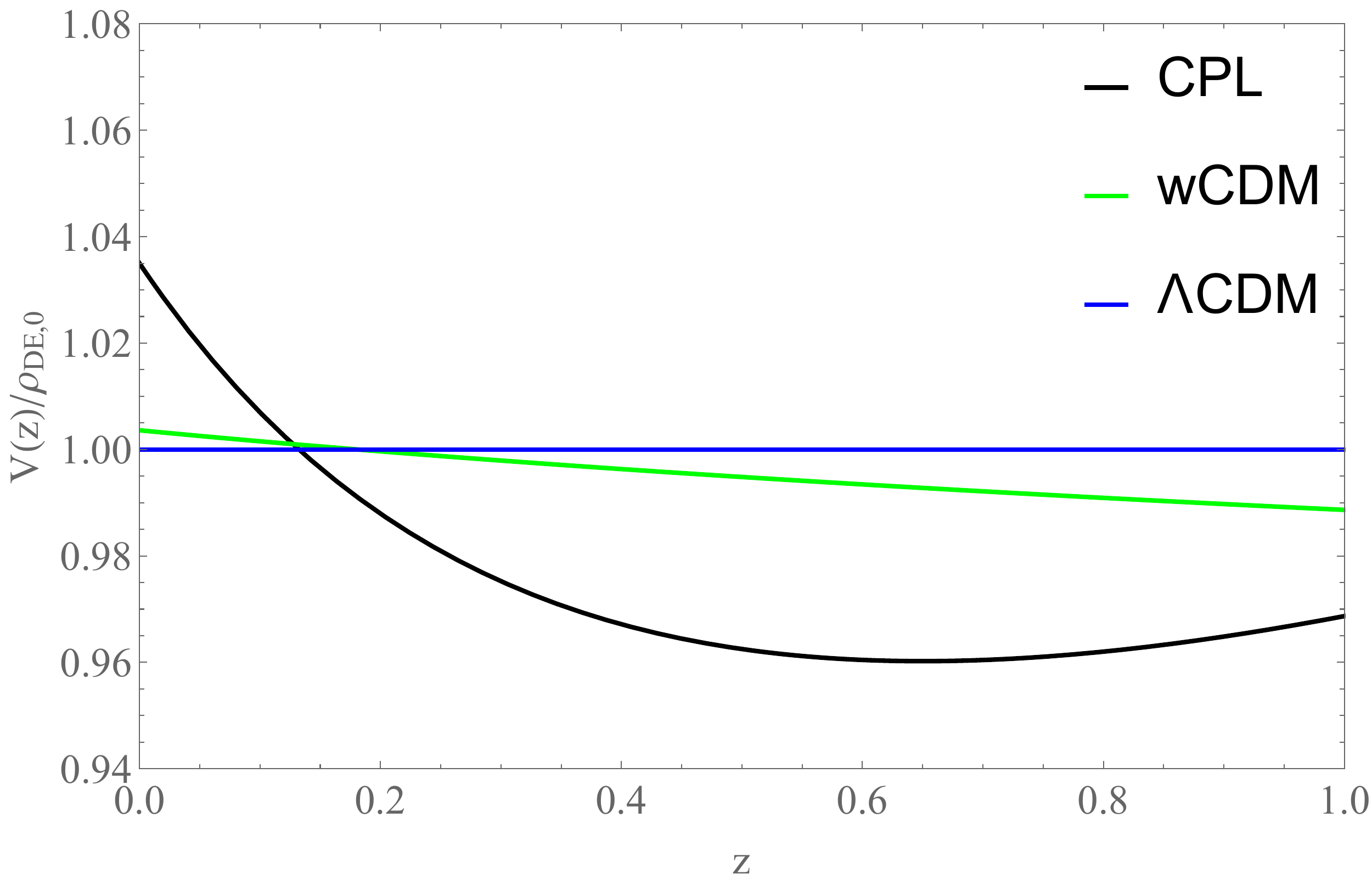} 
         \caption{}
         \label{fig:rec11}
     \end{subfigure}
     \hfill
     \begin{subfigure}[b]{6.1cm}
     \captionsetup{justification=centering}
         \includegraphics[width=6.1cm]{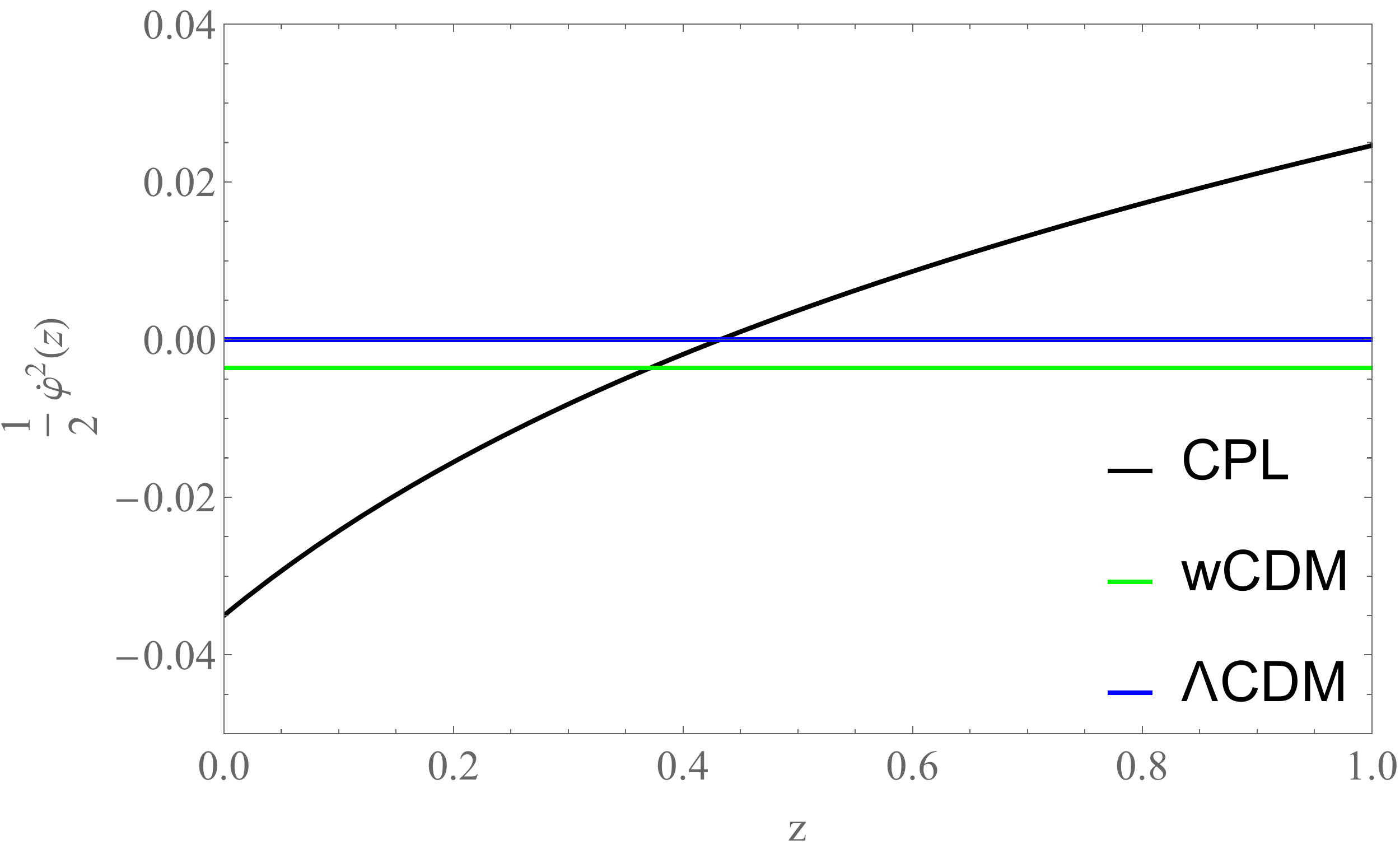}
         \caption{}
         \label{fig:rec22}
     \end{subfigure}
     \begin{subfigure}[b]{12cm}
     \captionsetup{justification=centering}
         \includegraphics[width=12cm]{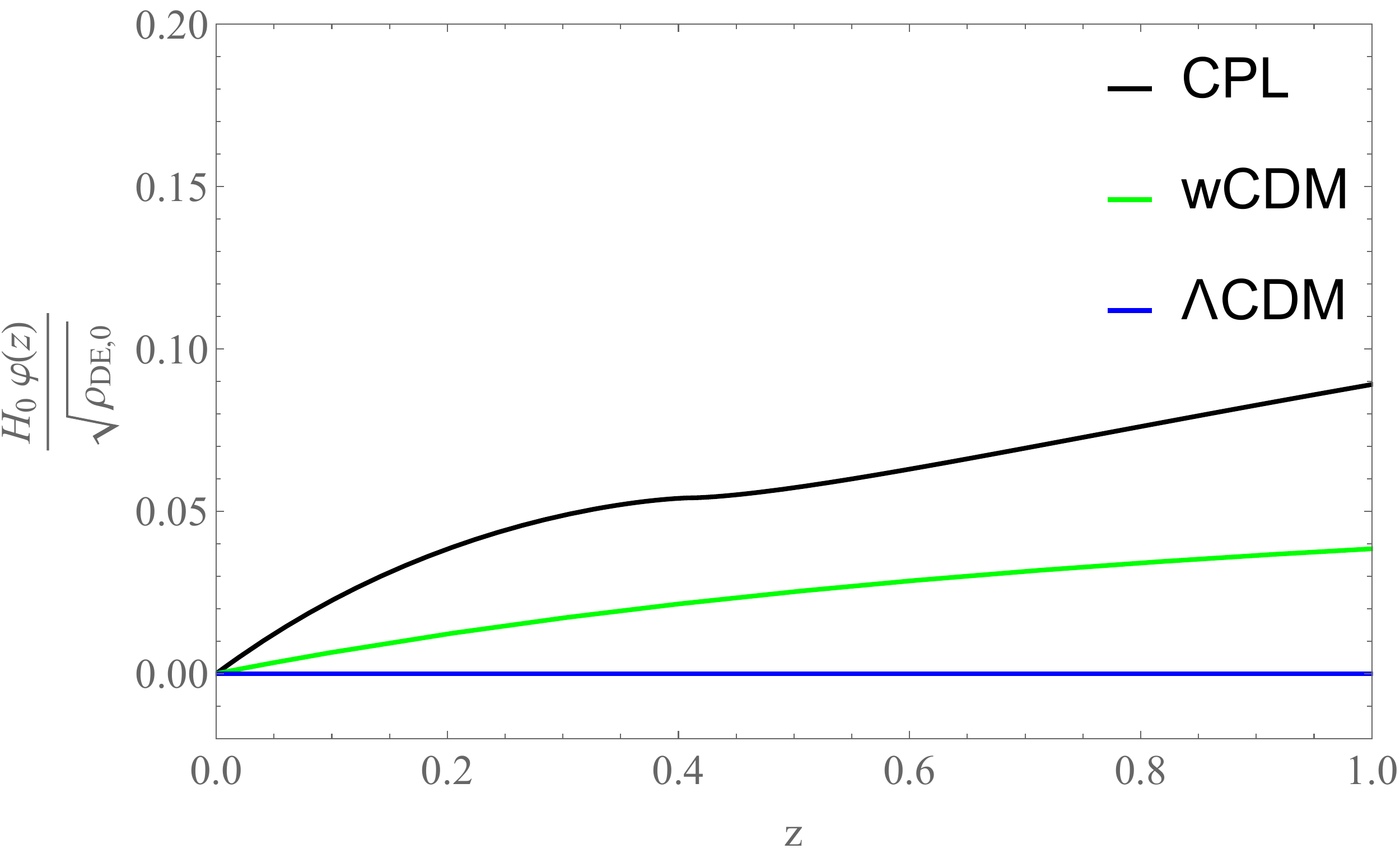}
         \caption{}
         \label{fig:rec33}
     \end{subfigure}
     \begin{subfigure}[b]{6cm}
     \captionsetup{justification=centering}
         \includegraphics[width=6cm]{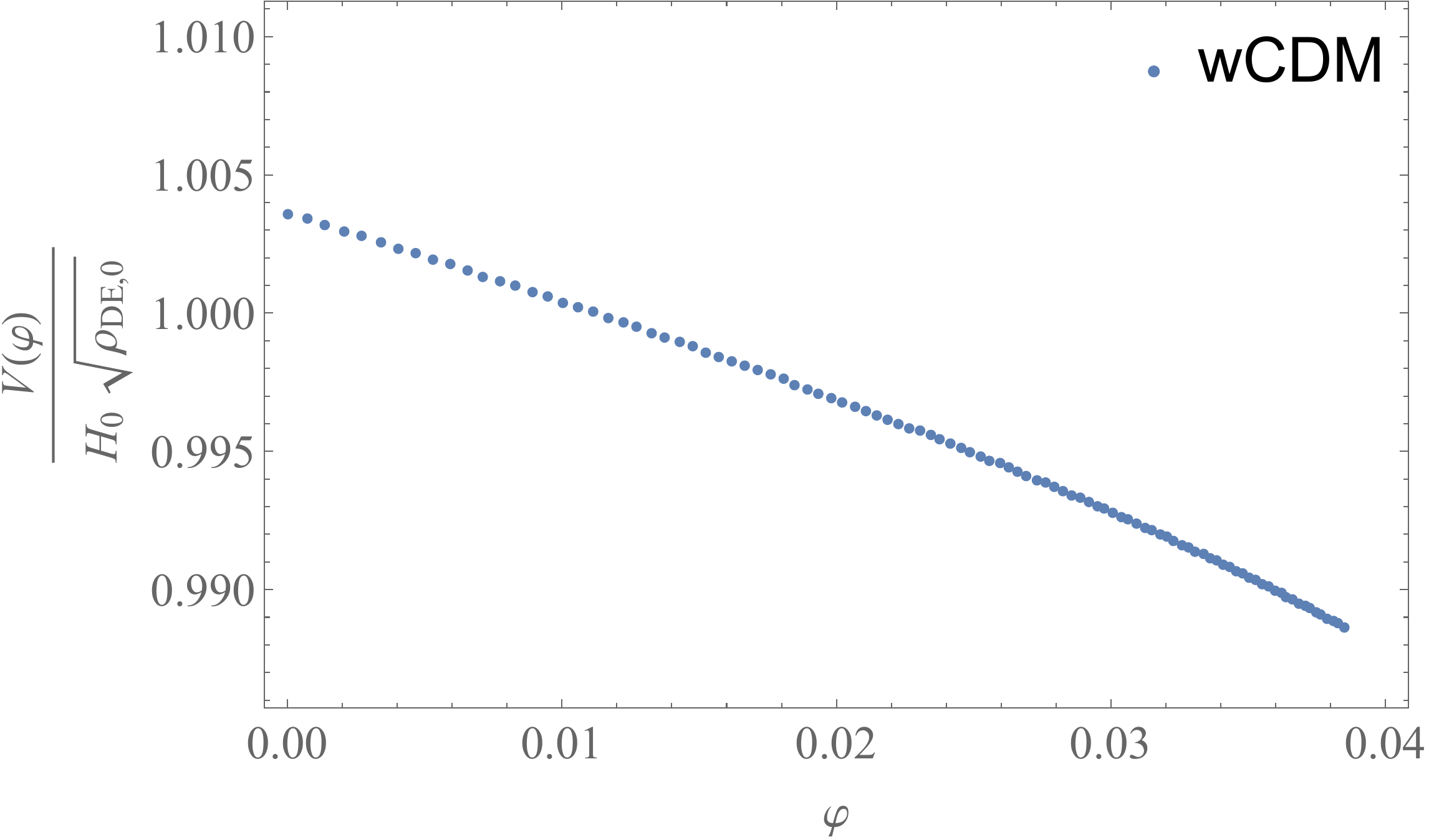}
         \caption{}
         \label{fig:vphiww}
     \end{subfigure}
     \hfill
     \begin{subfigure}[b]{6cm}
     \captionsetup{justification=centering}
         \includegraphics[width=6cm]{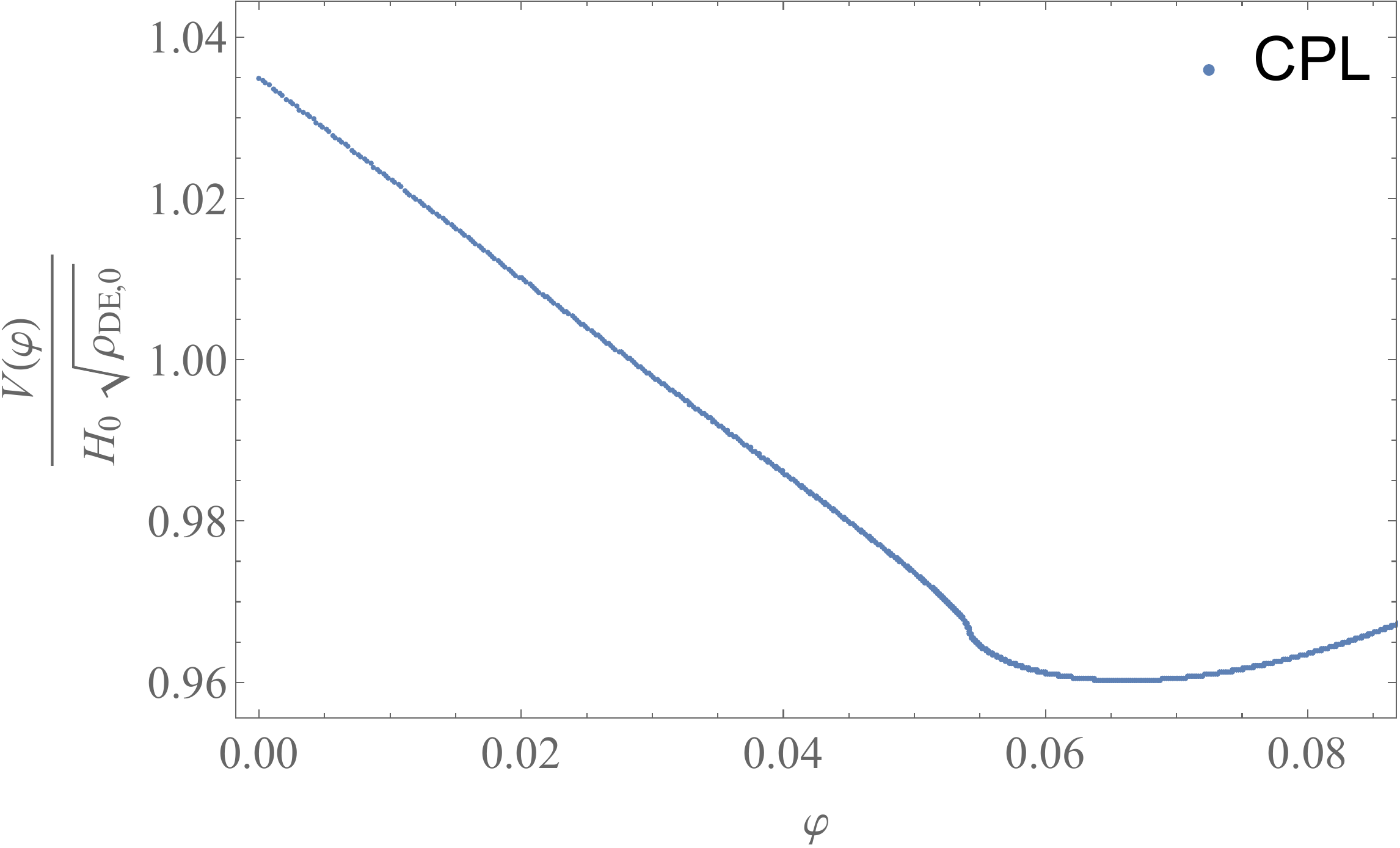}
         \caption{}
         \label{fig:vphicpll}
     \end{subfigure}
        \caption{{Plots} for the field reconstruction: (\textbf{a}) Plot of the potential  $V(z)$ in terms of the redshift z for the three models considered. (\textbf{b}) Plot of the kinetic term $\Dot{\phi}^2/2$ in terms of the redshift z for the three models considered. (\textbf{c}) Plot of the field $\phi(z)$ in terms of the redshift z for the three models considered. (\textbf{d}) Plot of the potential of the $wCDM$ model in terms of the field $\phi$. (\textbf{e}) Plot of the potential $CPL$ model in terms of the field $\phi$.}
        \label{fig: recon_002}
\end{figure}

\section{Discussion and Conclusions}\label{sec6}
\textls[-35]{We have presented constraints of the parameters of some generic dark energy parameterizations}, including \lcdm, $wCDM$ and CPL. In~the model parameters, we  included the SnIa absolute magnitude $M$. In~order to constrain these parameters, we  used the  SnIa Pantheon dataset, CMB shift parameters and BAO data. We  also included the local determination of $H_0$ as a data point and  demonstrated that the Hubble tension persists for these parameterizations even after the local $H_0$ point is included in the data. This tension manifests itself in three~ways:
\begin{itemize}
    \item The best-fit value of $H_0$ in the context of all these models is not consistent with the local determination of $H_0$ shown in Equation~(\ref{hr20}).
    \item The best-fit value of the SnIa absolute magnitude $M$ is not consistent with the value of $M$ determined by the local Cepheid calibrators.
    \item The quality of fit of all these parameterizations becomes significantly worse when the local determination of the $H_0$ point is included.
\end{itemize}

We conclude that these dark energy parameterizations are unable to resolve the Hubble tension. {Another argument adding to the inability of quintessence models to ease the Hubble tension can be found in~\cite{Banerjee:2020xcn}. It is stated that, being a $w>-1$ model (excluding the cosmological constant), it can only make the Hubble tension worse. In~general, late time models trying to resolve the cosmological tensions do not succeed, as~can also be seen in~\cite{Alestas:2021xes}, where it is shown that late time approaches worsen the growth tension and in~\cite{Alestas:2020mvb,Yang:2021flj} where they worsen the fit in~comparison with the $\Lambda CDM$, or even the Hubble tension when BAO data are used.} However, {since} the Hubble tension also manifests itself as an $M$-tension~\cite{Efstathiou:2021ocp,Camarena:2021jlr}, a~transition of $M$ {can be used} for these models to fully resolve the Hubble tension, as shown in Figure~\ref{fig1} and discussed in detail in~\cite{Marra:2021fvf,Alestas:2020zol}. 

In addition to finding the best-fit form of $H(z)$ corresponding to these parametrizations, we have reconstructed the quintessence Lagrangian that would {reproduce} the observed form of $H(z)$. Due to the phantom nature of the best-fit dark energy equation-of-state parameter, we  {found} that the reconstructed Lagrangian has a negative kinetic term and is therefore plagued with instabilities. {A possible extension of this work is the consideration of more parameterizations of the equation-of-state parameter $w(z)$ and the comparison of the quality of fit provided by each one, using the most recent cosmological data. A~few interesting parameterizations include the linear~\cite{Cooray:1999da}: $w(z) = w_0 + w_a z$, the~Alcaniz--Barbosa~\cite{Barboza:2008rh}: $w(z) = w_0 + w_a\frac{z(1+z)}{1+z^2}$, the~sqrt~\cite{Pantazis:2016nky}: $w(z) = w_0 + w_a\frac{z}{\sqrt{1+z^2}}$, the~Sine~\cite{Lazkoz:2010gz}: $w(z) = w_0 + w_a \sin{z}$ and the Jassal--Bagna--Padmanabhan~\cite{Jassal:2004ej}: $w(z) = w_0 + w_a\frac{z}{(1+z)^2}$. Another} interesting extension of the present analysis is the consideration of modified gravity scalar tensor Lagrangians, which can accommodate more naturally  the phantom behavior of the Hubble expansion without instabilities~\cite{Perivolaropoulos:2005yv}.

\vspace{6pt} 

\supplementary{{The numerical analysis files for the reproduction of the figures are } available online at the Github repository: \url{https://github.com/TasosTheodoropoulos/M_crisis}}


\authorcontributions{{Conceptualization, L.P.; Methodology, A.T; Software, A.T.;  Writing – Original Draft Preparation, A.T.; Supervision, L.P.}}

\funding{{LP's research is co-financed by Greece and the European Union (European Social Fund - ESF) through the Operational Programme "Human Resources Development, Education and Lifelong Learning 2014-2020" in the context of the project  MIS 5047648.}}

\dataavailability{The SNIa data used in this paper are from the Pantheon dataset, also used in the analysis~\cite{Scolnic:2017caz} and can be found in the Github repository~\cite{Pantheon_data_01} and on the website~\cite{Pantheon_data_02}. The~Pantheon dataset consists of 1048 data points coming from six different probes and covering the redshift range $0.01 < z < 2.3$. This publicly available dataset contains the name of each SNIa, its redshift both in the CMB and the  heliocentric frame and the observed apparent magnitude $m_{obs}$ along with its corresponding error $\sigma_{m_{obs}}$. 
The apparent magnitude $m_{obs}$ is reported after the application of a K-correction and a correction in terms of the light curve width along with some other corrections due to biases from simulations of the SNIa~\cite{Kazantzidis:2020tko}. 
The BAO observational data that we will use in this analysis are the data from 6dFGS and WiggleZ found in~\cite{Escamilla-Rivera:2016qwv}, from~SDSS (Data Release 7 and 11) found in~\cite{Ross:2014qpa,Anderson:2013zyy} and from Ly-$\alpha$ measurements found in~\cite{Agathe:2019vsu}. 
For the CMB observational data, we use the central values for $R$,$l_A$ and $\Omega_bh^2$ from Planck 2018 for a flat universe. The~data can be found in~\cite{Zhai_2019} and are expressed in terms of a data vector and a covariance~matrix. 
Lastly, the~numerical data files for the reproduction of the figures can be found in the \href{https://github.com/TasosTheodoropoulos/M_crisis}{M\_crisis} Github repository under the MIT license.} 

\acknowledgments{We thank Lavrentios Kazantzidis and George Alestas for useful discussions and their help with the~code.}

\conflictsofinterest{{The authors declare no conflict of interest. The funders had no role in the design of the study; in the collection, analyses, or interpretation of data; in the writing of the manuscript, or in the decision to publish the~results }}
\abbreviations{The following abbreviations are used in this manuscript:\\

\noindent 
\begin{tabular}{@{}ll}
FRW & Friedman--Robertson--Walker\\
DE & dark energy\\
SN & supernova\\
SNe & supernovae\\
\end{tabular}
 
\noindent 
\begin{tabular}{@{}ll} 
SNIa & supernova of Type Ia\\
SNeIa & supernovae of Type Ia\\
BAO & baryonic acoustic oscillations\\
CMB & cosmic microwave background\\
CDM & cold dark matter\\
CPL & Chevalier--Polarski--Linder\\
PDL & phantom divide line
\end{tabular}}

\appendixtitles{yes} 
\appendixstart
\appendix
\section{Notation---Cosmology~Basics}\label{section:notation}
\unskip
\subsection{FRW~Metric}
The cosmic metric is well approximated by the Friedmann--Robertson--Walker metric~\cite{Baumann_cosmology,Hobson:2006se}:
\be
ds^2 = dt^2-a^2(t)\left[\frac{dr^2}{1-kr^2}+r^2(d\theta^2+sin^2\theta d\phi^2)\right]
\ee
where
\begin{itemize}

  \item The curvature parameter $k$ can be $-$1, 0, 1. 
  \item $a(t)$ is the cosmic scale factor.
  \item The cosmic time t is the proper time measured by a free-falling observer.
  \item The coordinates $r,\theta,\phi$ are comoving coordinates.
\end{itemize}

Another form of the FRW metric is:
\be\label{eq:1.2}
ds^2 = dt^2-a^2(t)\left[d\chi^2+S^2(\chi)(d\theta^2+sin^2\theta d\phi^2)\right]
\ee
where: \hspace{0.5cm}
\be\label{eq:1.3}
S^2(\chi) =
    \begin{cases}
      sin\chi &    , k=1 \\
      \chi       &   , k=0 \\
      \sinh\chi &   ,k=-1
    \end{cases} 
    \ee
 \noindent
 \begin{itemize}
\item The time is again the proper time measured by a free-falling observer, while $\chi$ is a new radial coordinate (in which $r,\theta,\phi$ is a comoving coordinate).
\end{itemize}
The cosmic dynamics is determined by the Friedmann~equations

\underline{\textbf{{Friedmann Equation:}}}\noindent
\be
\left(\frac{\dot{a}}{a}\right)^2 = \frac{8 \pi G}{3}\rho -\frac{k}{a^2}
\ee

\underline{\textbf{{Friedmann Acceleration Equation:}}}
\be
\frac{\ddot{a}}{a} = -\frac{4 \pi G}{3}\left(\rho+3p\right)
\ee
\noindent where $\rho$ and $p$ correspond to the sum of all contributions to the energy density and pressure of the contents of the~universe.

The Friedmann equation in terms of the Hubble parameter takes the form:

\be
H^2 =\frac{8 \pi G}{3}\rho -\frac{k}{a^2}
\ee

\be 
H = \frac{\dot{a}}{a}
\ee

The critical density is defined for a given value of the Hubble parameter as 
\be
H^2 =\frac{8 \pi G}{3}\rho_{crit}
\ee

The quantities evaluated at present are denoted by a subscript `0'. For~example, the~age of the Universe evaluated today is expressed as $t = t_0$ and
\be
\rho_{crit,0} = \frac{3H_0^2}{8 \pi G}
\ee
is the critical density~today.

The dimensionless density parameters are defined as follows:
\be
\Omega_{I,0} \equiv \frac{\rho_{I,0}}{\rho_{crit,0}}
\ee
where the index $I$ corresponds to a given density component (e.g., matter, radiation or dark energy). Thus, the~curvature density parameter is
\be
\Omega_{k,0} \equiv -\frac{k}{(a_0 H_0)^2}
\ee

Using the stress-energy tensor of a perfect fluid as seen by a comoving observer:
\be
T^{\mu}_{\nu} = \begin{pmatrix}
\rho & 0 & 0 & 0 \\
0 & -p & 0 & 0 \\
0 & 0 & -p& 0 \\
0 & 0 & 0& -p 
\end{pmatrix}
\ee
and the covariant conservation equation for it:
\be
\nabla_{\mu}T^{\mu}_{\nu} = 0
\ee
the continuity equation is then obtained as
\be\label{eq:1.14}
\dot{\rho}+3\frac{\dot{a}}{a}(\rho + p) = 0
\ee
\noindent
\paragraph{{Constant Equation of State}}
\noindent
Introducing the constant equation of state: $w = p/\rho$, which can be used to parametrize most cosmological fluids, it is easy to show that $\rho$ scales as: 
\be
\rho \propto a^{-3(1+w)}
\ee

 In the case of cold dark matter (w = 0), radiation (w = 1/3) and vacuum energy or cosmological constant $\Lambda$ (w = $-$1), we obtain:
\be
\rho \propto
    \begin{cases}
      a^{-3} &     matter \\
      a^{-4} &   radiation  \\
      a^{0} &   vacuum 
    \end{cases} 
\ee
\vspace{-6pt}
\begin{figure}[H]
              \includegraphics[width=13cm]{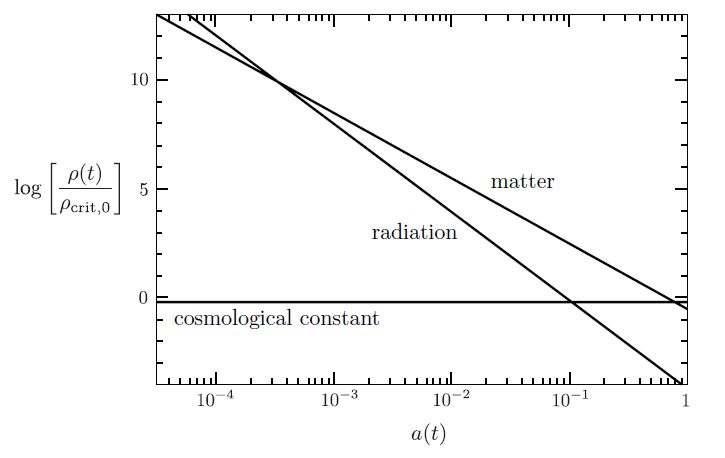}
         \caption{Evolution of the energy densities in the Universe. {{Adapted from}}: \cite{Baumann_cosmology}.}
         \label{fig:Energy densities evolution}
\end{figure}
\paragraph{{Dynamical} Equation of State}

Introducing a time dependent equation of state:
\be
w(t) = \frac{\rho(t)}{p(t)}
\ee
in Equation~(\ref{eq:1.14})  we obtain:
\be
\frac{d\rho(t)}{dt} = -3\frac{\dot{a}}{a}\left[ \rho(t)+p(t)\right]
\ee
which by dividing both sides of the equation by $\rho(t)$ takes the form:
\be
\frac{\dot{\rho}(t)}{\rho(t)} = -3\frac{\dot{a}}{a}\left[1+\frac{p(t)}{\rho(t)} \right] = -3\frac{1}{a}\frac{da}{dt}\left[1+w(t) \right]
\ee

Thus:
\be
\frac{d\rho}{\rho} = -3\left[1+w(a)\right]\frac{da}{a}
\ee
which we can integrate as follows:
\be
\int_{\rho_0}^\rho\frac{d\rho'}{\rho'} = \int_{a_0}^a-3\left[1+w(a')\right]\frac{da'}{a'} \Rightarrow\ln(\rho) - \ln(\rho_0) = \int_{1}^a-3\left[1+w(a')\right]\frac{da'}{a'}
\ee

Here, we can make the substitution:
\be
a = \frac{1}{1+z}
\ee
which corresponds to the new differential:
\be
da'= d\left( \frac{1}{1+z}\right) = -(1+z)^{-2}d(1+z) =-\frac{dz}{(1+z)^2} 
\ee

Thus:
\be
\ln\frac{\rho}{\rho_0} = \int_{0}^{z}3\left[1+w(z')\right]\frac{dz'}{(1+z')^2}(1+z') = \int_0^z3\frac{1+w(z')}{1+z'}dz'
\ee

Finally we obtain:
\be\label{eq:1.25}
\rho = \rho_0 \exp\left[ \int_0^z3\frac{1+w(z')}{1+z'}dz'\right]
\ee
\subsubsection{Single Component~Universe}

For a flat single component universe, the~Friedmann equation can take the form: \be
\frac{\dot{a}}{a} = H_0 \sqrt{\Omega_I}a^{-\frac{3}{2}(1+w_I)}
\ee

Solving in terms of $a$, we find that it scales as:
\be
a(t) \propto
    \begin{cases}
      t^{2/3} &     Matter\hspace{2mm} Domination\\
      t^{1/2} &   Radiation \hspace{2mm} Domination \\
      e^{Ht} &   \Lambda \hspace{2mm} Domination 
    \end{cases} 
\ee

\subsubsection{Two-Component~Universe}
There are many ways to choose two components to study, but~an interesting choice is a universe where matter and radiation densities are comparable.

We denote the value of the scale factor when matter and radiation were equally important as:
\be
a_{eq} \equiv \frac{\Omega_r}{\Omega_m}
\ee
which was shortly before the cosmic microwave background was~released.

We can write the total energy density at that time in the form:
\be
\rho \equiv \rho_m + \rho_r = \frac{\rho_{eq}}{2}\left[\left( \frac{a_{eq}}{a}\right)^3 + \left( \frac{a_{eq}}{a}\right)^4\right]
\ee

To help with the calculations, we introduce conformal time:
\be
\tau = \frac{dt}{a(t)}
\ee

Using the above the Friedmann Equations, take the form:
\be
(a')^2 = \frac{8 \pi G}{3}\rho a^4
\ee
\be
a'' = \frac{4 \pi G}{3}(\rho-3p)a^3
\ee

with primes denoting the derivatives with respect to conformal~time.

Solving for $a$, we obtain:
\noindent
\be
a(\tau) = a_{eq}\left[\left( \frac{\tau}{\tau_\ast}\right)^2+2 \left( \frac{\tau}{\tau_\ast}\right)\right]
\ee
where:

\be
\tau_\ast \equiv \left( \frac{\pi
G}{3}\rho_{eq}a_{eq}^2\right)^{-1/2}
\ee

\noindent

\subsubsection{Multi-Component~Universe}

The general form of the Friedmann equations for multiple components is:
\be\label{eq:1.26}
\left(\frac{\dot{a}}{a}\right)^2 = \frac{8 \pi G}{3}\sum_{i}\rho_i -\frac{k}{a^2}
\ee
\be
\frac{\ddot{a}}{a} = -\frac{4 \pi G}{3}\sum_{i}\left(\rho_i+3p_i\right)
\ee

The equations of state are: $p_i = w_i\rho_i$, where $w_i$ is the dimensionless equation-of-state parameter of the i-th~component.

Now, considering a universe with matter, radiation, curvature and a cosmological constant ($\Lambda$), we obtain:
\be\label{eq:1.28}
H^2 = H_0^2\left[\Omega_{r,0}a^{-4} + \Omega_{m,0}a^{-3} + \Omega_{\Lambda,0} + \Omega_{k,0}a^{-2}\right]
\ee
while the density parameters at the present time satisfy the relation: \be
\Omega_{m,0}+\Omega_{r,0}+\Omega_{k,0}+\Omega_{\Lambda,0} = 1
\ee
\subsection{Redshift~z}
\begin{figure}[H]
         \includegraphics[width=10cm]{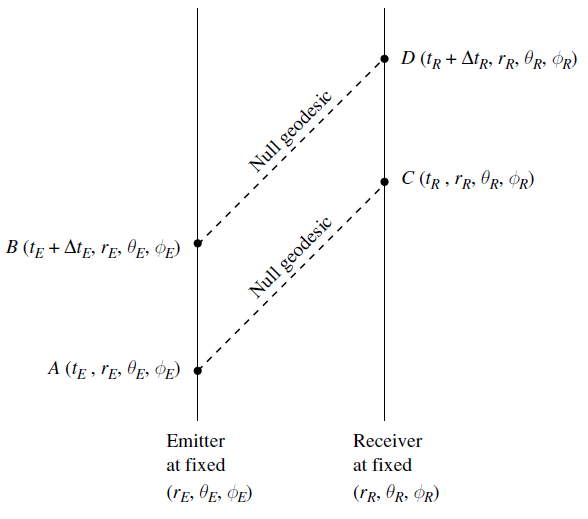}
         \caption{Diagram showing two null (photon) geodesics. In~our case where the photons are emitted radially, we have $\theta_E = \theta_R $ and $\phi_E = \phi_R $. {{Adapted from}}~\cite{Hobson:2006se}.}
         \label{fig:Redshift}
\end{figure}

Consider a photon emitted (radially) from a distant galaxy at time $t = t_E$ and received by us on earth at $t = t_R$. Along the photon path, we have $ds = d\theta = d\phi = 0$, so from the FRW metric for a flat universe Equation~(\ref{eq:1.2}),  we obtain:
\be
c^2dt^2 = a^2(t)d\chi^2 \Rightarrow \frac{c^2dt^2}{a^2(t)} = d\chi^2
\ee

Since we have an incoming photon:
\be
\frac{cdt}{a(t)} = -d\chi\Rightarrow \int_{t_E}^{t_R}\frac{cdt}{a(t)} = -\int_{\chi_E}^{\chi_R}d\chi
\ee

The photon is emitted at a comoving distance $\chi_E$ and is observed here on Earth at a comoving distance $\chi_R = 0$:
\be\label{eq:1.51}
\int_{t_E}^{t_R}\frac{cdt}{a(t)} = \int_{0}^{\chi_E}d\chi
\ee

Now, consider another photon emitted at time $t = t_E + \delta t_E$ and received at $t = t_R + \delta t_R$. For~that photon:
\be\label{eq:1.52}
\int_{t_E + \delta t_E}^{ t_R + \delta t_R}\frac{cdt}{a(t)} = \int_{0}^{\chi_E}d\chi
\ee

Combining Equations~(\ref{eq:1.51})  and~(\ref{eq:1.52}),  we obtain:
\be
\int_{t_E + \delta t_E}^{ t_R + \delta t_R}\frac{cdt}{a(t)} = \int_{t_E}^{t_R}\frac{cdt}{a(t)}
\ee

Adding $\int_{t_E}^{t_E + \delta t_E}\frac{cdt}{a(t)}$ in both sides of the equation:
\be
\int_{t_E}^{ t_R + \delta t_R}\frac{cdt}{a(t)} = \int_{t_E}^{t_R}\frac{cdt}{a(t)}+\int_{t_E}^{t_E + \delta t_E}\frac{cdt}{a(t)}
\ee
which gives:
\be
\int_{t_R}^{ t_R + \delta t_R}\frac{cdt}{a(t)} =
\int_{t_E}^{ t_E + \delta t_E}\frac{cdt}{a(t)}
\ee

If $\delta t_R$ is really small, we can consider a(t) to be constant. Therefore, the above equation gives:
\be
\frac{\delta t_R}{a(t_R)} = \frac{\delta t_E}{a(t_E)} \Rightarrow \frac{\delta t_R}{\delta t_E)} = \frac{a(t_R)}{a(t_E)}
\ee

Thus, for~the redshift, we obtain:
\be
z \equiv \frac{\lambda_R - \lambda_E}{\lambda_E} \Rightarrow 1+z = \frac{\lambda_R}{\lambda_E}
\ee
which by using: $c = l\nu$ takes the form:
\be
1+z = \frac{\frac{1}{\nu_R}}{\frac{1}{\nu_E}} = \frac{\nu_E}{\nu_R}
\ee

Finally:
\be
1+z = \frac{\nu_E}{\nu_R} = \frac{\frac{1}{\delta t_E}}{\frac{1}{\delta t_R}} = \frac{\delta t_R}{\delta t_E} = \frac{a(t_R)}{a(t_E)}
\ee

In cosmology we define the redshift parameter z as the fractional shift in wavelength of a photon emitted by a distant galaxy at time $t$ and observed on Earth today at $t_0$.
Thus, by~replacing $t_R$ with $t_0$ and $t_E$ simply with $t$, we obtain:
\be
z \equiv \frac{\lambda_0 - \lambda}{\lambda}\Rightarrow
1+z\equiv \frac{\lambda_0}{\lambda} = \frac{\nu}{\nu_0}
\ee
which gives:
\be
1+z = \frac{a(t_0)}{a(t)} = \frac{1}{a(t)}
\ee
using that:\hspace{2mm}$a(t_0) = a_0 = 1$.\\

To better understand the concept of redshift and scale factor, consider the following example. When we observe a galaxy with a redshift $z=2$, we are observing it as it was at time $t = t_E$ when the Universe had a scale factor $a(t_E) = 1/3$, and thus, it was 1/3 of its present~size.
\subsection{Hubble's~Law}
\subsubsection{Edwin~Hubble}

According to Hubble's Law, named after Edwin Hubble, who discovered the expansion of the Universe in 1929, galaxies appear to recede from us with a recession speed proportional to their distance from us~\cite{Hobson:2006se}:
\be
u = H_0d
\ee

The proportionality constant $H_0$ is called the Hubble constant and measures the current expansion of the~Universe.
         \label{fig:Hubble Law}
\unskip

\subsubsection{Hubble~Constant}

The expansion of the Universe at any time t is measured by the Hubble parameter:
\be
H = \frac{\dot{a}(t)}{a(t)}
\ee
so we can write $H_0$ as:
\be
H_0 = \frac{\dot{a}(t_0)}{a(t_0)}
\ee

Measurements of the Hubble constant have a lot of uncertainties, so it is conventional to define:
\be
H_0 \equiv 100\cdot h\cdot kms^{-1}Mpc^{-1}
\ee
with h being a dimensionless parameter that helps us keep track of how uncertainties in $H_0$ propagate into other cosmological parameters.
As h is dimensionless, we can discern the dimensions of the Hubble constant: $\rm kms^{-1}Mpc^{-1}$. Furthermore, we can use the inverse of the Hubble constant to obtain a rough approximation of the age of the~Universe.
\subsubsection{Physical Density~Parameters}

Now that we have defined the dimensionless Hubble parameter $h$, we can also define the physical density parameters as:
\be
\omega_{l,0} \equiv \Omega_{l,0} h^2 
\ee

Thus, for~example, $\omega_m = \Omega_m\cdot h^2 $ is the physical matter density parameter and  $\omega_k = \Omega_k\cdot h^2 $ is the physical curvature density parameter.
\subsubsection{Derivation}

Consider a photon coming from a nearby galaxy emitted at time $t$ and received today at $t = t_0$. Since the galaxy is nearby $(t-t_0 <<)$, we can expand the scale factor $a(t)$ around $t_0$ as:
\be\label{eq:1.36}
a(t) = a(t_0 -(t_0-t)) = a(t_0) - (t_0 - t) \dot{a}(t)|_{t = t_0} + \dots
\ee
which gives:
\be
a(t) = a(t_0)\left[ 1 - (t_0 - t)H_0 + \dots\right]
\ee

Now, along the photons path: $ds = d\theta = d\phi = 0$. Thus, Equation~(\ref{eq:1.2})  gives:
\be
\chi = \int_t^{t_0}\frac{cdt'}{a(t')} 
\ee
while also considering a flat universe. Using Equation~(\ref{eq:1.36}),  we obtain:
\be
\chi = \int_t^{t_0}ca^{-1}(t_0)\left[ 1 - (t_0 - t)H_0\right]^{-1}dt'
\ee

Using the fact that $t-t_0 <<$ and $(a+x)^n \simeq (a+nx)$:
\be
\chi \simeq \int_t^{t_0}ca^{-1}(t_0)\left[ 1 + (t_0 - t)H_0\right]dt'
\ee
which, by~evaluating the integral, gives:
\be
\chi = ca^{-1}\left[t_0 - t + \frac{(t_0 - t)^2}{2}H_0\right]
\ee

To prove the final equation, we need to express $t - t_0$, commonly called look-back time, in~terms of the redshift z. However,~it is easier to expand z in terms of $t-t_0$ and then invert the relation.
\be
z = \frac{a(t_0)}{a(t)} - 1 = \left[ 1 - (t_0 - t)H_0\right]^{-1} - 1
\ee

As we implemented before, as~$t-t_0 <<$:
\be
z \simeq \left[ 1 - (t_0 - t)H_0\right] - 1 = (t_0 - t)H_0
\ee

Thus, by~inverting it, we obtain:
\be
t_0-t \simeq \frac{z}{H_0}
\ee

Thus:
\be
\chi \simeq \frac{c}{a(t_0)}\left[ \frac{z}{H_0} + \frac{\left(\frac{z}{H_0}\right)^2}{2}H_0\right] = \frac{cz}{a(t_0)H_0}+\frac{c}{a(t_0)H_0}z^2
\ee

In first order to z:
\be
\chi \simeq \frac{cz}{a(t_0)H_0} \Rightarrow a(t_0)\chi \simeq \frac{c}{H_0}z \Rightarrow H_0 \cdot a(t_0) \chi \simeq cz
\ee

Considering z as a ``Doppler shift'' due to a recession velocity $u$:
\be
u = cz
\ee

The distance to the emitting galaxy solely due to cosmic expansion at $t = t_0 $ is:
\be
d = a(t_0)\chi
\ee

Later we will see that this is the physical distance $r_p$.
Thus:
\be
u \simeq H_0 \cdot d\hspace{1cm} \text{for}\hspace{0.25cm} z<<.
\ee
\subsection{Distances in~Cosmology}

In cosmology, there are many different ways to specify the distance between two points in space.
Let us start by assuming that we sit at $(t_0,0,0,0)$, and we observe a remote comoving object emitting light from $(t_e,r,\theta,\phi)$ and the geometry of spacetime is described by the line element:
\be\label{eq:1.54}
ds^2 = dt^2-a^2(t)\left[d\chi^2+S^2(\chi)(d\theta^2+sin^2\theta d\phi^2)\right]
\ee
\subsubsection{Theoretically defined~Distances}
This kind of distance cannot be observed, but it is very useful in defining observable distances.
\paragraph{{Metric Distance}}
The distance multiplying the solid angle $d\Omega^2 = d\theta^2+sin^2\theta d\phi^2$ is the metric distance:%
\be\label{eq:1.39}
d_m = S(\chi)
\ee
which means that for a flat universe, the metric distance is simply equal to the comoving distance $\chi$:
\be
d_m = \chi
\ee
 distance is not observable but is useful in defining observable~distances.
 \begin{figure}[H]
         \includegraphics[width=12cm]{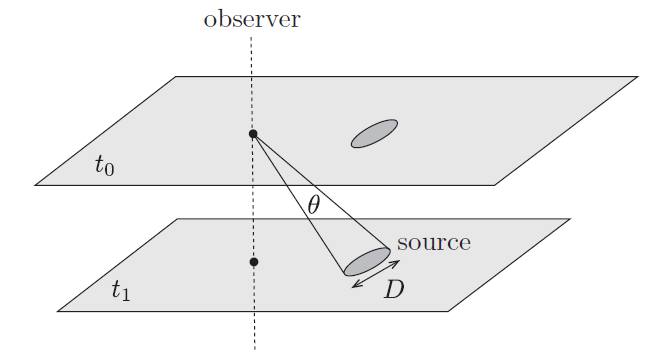}
         \caption{ Figure that helps with the definition of  angular diameter distance. {Adapted from}: \cite{Baumann_cosmology}}
         \label{fig:Angular Diameter Distance}
     \end{figure}
\noindent
\paragraph{{Comoving}/Coordinate Distance}%
The comoving distance between two objects in the Universe is the distance between them that remains constant with the epoch, when they are both moving only with the Hubble flow, i.e.,  they move solely due to the expansion of the~Universe.

Setting $ds=0$ and $d\Omega = 0$ in Equation~(\ref{eq:1.36}), we obtain:
\be
d_{co} = \int_{t_e}^{t_0}\frac{dt'}{a(t')} = \int_{a_e}^{a_0}\frac{da}{a\dot{a}} = \int_{0}^{z}\frac{dz'}{a_0H(z')} =\frac{1}{a_0H_0} \int_{0}^{z}\frac{dz'}{E(z')}
\ee
where: \be
E(z) \equiv \frac{H(z)}{H_0}
\ee
is the dimensionless Hubble~parameter.
\subsubsection{Observable~Distances}
\paragraph{{Physical Distance}}
It is the actual proper distance $r_p$ between two objects in the Universe that can be measured by a physical ruler. It is related to the comoving distance $d_{co}(z)$ through:
\be
r_p(z) = a(t)d_{co}(z)
\ee

\paragraph{Angular Diameter Distance}
Let us assume that we observe an object at a comoving distance $\chi$ and that the photons that we observe today were emitted at time $t_1$. Assuming that the object has a known physical size D and someone on Earth measures its angular size to be $\delta \theta$, we can define the angular diameter distance as:
\be
d_A = \frac{D}{\delta \theta}
\ee
which for small $\delta \theta \ll 1$ is the Euclidean formula for its~distance.

Through the FRW metric, we can find the relation between the physical transverse size of the object and its angular size on the sky:
\be
D = a(t_1)S(\chi)\delta \theta = \frac{d_m}{1+z}\delta \theta
\ee

Hence:
\be\label{eq:1.46}
d_A = \frac{d_m}{1+z}
\ee

The angular diameter distance is really useful because of objects of known physical size D, which we call ``standard rulers'' as, for~example, the~fluctuations in the~CMB.
\begin{figure}[H]
         \includegraphics[width=12cm]{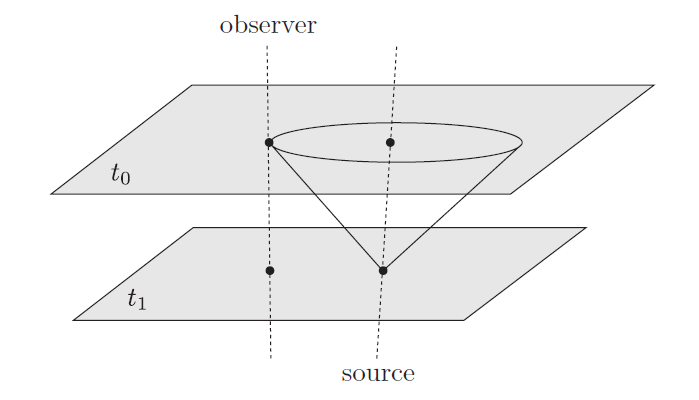}
         \caption{Figure that helps with the definition of Luminosity distance. {Adapted from}: \cite{Baumann_cosmology}.}
         \label{fig:Luminosity distance}
     
     \end{figure}

\paragraph{Luminosity Distance}
Let us consider a luminous cosmological object at a fixed comoving distance $\chi$, which emits at an absolute luminosity \emph{L}. In~a static Euclidean space, the~relation between the absolute luminosity and the observed flux \emph{F} would be:
\be
F = \frac{L}{4\pi \chi^2}
\ee
since the power radiated by the luminous object is distributed in the spherical surface with radius $\chi$.
In an FRW spacetime, however, the~result must be modified. This happens~because:
\begin{itemize}

  \item At the time $t_0$ that we observe the light from the object, the~proper area of a sphere drawn around a supernova and passing through the Earth is $4\pi d_m^2$.
  \item The rate at which we detect photons from the object is reduced compared to the rate that they are emitted, by~the redshift factor: $ \frac{a(t_0)}{a(t)} = 1+z$.
  \item The energy of the photons is also being redshifted, so the energy that we observe them to have is reduced compared to the one they had when they were emitted by the same redshift factor: $1+z$.
\end{itemize}

Thus, we obtain the formula: 
\be
F = \frac{L}{4\pi d_m^2}\frac{1}{(1+z)^2}
\ee

We define the luminosity distance to be:
\be
d_L \equiv \sqrt{\frac{L}{4\pi F}}
\ee
so that:
\be
F = \frac{L}{4\pi d_m^2}\frac{1}{(1+z)^2}\equiv \frac{L}{4\pi d^2_L}
\ee

Hence, we find that:
\be
d_L = d_m(1+z)
\ee

The luminosity distance proves to be very useful because of objects called standard candles, which are objects of "known" absolute luminosity, or~more appropriately of an absolute luminosity that we can estimate independently of their distance and apparent luminosity. Such objects are variable stars called cepheids or a special type of supernovae (SN) called Type Ia~supernovae.
\begin{figure}[H]
     \begin{subfigure}[b]{4.5cm}
     \captionsetup{justification=centering}
         \centering
         \includegraphics[width=4.5cm]{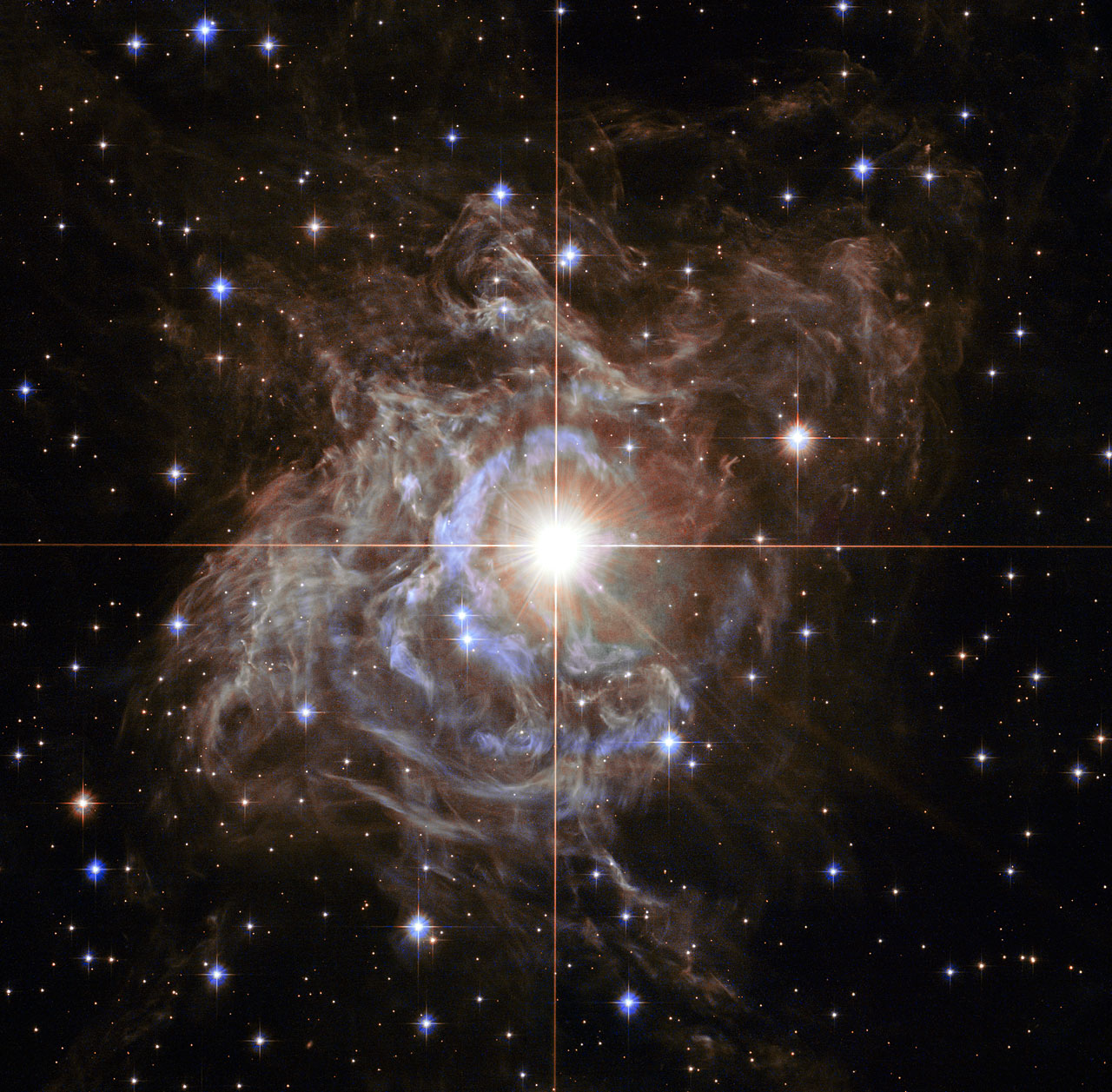}
         \caption{}
         \label{fig:RS Puppis Cepheid}
     \end{subfigure}
     \hfill
     \begin{subfigure}[b]{8cm}
     \captionsetup{justification=centering}
         \includegraphics[width=8cm]{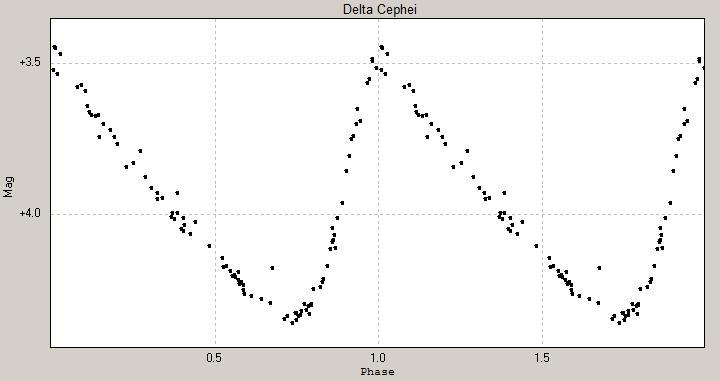}
         \caption{}
         \label{fig:Cepheid Light Curve}
     \end{subfigure}
        \caption{(\textbf{a}) This Hubble image shows RS Puppis, a~type of variable star known as a Cepheid variable. {Adapted from}: \cite{hubble_rsrup}. (\textbf{b}) Phase lightcurve of variable star Delta Cepheid. {Adapted} \mbox{from: \cite{hubble_ceplig}.}}
        \label{fig:Cepheid images}
\end{figure}

\paragraph{Distance Modulus}
The apparent magnitude m of an astronomical object is defined as: 
\be
m = -2.5 \log_{10}\left( \frac{F}{F_{ref}}\right)
\ee
where \emph{F} is the apparent flux of the object, and $F_{ref}$ is a reference~flux.

The absolute magnitude M is defined as the apparent magnitude the object would have if it was 10 pc away from the observer:
\be
M = -2.5\log_{10}\left( \frac{F_{10pc}}{F_{ref}}\right)
\ee

The distance modulus $\mu$ is defined as the difference between them and after some calculations can be expressed as:
\be\label{eq:1.87}
\mu \equiv m - M = 5\log_{10}\left(\frac{d_L}{10pc}\right)
\ee
where $d_L$ is the luminosity~distance.

If the distance is expressed in Mpc, then we can write $\mu$ in the form:
\be\label{eq:1.55}
\mu = 5log_{10}d_L + 25
\ee

This kind of distance is very useful when we use data from supernovae.
\section{Theoretical~Background}
As we discussed earlier, in~1929, Hubble showed that the Universe is expanding, as~every other galaxy appeared receding from us with a recession speed analogous to its distance. The~proportionality constant is $H_0 = 100\,\rm 
h\,kms^{-1}\,Mpc^{-1}$.
We have already talked about observable distances, and thus, we can identify that the distance that was measured and used in Hubble's Law is actually the physical distance $r_p$, so:
\be
u = H_0\cdot r_p
\ee

However, some later results taken by probing supernovae in high redshifts showed that this linear relation does not hold anymore, and it seems that the expansion is speeding up. Thus, the~latest data imply that we live not only in an expanding universe, but~in a universe with an accelerating~expansion.

\subsection{Dark~Energy}
In a universe described by general relativity and that is matter-dominated (as it was thought to be by cosmologists in the last century), one would expect the expansion to be slowing down due to the influence of gravity. Therefore, the second derivative of the expansion was named the deceleration~parameter:
\begin{figure}[H]
         \includegraphics[width=7cm]{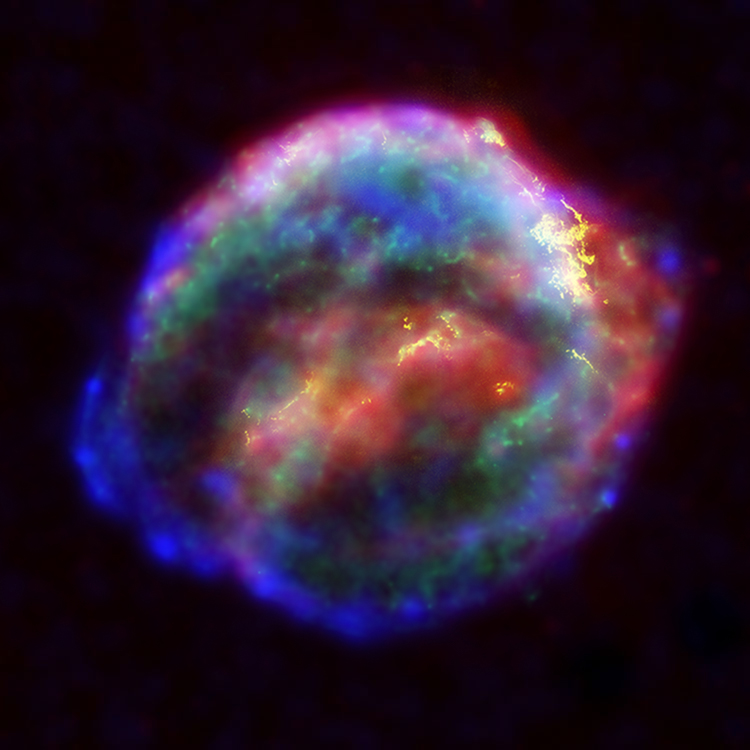}
         \caption{The remnants of Supernova 1604 or Kepler's Supernova. It was a Type Ia supernova that occurred in our galaxy in the constellation of Ophiuchus. It was observable by the naked eye and was named by Johannes Kepler who described it in: De Stella Nova. {Adapted from}: \cite{kepler_sup}.}
         \label{fig:Kepler's Supernova}
\end{figure}
\be
q(z) \equiv - \frac{\ddot{a}}{aH^2}
\ee

We can see that matter can only lead to decelerating expansion using the acceleration equation and its equation of state:
\be
\frac{\ddot{a}}{a} = -\frac{4 \pi G}{3}\rho_m 
\ee
\be
w_m =\frac{p_m}{\rho_m} = 0
\ee

As does radiation:
\be
\frac{\ddot{a}}{a} = -\frac{4 \pi G}{3}\left(2\rho_r \right)
\ee
\be
w_m =\frac{p_r}{\rho_r} = \frac{1}{3}
\ee
since $a(t)$, $\rho_m$ and $\rho_r$ are positive.
In the context of general relativity, the~only way we can obtain an accelerating expansion is by assuming that there is an additional component in our Universe called ``dark~energy''. 
In order to have an equation that describes the universe as it is, we need dark energy to have~\cite{Perivolaropoulos:2006ce}:
\begin{itemize}

  \item A positive energy density $(\rho_X > 0)$, assuming that the universe is flat.
  \item A negative pressure $(p_X<0)$, which  can cancel out gravity and potentially lead to accelerating expansion.
\end{itemize}

Adding the dark energy term in the Friedmann acceleration equation, yields:
\be
\frac{\ddot{a}}{a} = -\frac{4 \pi G}{3}\left(\rho_m+\rho_X + 3p_X\right) = -\frac{4 \pi G}{3}\left[\rho_m+\rho_X\left(1 + 3w\right)\right]
\ee
with:
\be\label{eq:2.7}
p_X = w\rho_X
\ee 

This means that if we want to end up with an accelerating expansion of the universe, we need at least:
\be
w<-\frac{1}{3}
\ee

This results in a form of repulsive gravity.

Now, using the Friedmann Equation~(\ref{eq:1.26})  for a flat universe having only matter and dark energy, we obtain:
\be\label{eq:2.10}
H^2 = \frac{8\pi G}{3}\left[ \rho_m\left(\frac{a_0}{a}\right)^3+\rho_{X}(a)\right] = H_0^2\left[\Omega_{m}a^{-3} + \Omega_{X}\right] 
\ee
where:
\begin{itemize}
  \item $\Omega_{m}$ is the matter density parameter.
  \item $\Omega_{X}(z)$ is the dark energy density parameter.
\end{itemize}

In terms of the redshift $z$, Equation~(\ref{eq:2.10})  takes the form:
\be
H^2 = H_0^2\left[\Omega_{m}(1+z)^{3} + \Omega_{X}(z)\right]
\ee

\section{Data~Analysis}\label{section:datanal}
\subsection{Useful~Functions}
\subsubsection{Gamma~Function}
\begin{figure}[H]
     \begin{subfigure}[b]{6cm}
     \captionsetup{justification=centering}
         \includegraphics[width=6cm]{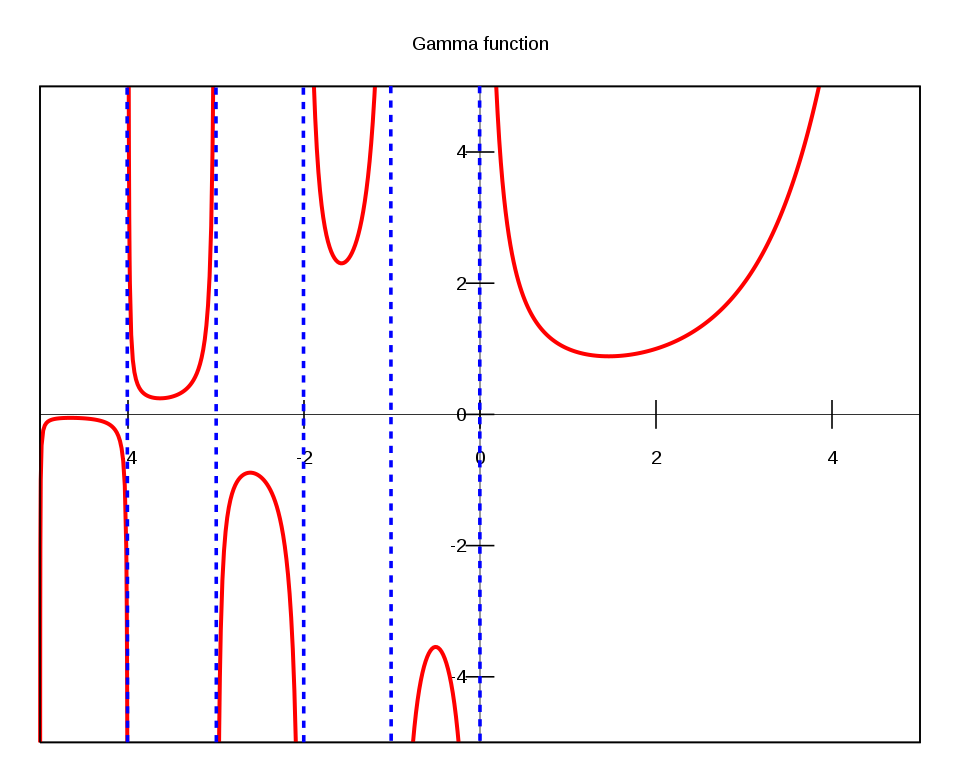}
         \caption{}
         \label{fig:Gamma_func}
     \end{subfigure}
     \begin{subfigure}[b]{6.4cm}
     \captionsetup{justification=centering}
         \includegraphics[width=6.4cm]{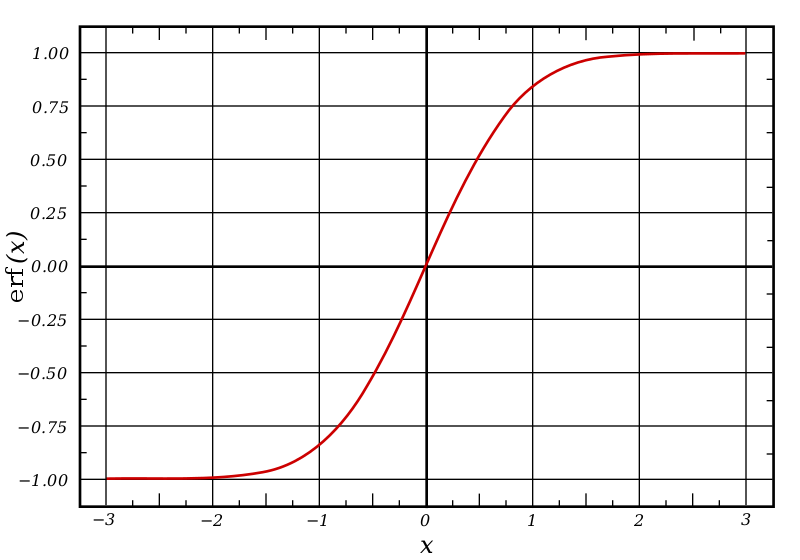}
         \caption{}
         \label{fig:Error_func}
     \end{subfigure}
        \caption{(\textbf{a}) Plot of the gamma function along the real axis. {Adapted from}: \cite{gamma_func}. (\textbf{b}) Plot of the error function. {Adapted from}: \cite{error_func}.}
        \label{fig:Func_plots_01}
\end{figure}

The gamma function is defined using the integral:
\be
\Gamma(z) = \int_0^\infty t^{z-1}e^{-t}dt
\ee

It satisfies the recurrence relation:
\be
\Gamma(z+1) = z\Gamma(z)
\ee
and if $z$ is a natural number $n\in \mathbb{N}$,x then:
\be
\Gamma(n+1) = n!
\ee

\paragraph{Incomplete Gamma Function}
The incomplete gamma function is defined using the gamma function as:
\be
P(a,x) \equiv \frac{\gamma(a,x)}{\Gamma(a)}\equiv\frac{1}{\Gamma(a)}\int_0^x e^{-t}t^{a-1}dt
\ee
for $a>0$. It is also common to use $Q(a,x)$, the~compliment of $P(a,x)$, which is also called an incomplete gamma function:
\be
Q(a,x) \equiv 1- P(a,x)\equiv \frac{1}{\Gamma(a)}\int_x^\infty e^{-t}t^{a-1}dt
\ee
for $a>0$. These two functions have the limiting values:

\be
P(a,0) = 0
\ee
\be
\lim_{x \to \infty} P(a,x) = 1
\ee
\be
Q(a,0) = 1
\ee
\be
\lim_{x \to \infty} Q(a,x) = 0
\ee

Except the integral form for $\gamma(a,x)$, there is a series development:
\be
\gamma(a,x) = e^{-x}x^a\sum_{n = 0}^\infty\frac{\Gamma(a)}{\Gamma(a+1+n)}x^n
\ee
\paragraph{Error Function}
The error function is a special case of the incomplete gamma function and is \mbox{defined as:}
\be
erf(x) = P\left(\frac{1}{2},x^2 \right) = \frac{2}{\sqrt{\pi}}\int_0^x e^{-t^2}dt
\ee

The error function has the following properties:

     \be 
     erf(0) = 0
     \ee
     \be
     \lim_{x \to \infty}erf(x) = 1 
     \ee
     \be
     erf(-x) = -erf(x)
     \ee

\subsubsection{Useful~Distributions}
\paragraph{Normal or Gaussian Distribution}
The probability density function $p(x)$ for a Gaussian distribution with mean $\mu$ and standard deviation $\sigma$ is:
\be
p(x) = \frac{1}{\sqrt{2\pi\sigma}}\text{exp}\left( -\frac{1}{2}\left[ \frac{x-\mu}{\sigma}\right]^2\right)
\ee

The variance of this distribution is $\sigma^2$.
\begin{figure}[H]
     \begin{subfigure}[b]{7cm}
     \captionsetup{justification=centering}
         \centering
         \includegraphics[width=7cm]{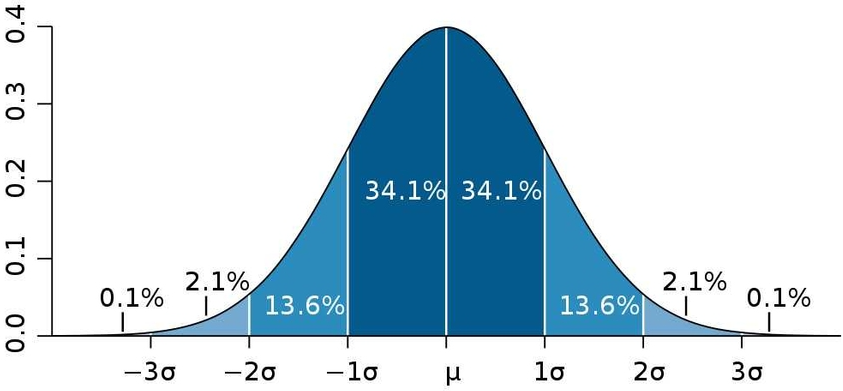}
         \caption{}
         \label{fig:Gaussian_distr}
     \end{subfigure}
     \begin{subfigure}[b]{6cm}
     \captionsetup{justification=centering}
         \includegraphics[width=6cm]{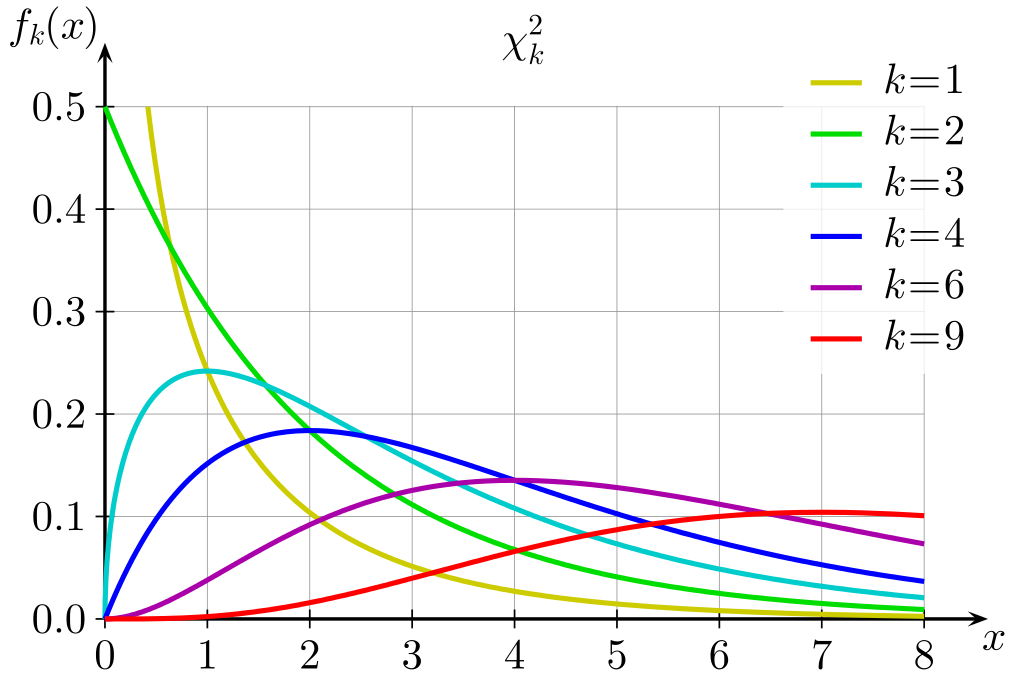}
         \caption{}
         \label{fig:Chi_distr}
     \end{subfigure}
        \caption{(\textbf{a}) Probability density function for the Gaussian distribution. {Adapted from}: \cite{sigma_MIT}.  (\textbf{b}) Probability density function for the chi-square distribution. {Adapted from}: \cite{chisq_dist}.}
        \label{fig:Distr_plots_01}
\end{figure}
\paragraph{\texorpdfstring{Chi-Square or $\chi^2$ Distribution}{}}
The $\chi^2$ distribution has a single parameter $\nu>0$, which controls both the location and the width of its peak. In~most applications $\nu\in \mathbb{N}$ and is reffered to as the number of degrees of freedom~\cite{10.5555/1403886}.

The probability density function for the $\chi^2$ distribution is:
\be
p(\chi^2) = \frac{1}{2^{\frac{1}{2}\nu}\Gamma(\frac{1}{2}\nu)}\left(\chi^2\right)^{\frac{1}{2}\nu-1}\text{exp}\left( -\frac{1}{2}\chi\right)d\chi^2 \hspace{1cm} ,\chi^2>0
\ee

Here, $\chi^2$ is used as the variable, since this is the independent variable and not $\chi$.

The mean value and the variance of the distribution~is:
\begin{itemize}
    \item Mean$\{\chi^2(\nu)\} = \nu$
    \item Var$\{\chi^2(\nu)\} = 2\nu$
\end{itemize}

The $\chi^2$ distribution is actually a special case of the gamma distribution and thus its cumulative distribution can be written in terms of the incomplete gamma function~\cite{10.5555/1403886}:
\be
P(<\chi^2)\equiv\int_0^{\chi^2}p\left(\chi^{2'}\right)d\chi^{2'} = P\left(\frac{\nu}{2},\frac{\chi^2}{2} \right)
\ee

\subsubsection{Derivation of $\Delta \chi^2$ for given confidence region in parameter space.}
\label{subsection:Derivation_01}
\vspace{-6pt}
$$ P(<\Delta\chi^2) = P(<\chi_2^2) - P(<\chi_1^2) = P\left(\frac{\nu}{2},\frac{\chi^2_2}{2}\right) - P\left(\frac{\nu}{2},\frac{\chi^2_1}{2}\right) =  $$
$$ = \frac{1}{\Gamma\left( \frac{\nu}{2}\right)}\int_0^{\frac{\chi^2_2}{2}} e^{-t}t^{\frac{\nu}{2}-1}dt-\frac{1}{\Gamma\left( \frac{\nu}{2}\right)}\int_0^{\frac{\chi^2_1}{2}} e^{-t}t^{\frac{\nu}{2}-1}dt = \frac{1}{\Gamma\left( \frac{\nu}{2}\right)}\int_0^{\frac{\Delta\chi^2}{2}} e^{-t}t^{\frac{\nu}{2}-1}dt = $$
\be
 = 1 - Q\left( \frac{\nu}{2},\frac{\Delta\chi^2}{2}\right) = 1-\frac{\Gamma\left(\frac{\nu}{2},\frac{\Delta\chi^2}{2} \right)}{\Gamma\left(\frac{\nu}{2}\right)}
\ee
for $\Delta\chi^2 = 1$ and $\nu = 1$:
\be
P = 1-\frac{\Gamma\left(\frac{1}{2},\frac{1}{2} \right)}{\Gamma\left(\frac{1}{2}\right)} = 1- \frac{0.562418}{1.77245} = 0.6827 \approx 0.683
\ee
for $\Delta\chi^2 = 9.72$ and $\nu = 4$:
\be
P = 1-\frac{\Gamma\left(\frac{4}{2},\frac{9.72}{2} \right)}{\Gamma\left(\frac{4}{2}\right)} = 1- \frac{0.00454178}{1} = 0.95458 \approx 0.9545
\ee

All the other values in  table \ref{tab: Delta-chi_square_01} can be obtained the same~way.

A useful way to obtain these values is through Mathematica by using the code in Figure~\ref{fig:Mathematica_dchi}.
\begin{figure}[H]
         \includegraphics[width=12cm]{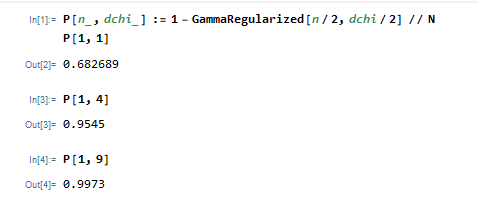}
         \caption{Mathematica code for the derivation of {Table}~\ref{tab: Delta-chi_square_01}, where n is the number of degrees of freedom and dchi is the value of $\Delta\chi^2$.}
         \label{fig:Mathematica_dchi}
\end{figure}
\vspace{-6pt}
\section{\texorpdfstring{Proof of \boldmath{$dg=gg^{ab}dg_{ab}$}}{}}\label{section:proof1}

At first, we assume the diagonal matrix:
\be
A = \begin{pmatrix}
a_0 & 0 & 0 & \dots\\
0 & a_1 & 0 & \dots  \\
\vdots & \vdots &  \ddots & \vdots \\
0 & 0 & \dots & a_{n}
\end{pmatrix}
\ee

Then:
\be
e^A = \begin{pmatrix}
e^{a_0} & 0 & 0 & \dots\\
0 & e^{a_1} & 0 & \dots  \\
\vdots & \vdots &  \ddots & \vdots \\
0 & 0 & \dots & e^{a_{n}}
\end{pmatrix}
\ee

By computing the determinant of the matrix $e^A$, we obtain:
\be
\det(e_A) =e^{a_0}\cdot e^{a_1}\cdots e^{a_n}=e^{a_0+a_0+\dots a_n} = e^{tr(A)}
\ee

Assuming a new matrix B with:
\be
B = e^A \Rightarrow A = \ln{B}
\ee
we obtain:
\be
\det (B) = e^{tr(\ln{B})}\Rightarrow\ln{\det B} = tr(\ln{B})\Rightarrow d[\ln{\det B}] = d[tr(\ln{B})]
\ee
which gives:
\be
\frac{d(\det B)}{\det B} = tr\left( \frac{dB}{B}\right)
\ee

In this case we have: $B = g_{ab}\Rightarrow B^{-1} = g^{ab}$ and $\det B = g$. Thus:
\be
\frac{dg}{g} = g^{ab}dg_{ab}\Rightarrow dg = gg^{ab}dg_{ab}
\ee
\section{Derivation of the Klein-Gordon~Equation}\label{section:klein}
The general scalar field action in Riemann spacetime is:
\be
S_\phi = \int\sqrt{-g}\mathcal{L}_\phi(\phi,\partial_\alpha\phi)d^4x
\ee

In a spacetime with signature $(+,-,-,-)$, the~Lagrangian density of $\phi$ is:
\be
\mathcal{L}_\phi = \frac{1}{2}g^{\mu\nu}\partial_\mu\phi\partial_\nu\phi - V(\phi)
\ee

For any region $\Omega$, we consider variation of the field:
\be
\phi(x)\longrightarrow \phi(x) + \delta(x)
\ee
which vanishes on the surface $\Gamma(\Omega)$, bounding the region $\Omega$.\\
\be
\delta\phi(x) = 0 \hspace{0.25cm} \text{on}\hspace{0.25cm} \Gamma(\Omega)
\ee
\be
\delta S(\Omega) = \delta\left( \int_\Omega \sqrt{-g}\mathcal{L}_\phi(\phi,\partial_a\phi)d^4x\right) = \int_\Omega\left[\frac{\partial(\sqrt{-g}\mathcal{L}_\phi)}{\partial\phi}\delta\phi+\frac{\partial(\sqrt{-g}\mathcal{L}_\phi)}{\partial(\partial_a \phi)}\delta(\partial_a\phi)\right]
\ee
\be
\frac{\partial(\sqrt{-g}\mathcal{L}_\phi)}{\partial(\partial_a \phi)}\delta(\partial_a\phi) = \frac{\partial}{\partial x^a}\left[\frac{\partial(\sqrt{-g}\mathcal{L}_\phi)}{\partial(\partial_a \phi)}\delta\phi\right]-\frac{\partial}{\partial x^a}\left[\frac{\partial(\sqrt{-g}\mathcal{L}_\phi)}{\partial(\partial_a \phi)}\right]\delta\phi
\ee
$$
\delta S(\Omega) = \int_\Omega\left[\frac{\partial(\sqrt{-g}\mathcal{L}_\phi)}{\partial\phi}\delta\phi-\frac{\partial}{\partial x^a}\left(\frac{\partial(\sqrt{-g}\mathcal{L}_\phi)}{\partial(\partial_a \phi)}\right) \right]\delta\phi dx^4+
$$
\be
+\int_\Omega\left[\frac{\partial}{\partial x^a}\left(\frac{\partial(\sqrt{-g}\mathcal{L}_\phi)}{\partial(\partial_a \phi)}\right)\delta\phi \right] dx^4
\ee

Using the divergence theorem, we obtain:
\be
\int_\Omega \partial_a\left[\left(\frac{\partial(\sqrt{-g}\mathcal{L}_\phi)}{\partial(\partial_a \phi)}\right)\delta\phi\right]d^4x = \int_{S = \Gamma(\Omega)}\frac{\partial\sqrt{-g}\mathcal{L}_\phi}{\partial(\partial_a\phi)}\delta\phi\cdot ndS = 0
\ee
since $\delta\phi = 0 $ on $S = \Gamma(\Omega)$.

For $\delta S(\Omega) = 0 $, we derive the Euler--Lagrange equation:
\be
\delta S(\Omega) = 0 \Rightarrow \int_\Omega\left[\frac{\partial(\sqrt{-g}\mathcal{L}_\phi)}{\partial\phi}\delta\phi-\frac{\partial}{\partial x^a}\left(\frac{\partial(\sqrt{-g}\mathcal{L}_\phi)}{\partial(\partial_a \phi)}\right) \right]\delta\phi dx^4 = 0
\ee
Thus,
\be
\frac{\partial(\sqrt{-g}\mathcal{L}_\phi)}{\partial\phi}\delta\phi-\frac{\partial}{\partial x^a}\left(\frac{\partial(\sqrt{-g}\mathcal{L}_\phi)}{\partial(\partial_a \phi)}\right) = 0
\ee
Calculating each term individually, we obtain:
\be
\frac{\partial}{\partial\phi}(\sqrt{-g}\mathcal{L}_\phi) = \sqrt{-g}\frac{\partial}{\partial\phi}\mathcal{L}_\phi  = \sqrt{-g}\frac{\partial}{\partial\phi}\left( \frac{1}{2}g^{\mu\nu}\partial_\mu\phi\partial_\nu\phi-V(\phi)\right) = -\sqrt{-g}\frac{\partial V}{\partial\phi}
\ee

\begin{align*}
\partial_\lambda\left[\frac{\partial(\sqrt{-g}\mathcal{L}_\phi)}{\partial(\partial_\lambda \phi)}\right] &= \partial_\lambda\left[\frac{\partial}{\partial(\partial_\lambda \phi)}\left(\sqrt{-g}\frac{1}{2}g^{\mu\nu}\partial_\mu\phi\partial_\nu\phi-\sqrt{-g}V(\phi) \right) \right] = \\
&=\partial_\lambda\left[ \sqrt{-g}\frac{1}{2}g^{\mu\nu}(\partial_\nu\phi\delta_{\mu\lambda}+ \partial_\mu\phi\delta_{\nu\lambda})\right] = \\ &= \partial_\mu\left( \sqrt{-g}g^{\mu\nu}\partial_\nu\phi\right) = \\ &= \partial_\mu\left( \sqrt{-g}\right)g^{\mu\nu}\partial_\nu\phi + \partial_\mu\left(g^{\mu\nu} \right)\sqrt{-g}\partial_\nu\phi + \sqrt{-g}g^{\mu\nu}\partial_\mu\partial_\nu\phi = \\&=
\frac{1}{2\sqrt{-g}}\partial_\mu(-g)g^{\mu\nu}\partial_\nu\phi+\sqrt{-g}\partial_\mu(g^{\mu\nu})\partial_\nu\phi+\sqrt{-g}g^{\mu\nu}\partial_\mu\partial_\nu\phi = \\&=
\sqrt{-g}\left(\frac{1}{2g}\partial_\mu(g)g^{\mu\nu}\partial_\nu\phi+\partial_\mu(g^{\mu\nu})\partial_\nu\phi + g^{\mu\nu}\partial_\mu\partial_\nu\phi \right)
\end{align*}

Using the FRW metric:
\be
g_{\mu\nu} = \begin{pmatrix}
1 & 0 & 0 & 0 \\
0 & -a^2(t) & 0 & 0 \\
0 & 0 & -a^2(t)& 0 \\
0 & 0 & 0& -a^2(t) 
\end{pmatrix}
\ee
with:
\be
det(g_{\mu\nu}) = g = g_{\mu\nu} = \begin{vmatrix}
1 & 0 & 0 & 0 \\
0 & -a^2(t) & 0 & 0 \\
0 & 0 & -a^2(t)& 0 \\
0 & 0 & 0& -a^2(t) 
\end{vmatrix} =  -a^6(t)
\ee

Thus:
\be
\partial_0g = \frac{\partial}{\partial t}(-a^6(t)) = -6a^5(t)\frac{d}{dt}(a(t)) = -6a^5\Dot{a}
\ee
\be
\partial_ig = \frac{\partial}{\partial x^i}(a^6(t)) = 0
\ee
\be
\partial_\mu g^{\mu\nu} = \partial_\mu g^{\mu\mu} = \frac{\partial(1)}{\partial t}+\frac{\partial(a^2(t))}{\partial x^1}+\frac{\partial(a^2(t))}{\partial x^2}+\frac{\partial(a^2(t))}{\partial x^3} = 0
\ee
since the metric is diagonal. Furthermore:

\noindent
\be
\partial_0\phi = \Dot{\phi}
\ee
\noindent
\be
\partial_i\phi = \frac{\partial}{\partial x^i}(\phi(t)) = 0
\ee

Thus,
\begin{align*}
\partial_\lambda\left[\frac{\partial}{\partial(\partial_\lambda\phi)}(\sqrt{-g}\mathcal{L}_\phi) \right] &= \sqrt{-g}\left(\frac{1}{2}\partial_\mu(g)g^{\mu\nu}\partial_\nu\phi+\partial_\mu(g^{\mu\nu})\partial_\nu\phi+g^{\mu\nu}\partial_\mu\partial_\nu\phi \right) = \\
&= \sqrt{a^6}\left(-\frac{1}{2a^6}\partial_0(g)g^{0\nu}\partial_\nu\phi+g^{\mu0}\partial_\mu\partial_0\phi \right) = \\ &= a^3\left(-\frac{1}{2a^6}\partial_0(g)g^{00}\partial_0\phi +g^{00}\partial_0\partial_0\phi\right) = \\ &=
a^3\left( -\frac{1}{2a^6}(-6a^5\Dot{a})\Dot{\phi}+\Ddot{\phi}\right) = \\ &=
a^3\left( 3\frac{a}{\Dot{a}}\Dot{\phi}+\Ddot{\phi}\right)
\end{align*}

Thus, lastly, we have:
\be
-\sqrt{-g}\frac{\partial V}{\partial \phi} - a^3\left( 3\frac{a}{\Dot{a}}\Dot{\phi}+\Ddot{\phi}\right) = -a^3\left(\frac{\partial V}{\partial\phi}+3\frac{a}{\Dot{a}}\Dot{\phi}+\Ddot{\phi}\right) =0
\ee
and finally:
\be
\Ddot{\phi}+3\frac{a}{\Dot{a}}\Dot{\phi}+\frac{\partial V}{\partial\phi} = 0
\ee

\end{paracol}
\reftitle{References}


\begin{thebibliography}{999}

\bibitem[Aghanim \em{et~al.}(2020)Aghanim et~al.]{Aghanim:2018eyx}
Aghanim, N.; Akrami, Y.; Ashdown, M.; Aumont, J.; Baccigalupi, C.; Ballardini, M.; Bandy, J.A.; Barreiro, R.B.; Bartolo, N.; \mbox{Basak, S.;} et al. 
\newblock {Planck 2018 results. VI. Cosmological parameters}.
\newblock {\em Astron. Astrophys.} {\bf 2020}, {\em 641},~A6,
 doi:10.1051/0004-6361/201833910.

\bibitem[Riess \em{et~al.}(2020)Riess, Casertano, Yuan, Bowers, Macri, Zinn,
  and Scolnic]{Riess:2020fzl}
Riess, A.G.; Casertano, S.; Yuan, W.; Bowers, J.B.; Macri, L.; Zinn, J.C.;
  Scolnic, D.
\newblock {Cosmic Distances Calibrated to 1\% Precision with Gaia EDR3
  Parallaxes and Hubble Space Telescope Photometry of 75 Milky Way Cepheids
  Confirm Tension with LambdaCDM} \emph{ Astrophys. J. Lett.} {\bf 2020}, \emph{908}, L6.

\bibitem[Camarena and Marra(2020{\natexlab{a}})]{Camarena:2019rmj}
Camarena, D.; Marra, V.
\newblock {A new method to build the (inverse) distance ladder}.
\newblock {\em Mon. Not. R. Astron. Soc.} {\bf 2020}, {\em 495},~2630--2644,  doi:10.1093/mnras/staa770.

\bibitem[Camarena and Marra(2020{\natexlab{b}})]{Camarena:2019moy}
Camarena, D.; Marra, V.
\newblock {Local determination of the Hubble constant and the deceleration
  parameter}.
\newblock {\em Phys. Rev. Res.} {\bf 2020}, {\em 2},~013028,  doi:10.1103/PhysRevResearch.2.013028.

\bibitem[Scolnic \em{et~al.}(2018)Scolnic et~al.]{Scolnic:2017caz}
Scolnic, D.M.; Jones, D.O.; Rest, A.; Pan, Y.C.; Chornock, R.; Foley, R.J.; Huber, M.E.; Kessler, R.; Narayan, G.; Riess, A.G.; et al. 
\newblock {The Complete Light-curve Sample of Spectroscopically Confirmed SNe
  Ia from Pan-STARRS1 and Cosmological Constraints from the Combined Pantheon
  Sample}.
\newblock {\em Astrophys. J.} {\bf 2018}, {\em 859},~101,  doi:10.3847/1538-4357/aab9bb.

\bibitem[Alestas \em{et~al.}(2020)Alestas, Kazantzidis, and
  Perivolaropoulos]{Alestas:2020mvb}
Alestas, G.; Kazantzidis, L.; Perivolaropoulos, L.
\newblock {$H_0$ tension, phantom dark energy, and cosmological parameter
  degeneracies}.
\newblock {\em Phys. Rev. D} {\bf 2020}, {\em 101},~123516, doi:10.1103/PhysRevD.101.123516.

\bibitem[Alestas and Perivolaropoulos(2021)]{Alestas:2021xes}
Alestas, G.; Perivolaropoulos, L.
\newblock {Late time approaches to the Hubble tension deforming $H(z)$, worsen
  the growth tension}.
\newblock {\em Mon. Not. R.  Astron. Soc.} {\bf 2021}, {\em 504},~3956, doi:10.1093/mnras/stab1070.

\bibitem[Frieman \em{et~al.}(2008)Frieman, Turner, and Huterer]{Frieman:2008sn}
Frieman, J.; Turner, M.; Huterer, D.
\newblock {Dark Energy and the Accelerating Universe}.
\newblock {\em Ann. Rev. Astron. Astrophys.} {\bf 2008}, {\em 46},~385--432, doi:10.1146/annurev.astro.46.060407.145243.

\bibitem[Chevallier and Polarski(2001)]{Chevallier:2000qy}
Chevallier, M.; Polarski, D.
\newblock {Accelerating universes with scaling dark matter}.
\newblock {\em Int. J. Mod. Phys. D} {\bf 2001}, {\em 10},~213--224, doi:10.1142/S0218271801000822.

\bibitem[Linder(2003)]{Linder:2002et}
Linder, E.V.
\newblock {Exploring the expansion history of the universe}.
\newblock {\em Phys. Rev. Lett.} {\bf 2003}, {\em 90},~091301,  doi:10.1103/PhysRevLett.90.091301.

\bibitem[Farooq(2013)]{Farooq:2013syn}
Farooq, M.O.
\newblock {Observational Constraints on Dark Energy Cosmological Model
  Parameters}.
\newblock Ph.D. Thesis, Kansas State University, Manhattan, KS, USA,  2013.

\bibitem[Amendola and Tsujikawa(2015)]{Amendola:2015ksp}
Amendola, L.; Tsujikawa, S.
\newblock {\em {Dark Energy}: {Theory and Observations}}; Cambridge University
  Press: Cambridge, UK,  2015.

\bibitem[g29()]{g299sup}
G299 Type Ia Supernova.
\newblock
  Available online:  \url{https://www.nasa.gov/sites/default/files/thumbnails/image/g299.jpg} (accessed on 7  July 2020).

\bibitem[Signore and Puy(2001)]{Signore:2000mg}
Signore, M.; Puy, D.
\newblock {Supernova and cosmology}.
\newblock {\em New Astron. Rev.} {\bf 2001}, {\em 45},~409--423, doi:10.1016/S1387-6473(00)00163-9.



\bibitem[Wright and Li(2018)]{Wright:2017rsu}
Wright, B.S.; Li, B.
\newblock {Type Ia supernovae, standardizable candles, and gravity}.
\newblock {\em Phys. Rev. D} {\bf 2018}, {\em 97},~083505,  doi:10.1103/PhysRevD.\linebreak
97.083505.

\bibitem[Mazzali \em{et~al.}(2001)Mazzali, Nomoto, Cappellaro, Nakamura, Umeda,
  and Iwamoto]{Mazzali:2000gk}
Mazzali, P.A.; Nomoto, K.; Cappellaro, E.; Nakamura, T.; Umeda, H.; Iwamoto, K.
\newblock {Can differences in the nickel abundance in chandrasekhar mass models
  explain the relation between brightness and decline rate of normal type ia
  supernovae?}
\newblock {\em Astrophys. J.} {\bf 2001}, {\em 547},~988, doi:10.1086/318428.

\bibitem[Arnett(1982)]{Arnett:1982ioj}
Arnett, W.D.
\newblock {Type I supernovae. I---Analytic solutions for the early part of the
  light curve}.
\newblock {\em Astrophys. J.} {\bf 1982}, {\em 253},~785--797, doi:10.1086/159681.

\bibitem[Gaztanaga \em{et~al.}(2002)Gaztanaga, Garcia-Berro, Isern, Bravo, and
  Dominguez]{Gaztanaga:2001fh}
Gaztanaga, E.; Garcia-Berro, E.; Isern, J.; Bravo, E.; Dominguez, I.
\newblock {Bounds on the possible evolution of the gravitational constant from
  cosmological type Ia supernovae}.
\newblock {\em Phys. Rev. D} {\bf 2002}, {\em 65},~023506,  doi:10.1103/PhysRevD.65.023506.

\bibitem[Amendola \em{et~al.}(1999)Amendola, Corasaniti, and
  Occhionero]{Amendola:1999vu}
Amendola, L.; Corasaniti, P.S.; Occhionero, F.
\newblock {Time variability of the gravitational constant and type Ia
  supernovae.} \emph{arXiv}  {\bf 1999},  	arXiv:astro-ph/9907222.

\bibitem[Marra and Perivolaropoulos(2021)]{Marra:2021fvf}
Marra, V.; Perivolaropoulos, L.
\newblock {A rapid transition of $G_{\rm eff}$ at $z_t \simeq 0.01$ as a
  solution of the Hubble and growth tensions.} \emph{arXiv} {\bf 2021}, arXiv:2102.06012 66. 

\bibitem[Kazantzidis and Perivolaropoulos(2019)]{Kazantzidis:2019dvk}
Kazantzidis, L.; Perivolaropoulos, L.
\newblock {Is gravity getting weaker at low z? Observational evidence and
  theoretical implications.} \emph{arXiv} {\bf 2019}, arXiv:1907.03176.

\bibitem[Nugent \em{et~al.}(2002)Nugent, Kim, and Perlmutter]{Nugent:2002si}
Nugent, P.; Kim, A.; Perlmutter, S.
\newblock {K-Corrections and Extinction Corrections for Type Ia Supernovae}.
\newblock {\em Publ. Astron. Soc. Pac.} {\bf 2002}, {\em 114},~803--819,  doi:10.1086/341707.

\bibitem[Wright()]{bao_cart}
Wright, E.L.
\newblock Listening for the Size of the Universe.
\newblock  Available online: \url{http://www.astro.ucla.edu/~wright/BAO-cosmology.html} 
\newblock  (accessed on 7  July 2020).

\bibitem[Eisenstein and Hu(1998)]{Eisenstein_1998}
Eisenstein, D.J.; Hu, W.
\newblock {Baryonic features in the matter transfer function}.
\newblock {\em Astrophys. J.} {\bf 1998}, {\em 496},~605,  doi:10.1086/305424.

\bibitem[Castander()]{bao_figs_01}
Castander, F.J.
\newblock Baryon Acoustic Oscillations.
\newblock
 Available online:   \url{https://www.ias.u-psud.fr/Dark_energy/presentations/castanderBAO_081124.pdf} 
\newblock  (accessed on 7  July 2020).

\bibitem[Elisa~Ferreira()]{bao_anim}
Ferreira, E. Bryce, E.M.C.
\newblock Baryon Acoustic Oscillations.
\newblock
   Available online: \url{http://galaxies-cosmology-2015.wikidot.com/baryon-acoustic-oscillations} 
\newblock  (accessed on 7  July 2020).

\bibitem[Eisenstein \em{et~al.}(2005)Eisenstein, Zehavi, Hogg, Scoccimarro,
  Blanton, Nichol, Scranton, Seo, Tegmark, Zheng, and et~al.]{Eisenstein_2005}
Eisenstein, D.J.; Zehavi, I.; Hogg, D.W.; Scoccimarro, R.; Blanton, M.R.;
  Nichol, R.C.; Scranton, R.; Seo, H.; Tegmark, M.; Zheng, Z.; et~al. 
\newblock Detection of the Baryon Acoustic Peak in the Large‐Scale
  Correlation Function of SDSS Luminous Red Galaxies.
\newblock {\em  Astrophys. J.} {\bf 2005}, {\em 633},~560–574, doi:10.1086/466512.

\bibitem[Montesano()]{bao_figs_02}
Montesano, F.
\newblock The Full Shape of the Large-Scale Galaxy Power Spectrum: Modelling
  and Cosmological Implications.
\newblock
 Available online:   \url{https://www.imprs-astro.mpg.de/sites/default/files/2011_Montesano_Francesco.pdf} 
\newblock  (accessed on 7  July 2020).

\bibitem[Aubourg \em{et~al.}(2015)Aubourg et~al.]{Aubourg:2014yra}
Aubourg, E.; Bailey, S.; Bautista, J.; Beutler, F.; Bhardwaj, V.; Bizyaev, D.; Blanton, M.; Blomqvist, M.; Bolton, A.S.; Bovy, J.; et al.
\newblock {Cosmological implications of baryon acoustic oscillation
  measurements}.
\newblock {\em Phys. Rev. D} {\bf 2015}, {\em D92},~123516, doi:10.1103/PhysRevD.92.123516.

\bibitem[cmb()]{cmb_001}
Planck and the Cosmic Microwave Background.
\newblock
   Available online: \url{https://www.esa.int/Science_Exploration/Space_Science/Planck/Planck_and_the_cosmic_microwave_background} 
\newblock  (accessed on 7  July 2020).

\bibitem[Efstathiou and Bond(1999)]{Efstathiou:1998xx}
Efstathiou, G.; Bond, J.R.
\newblock {Cosmic confusion: Degeneracies among cosmological parameters derived
  from measurements of microwave background anisotropies}.
\newblock {\em Mon. Not. R. Astron. Soc.} {\bf 1999}, {\em 304},~75--97,  doi:10.1046/j.1365-8711.1999.02274.x.

\bibitem[Elgaroy and Multamaki(2007)]{Elgaroy:2007bv}
Elgaroy, O.; Multamaki, T.
\newblock {On using the CMB shift parameter in tests of models of dark energy}.
\newblock {\em Astron. Astrophys.} {\bf 2007}, {\em 471},~65, doi:10.1051/0004-6361:20077292.

\bibitem[Zhai and Wang(2019)]{Zhai_2019}
Zhai, Z.; Wang, Y.
\newblock Robust and model-independent cosmological constraints from distance
  measurements.
\newblock {\em J. Cosmol. Astropart. Phys.} {\bf 2019}, {\em
  2019},~5, doi:10.1088/1475-7516/2019/07/005.

\bibitem[cmb()]{cmb_002}
Planck Image Gallery.
\newblock Available online:  \url{https://www.cosmos.esa.int/web/planck/picture-gallery} 
\newblock  (accessed on 7  July 2020).

\bibitem[Hu and Sugiyama(1996)]{Hu:1995en}
Hu, W.; Sugiyama, N.
\newblock {Small scale cosmological perturbations: An Analytic approach}.
\newblock {\em Astrophys. J.} {\bf 1996}, {\em 471},~542--570,  doi:10.1086/177989.

\bibitem[Heavens(2009)]{Heavens:2009nx}
Heavens, A.
\newblock {Statistical techniques in cosmology}. \emph{arXiv} {\bf 2009}, arXiv:0906.0664

\bibitem[200(2003)]{2003theory}
{\em Theory And Problems Of Probability And Statistics (Schaum S Outline
  Series)}; McGraw-Hill Education (India) Pvt Limited: New York, NY, USA,  2003.

\bibitem[Verde(2010)]{Verde:2009tu}
Verde, L.
\newblock {Statistical methods in cosmology}.
\newblock {\em Lect. Notes Phys.} {\bf 2010}, {\em 800},~147--177, doi:10.1007/978-3-642-10598-2\_4.

\bibitem[Myung(2003)]{MYUNG200390}
Myung, I.J.
\newblock Tutorial on maximum likelihood estimation.
\newblock {\em J. Math. Psychol.} {\bf 2003}, {\em
  47},~90--100, doi:10.1016/S0022-2496(02)00028-7.

\bibitem[{Licia Verde}()]{Verde_stat}
{Verde, L}.
\newblock Statistical Techniques for Data Analysis in Cosmology.
\newblock  Available online: \url{https://www.ice.csic.es/personal/verde/verdeLecturesstat.pdf}  (accessed on 11 May  2021).

\bibitem[Press \em{et~al.}(2007)Press, Teukolsky, Vetterling, and
  Flannery]{10.5555/1403886}
Press, W.H.; Teukolsky, S.A.; Vetterling, W.T.; Flannery, B.P.
\newblock {\em Numerical Recipes 3rd Edition: The Art of Scientific Computing},
  3rd ed.; Cambridge University Press: Cambridge, MA, USA,  2007.

\bibitem[Kazantzidis and Perivolaropoulos(2020)]{Kazantzidis:2020tko}
Kazantzidis, L.; Perivolaropoulos, L.
\newblock {Hints of a Local Matter Underdensity or Modified Gravity in the Low
  $z$ Pantheon data}.
\newblock {\em Phys. Rev. D} {\bf 2020}, {\em 102},~023520, doi:10.1103/PhysRevD.102.023520.

\bibitem[Betoule \em{et~al.}(2014)Betoule et~al.]{SDSS:2014iwm}
Betoule, M.; Kessler, R.; Guy, J.; Mosher, J.; Hardin, D.; Biswas, R.; Astier, P.; El-Hage, P.; Konig, M.; Kuhlmann, S.; et al.
\newblock {Improved cosmological constraints from a joint analysis of the
  SDSS-II and SNLS supernova samples}.
\newblock {\em Astron. Astrophys.} {\bf 2014}, {\em 568},~A22,  doi:10.1051/0004-6361/201423413.

\bibitem[Conley \em{et~al.}(2010)Conley, Guy, Sullivan, Regnault, Astier,
  Balland, Basa, Carlberg, Fouchez, Hardin, and et~al.]{Conley_2010}
Conley, A.; Guy, J.; Sullivan, M.; Regnault, N.; Astier, P.; Balland, C.; Basa,
  S.; Carlberg, R.G.; Fouchez, D.; Hardin, D.; et~al.
\newblock Supernova constraints and systematic uncertainties from the first
  three years of the supernova legacy survey.
\newblock {\em  Astrophys. J. Suppl. Ser.} {\bf 2010}, {\em
  192},~1, doi:10.1088/0067-0049/192/1/1.

\bibitem[Riess \em{et~al.}(1998)Riess et~al.]{Riess:1998cb}
Riess, A.G.; Filippenko, A.V.; Challis, P.; Clocchiatti, A.; Diercks, A.; Garnavich, P.M.; Gilliland, R.L.; Hogan, C.J.; Jha, S.; \mbox{Kirshner, R.P.; et al. }
\newblock {Observational evidence from supernovae for an accelerating universe 
  and a cosmological constant}.
\newblock {\em Astron. J.} {\bf 1998}, {\em 116},~1009--1038, doi:10.1086/300499.

\bibitem[Perlmutter \em{et~al.}(1999)Perlmutter et~al.]{Perlmutter:1998np}
Perlmutter, S.; Aldering, G.; Goldhaber, G.; Knop, R.A.; Nugent, P.; Castro, P.G.; Deustua, S.; Fabbro, S.; Goobar, A.; Groom, D.E.
\newblock {Measurements of $\Omega$ and $\Lambda$ from 42 high redshift
  supernovae}.
\newblock {\em Astrophys. J.} {\bf 1999}, {\em 517},~565--586, doi:10.1086/307221.

\bibitem[Carroll(2001)]{Carroll:2000fy}
Carroll, S.M.
\newblock {The Cosmological constant}.
\newblock {\em Living Rev. Relativ.} {\bf 2001}, {\em 4},~1, doi:10.12942/lrr-2001-1.

\bibitem[Weinberg(1989)]{Weinberg:1988cp}
Weinberg, S.
\newblock {The Cosmological Constant Problem}.
\newblock {\em Rev. Mod. Phys.} {\bf 1989}, {\em 61},~1--23, doi:10.1103/RevModPhys.61.1.

\bibitem[Sahni(2002)]{Sahni:2002kh}
Sahni, V.
\newblock {The Cosmological constant problem and quintessence}.
\newblock {\em Class. Quant. Grav.} {\bf 2002}, {\em 19},~3435--3448, doi:10.1088/0264-9381/19/13/304.

\bibitem[P.J(1997)]{Princeton_01}
{Fitch,Val L., Marlow,Daniel R.  and Dementi,Margit A. E.  and Dyson,Freeman J.}
\newblock {\em Critical Problems in Physics}; Princeton University Press: Princeton, NJ, USA,
  1997.

\bibitem[Velten \em{et~al.}(2014)Velten, vom Marttens, and
  Zimdahl]{Velten:2014nra}
Velten, H.E.S.; vom Marttens, R.F.; Zimdahl, W.
\newblock {Aspects of the cosmological \textquotedblleft{}coincidence
  problem\textquotedblright{}}.
\newblock {\em Eur. Phys. J. C} {\bf 2014}, {\em 74},~3160, doi:10.1140/epjc/s10052-014-3160-4.

\bibitem[{Carter}(1974)]{1974IAUS...63..291C}
{Carter}, B.
\newblock {Large number coincidences and the anthropic principle in cosmology.}
\newblock  In \emph{Confrontation of Cosmological Theories with Observational Data}; 
  {Longair}, M.S., Ed.; Springer Science \& Business Media: Berlin/Heidelberg, Germany, 1974; Volume~63, \mbox{pp. 291--298.}

\bibitem[Perivolaropoulos and Skara(2021)]{Perivolaropoulos:2021jda}
Perivolaropoulos, L.; Skara, F.
\newblock {Challenges for $\Lambda$CDM: An update}. \emph{arXiv}  {\bf 2021}, arXiv:2105.05208.

\bibitem[Riess \em{et~al.}(2019)Riess, Casertano, Yuan, Macri, and
  Scolnic]{Riess:2019cxk}
Riess, A.G.; Casertano, S.; Yuan, W.; Macri, L.M.; Scolnic, D.
\newblock {Large Magellanic Cloud Cepheid Standards Provide a 1\% Foundation
  for the Determination of the Hubble Constant and Stronger Evidence for
  Physics beyond $\Lambda$CDM}.
\newblock {\em Astrophys. J.} {\bf 2019}, {\em 876},~85, doi:10.3847/1538-4357/ab1422.

\bibitem[Wong \em{et~al.}(2020)Wong et~al.]{Wong:2019kwg}
Wong, K.C.; Suyu, S.H.; Chen, G.C.-F.; E Rusu, C.; Millon, M.; Sluse, D.; Bonvin, V.; Fassnacht, C.D.; Taubenberger, S.; Auger, M.W.; et al. 
\newblock {H0LiCOW---XIII. A 2.4 per cent measurement of H0 from
  lensed quasars: 5.3\ensuremath{\sigma} tension between early- and
  late-Universe probes}.
\newblock {\em Mon. Not. R. Astron. Soc.} {\bf 2020}, {\em 498},~1420--1439, doi:10.1093/mnras/stz3094.
\bibitem[DiValentino(2021)]{DiValentino:2021izs}
Di Valentino, Eleonora and Mena, Olga and Pan, Supriya and Visinelli, Luca and Yang, Weiqiang and Melchiorri, Alessandro and Mota, David F. and Riess, Adam G. and Silk, Joseph.
\newblock{In the realm of the Hubble tension\textemdash{}a review of solutions.}
\newblock {\em Class. Quant. Grav..} {\bf 2021}, {\em 38},~153001, doi:10.1088/1361-6382/ac086d.

\bibitem[DiValentino(2020)]{DiValentino:2020zio}
Di Valentino, Eleonora and others.
\newblock{Snowmass2021 - Letter of interest cosmology intertwined II: The hubble constant tension.}
\newblock {\em Astropart. Phys..} {\bf 2021}, {\em 131},~102605, doi:10.1016/j.astropartphys.2021.102605.


\bibitem[Köhlinger \em{et~al.}(2017)Köhlinger et~al.]{Kohlinger:2017sxk}
Köhlinger, F.; Viola, M.; Joachimi, B.; Hoekstra, H.; Van Uitert, E.; Hildebrandt, H.; Choi, A.; Erben, T.; Heymans, C.; \mbox{Joudaki, S.;} \mbox{et al.} 
\newblock {KiDS-450: The tomographic weak lensing power spectrum and
  constraints on cosmological parameters}.
\newblock {\em Mon. Not. R. Astron. Soc.} {\bf 2017}, {\em 471},~4412--4435,  doi:10.1093/mnras/stx1820.

\bibitem[Joudaki \em{et~al.}(2018)Joudaki et~al.]{Joudaki:2017zdt}
Joudaki, S.; Blake, C.; Johnson, A.; Amon, A.; Asgari, M.; Choi, A.; Erben, T.; Glazebrook, K.; Harnois-Déraps, J.; Heymans, C.; \mbox{et al.}
\newblock {KiDS-450 + 2dFLenS: Cosmological parameter constraints from weak
  gravitational lensing tomography and overlapping redshift-space galaxy
  clustering}.
\newblock {\em Mon. Not. R. Astron. Soc.} {\bf 2018}, {\em 474},~4894--4924, doi:10.1093/mnras/stx2820.

\bibitem[Abbott \em{et~al.}(2018)Abbott et~al.]{Abbott:2017wau}
 Abbott, T.M.C.; Abdalla, F.B.; Alarcon, A.; Aleksić, J.; Allam, S.; Allen, S.; Amara, A.; Annis, J.; Asorey, J.; Avila, S.; et al.
\newblock {Dark Energy Survey year 1 results: Cosmological constraints from
  galaxy clustering and weak lensing}.
\newblock {\em Phys. Rev. D} {\bf 2018}, {\em 98},~043526, doi:10.1103/PhysRevD.98.043526.

\bibitem[Macaulay \em{et~al.}(2013)Macaulay, Wehus, and
  Eriksen]{Macaulay:2013swa}
Macaulay, E.; Wehus, I.K.; Eriksen, H.K.
\newblock {Lower Growth Rate from Recent Redshift Space Distortion Measurements
  than Expected from Planck}.
\newblock {\em Phys. Rev. Lett.} {\bf 2013}, {\em 111},~161301, doi:10.1103/PhysRevLett.111.161301.

\bibitem[Tsujikawa(2015)]{Tsujikawa:2015mga}
Tsujikawa, S.
\newblock {Possibility of realizing weak gravity in redshift space distortion
  measurements}.
\newblock {\em Phys. Rev. D} {\bf 2015}, {\em 92},~044029, doi:10.1103/PhysRevD.92.044029.

\bibitem[Johnson \em{et~al.}(2016)Johnson, Blake, Dossett, Koda, Parkinson, and
  Joudaki]{Johnson:2015aaa}
Johnson, A.; Blake, C.; Dossett, J.; Koda, J.; Parkinson, D.; Joudaki, S.
\newblock {Searching for Modified Gravity: Scale and Redshift Dependent
  Constraints from Galaxy Peculiar Velocities}.
\newblock {\em Mon. Not. R. Astron. Soc.} {\bf 2016}, {\em 458},~2725--2744, doi:10.1093/mnras/stw447.

\bibitem[Kazantzidis and Perivolaropoulos(2018)]{Kazantzidis:2018rnb}
Kazantzidis, L.; Perivolaropoulos, L.
\newblock {Evolution of the $f\sigma_8$ tension with the Planck15/$\Lambda$CDM
  determination and implications for modified gravity theories}.
\newblock {\em Phys. Rev. D} {\bf 2018}, {\em 97},~103503,  doi:10.1103/PhysRevD.97.103503.

\bibitem[Nesseris \em{et~al.}(2017)Nesseris, Pantazis, and
  Perivolaropoulos]{Nesseris:2017vor}
Nesseris, S.; Pantazis, G.; Perivolaropoulos, L.
\newblock {Tension and constraints on modified gravity parametrizations of
  $G_{\textrm{eff}}(z)$ from growth rate and Planck data}.
\newblock {\em Phys. Rev. D} {\bf 2017}, {\em 96},~023542,  doi:10.1103/PhysRevD.96.023542.

\bibitem{Nunes:2021ipq}
R.~C.~Nunes and S.~Vagnozzi,
Mon. Not. Roy. Astron. Soc. \textbf{505}, no.4, 5427-5437 (2021)
doi:10.1093/mnras/stab1613
[arXiv:2106.01208 [astro-ph.CO]].

\bibitem[Skara and Perivolaropoulos(2020)]{Skara:2019usd}
Skara, F.; Perivolaropoulos, L.
\newblock {Tension of the $E_G$ statistic and redshift space distortion data
  with the Planck---$\Lambda CDM$ model and implications for weakening
  gravity}.
\newblock {\em Phys. Rev. D} {\bf 2020}, {\em 101},~063521,  doi:10.1103/PhysRevD.101.063521.

\bibitem[L'Huillier \em{et~al.}(2017)L'Huillier, Shafieloo, and
  Kim]{LHuillier:2017ani}
L'Huillier, B.; Shafieloo, A.; Kim, H.
\newblock {Model-independent cosmological constraints from growth and
  expansion} \emph{Mon. Not. R. Astron. Soc. } {\bf 2017}, \emph{46}, 3263--3268.

\bibitem[Quelle and Maroto(2020)]{Quelle:2019vam}
Quelle, A.; Maroto, A.L.
\newblock {On the tension between growth rate and CMB data}.
\newblock {\em Eur. Phys. J. C} {\bf 2020}, {\em 80},~369, doi:10.1140/epjc/s10052-020-7941-7.

\bibitem{deAraujo:2021cnd}
J.~C.~N.~de Araujo, A.~De Felice, S.~Kumar and R.~C.~Nunes,
[arXiv:2106.09595 [astro-ph.CO]].

\bibitem{DiValentino:2020vvd}
E.~Di Valentino, L.~A.~Anchordoqui, \"O.~Akarsu, Y.~Ali-Haimoud, L.~Amendola, N.~Arendse, M.~Asgari, M.~Ballardini, S.~Basilakos and E.~Battistelli, \textit{et al.}
Astropart. Phys. \textbf{131}, 102604 (2021)
doi:10.1016/j.astropartphys.2021.102604
[arXiv:2008.11285 [astro-ph.CO]].

\bibitem[Arjona \em{et~al.}(2018)Arjona, Cardona, and Nesseris]{Arjona:2018jhh}
Arjona, R.; Cardona, W.; Nesseris, S.
\newblock {Unraveling the effective fluid approach for $f(R)$ models in the
  sub-horizon approximation}. \emph{Phys. Rev. D}  {\bf 2018}, \emph{99}, 43516.

\bibitem[Zhao \em{et~al.}(2019)Zhao, Zhou, and Chang]{Zhao:2019azy}
Zhao, D.; Zhou, Y.; Chang, Z.
\newblock {Anisotropy of the Universe via the Pantheon supernovae sample
  revisited}.
\newblock {\em Mon. Not. R. Astron. Soc.} {\bf 2019}, {\em 486},~5679--5689, doi:10.1093/mnras/stz1259.

\bibitem[Caldwell \em{et~al.}(1998)Caldwell, Dave, and
  Steinhardt]{Caldwell:1997ii}
Caldwell, R.R.; Dave, R.; Steinhardt, P.J.
\newblock {Cosmological imprint of an energy component with general equation of
  state}.
\newblock {\em Phys. Rev. Lett.} {\bf 1998}, {\em 80},~1582--1585,  doi:10.1103/PhysRevLett.80.1582.

\bibitem[Linder(2008)]{Linder:2007wa}
Linder, E.V.
\newblock {The Dynamics of Quintessence, The Quintessence of Dynamics}.
\newblock {\em Gen. Relativ. Gravit.} {\bf 2008}, {\em 40},~329--356, doi:10.1007/s10714-007-0550-z.

\bibitem[{C´eline Boehm,Julien Lesgourgues}()]{Boehm_dark}
{Boehm, C.; Lesgourgues, J}.
\newblock Dark Matter And Dark Energy.
\newblock  Available online: \url{https://lesgourg.github.io/courses/DMDE_EPFL.pdf} (accessed on 8 June 2020).

\bibitem[Mortonson \em{et~al.}(2009)Mortonson, Hu, and
  Huterer]{Mortonson:2009qq}
Mortonson, M.J.; Hu, W.; Huterer, D.
\newblock {Hiding dark energy transitions at low redshift}.
\newblock {\em Phys. Rev. D} {\bf 2009}, {\em 80},~067301,  doi:10.1103/PhysRevD.80.067301.

\bibitem[Huterer and Turner(1999)]{Huterer:1998qv}
Huterer, D.; Turner, M.S.
\newblock {Prospects for probing the dark energy via supernova distance
  measurements}.
\newblock {\em Phys. Rev. D} {\bf 1999}, {\em 60},~081301, doi:10.1103/PhysRevD.60.081301.

\bibitem[Copeland \em{et~al.}(2006)Copeland, Sami, and
  Tsujikawa]{Copeland:2006wr}
Copeland, E.J.; Sami, M.; Tsujikawa, S.
\newblock {Dynamics of dark energy}.
\newblock {\em Int. J. Mod. Phys. D} {\bf 2006}, {\em 15},~1753--1936, doi:10.1142/S021827180600\linebreak
942X.

\bibitem[Nakamura and Chiba(1999)]{Nakamura:1998mt}
Nakamura, T.; Chiba, T.
\newblock {Determining the equation of state of the expanding universe: Inverse
  problem in cosmology}.
\newblock {\em Mon. Not. R. Astron. Soc.} {\bf 1999}, {\em 306},~696--700,  doi:10.1046/j.1365-8711.1999.02551.x.

\bibitem[Rajvanshi and Bagla(2019)]{Rajvanshi:2019wmw}
Rajvanshi, M.P.; Bagla, J.S.
\newblock {Reconstruction of Dynamical Dark Energy Potentials: Quintessence,
  Tachyon and interacting models}.
\newblock {\em J. Astrophys. Astron.} {\bf 2019}, {\em 40},~44, doi:10.1007/s12036-019-9613-2.

\bibitem[Guo \em{et~al.}(2005)Guo, Ohta, and Zhang]{Guo:2005ata}
Guo, Z.K.; Ohta, N.; Zhang, Y.Z.
\newblock {Parametrization of quintessence and its potential}.
\newblock {\em Phys. Rev. D} {\bf 2005}, {\em 72},~023504, doi:10.1103/PhysRevD.72.023504.

\bibitem[Pantazis \em{et~al.}(2016)Pantazis, Nesseris, and
  Perivolaropoulos]{Pantazis:2016nky}
Pantazis, G.; Nesseris, S.; Perivolaropoulos, L.
\newblock {Comparison of thawing and freezing dark energy parametrizations}.
\newblock {\em Phys. Rev. D} {\bf 2016}, {\em 93},~103503,  doi:10.1103/PhysRevD.93.103503.

\bibitem[Scherrer(2015)]{Scherrer:2015tra}
Scherrer, R.J.
\newblock {Mapping the Chevallier-Polarski-Linder parametrization onto Physical
  Dark Energy Models}.
\newblock {\em Phys. Rev. D} {\bf 2015}, {\em 92},~043001,  doi:10.1103/PhysRevD.92.043001.

\bibitem[Bonilla(2020)]{Bonilla:2020wbn}
Bonilla, Alexander and Kumar, Suresh and Nunes, Rafael C..
\newblock{Measurements of $H_0$ and reconstruction of the dark energy properties from a model-independent joint analysis.}
\newblock {\em Eur. Phys. J. C.} {\bf 2021}, {\em 81},~127, doi:10.1140/epjc/s10052-021-08925-z.


\bibitem[Dabrowski(2015)]{Dabrowski:2014wia}
Dabrowski, M.P.
\newblock {Puzzles of dark energy in the Universe\textemdash{}phantom}.
\newblock {\em Eur. J. Phys.} {\bf 2015}, {\em 36},~065017,  doi:10.1088/0143-0807/36/6/065017.

\bibitem[Sbis\`a(2015)]{Sbisa:2014pzo}
Sbis\`a, F.
\newblock {Classical and quantum ghosts}.
\newblock {\em Eur. J. Phys.} {\bf 2015}, {\em 36},~015009, doi:10.1088/0143-0807/36/1/015009.

\bibitem[Wolf and Lagos(2019)]{Wolf:2019hzy}
Wolf, W.J.; Lagos, M.
\newblock {Cosmological Instabilities and the Role of Matter Interactions in
  Dynamical Dark Energy Models}.
\newblock {\em Phys. Rev. D} {\bf 2019}, {\em 100},~084035, doi:10.1103/PhysRevD.100.084035.

\bibitem[Banerjee \em{et~al.}(2021)Banerjee, Cai, Heisenberg, Colg\'ain,
  Sheikh-Jabbari, and Yang]{Banerjee:2020xcn}
Banerjee, A.; Cai, H.; Heisenberg, L.; Colg\'ain, E.O.; Sheikh-Jabbari, M.M.;
  Yang, T.
\newblock {Hubble sinks in the low-redshift swampland}.
\newblock {\em Phys. Rev. D} {\bf 2021}, {\em 103},~L081305, doi:10.1103/PhysRevD.103.L081305.

\bibitem{Yang:2021flj}
W.~Yang, E.~Di Valentino, S.~Pan, Y.~Wu and J.~Lu,
Mon. Not. Roy. Astron. Soc. \textbf{501}, no.4, 5845-5858 (2021)
doi:10.1093/mnras/staa3914
[arXiv:2101.02168 [astro-ph.CO]].

\bibitem[Efstathiou(2021)]{Efstathiou:2021ocp}
Efstathiou, G.
\newblock {To H0 or not to H0?} \emph{arXiv} {\bf 2021},  	arXiv:2103.08723

\bibitem[Camarena and Marra(2021)]{Camarena:2021jlr}
Camarena, D.; Marra, V.
\newblock {On the use of the local prior on the absolute magnitude of Type Ia
  supernovae in cosmological inference}.
\newblock {\em Mon. Not. R. Astron. Soc.} {\bf 2021}, {\em 504},~5164--5171,  doi:10.1093/mnras/stab1200.

\bibitem[Alestas \em{et~al.}(2020)Alestas, Kazantzidis, and
  Perivolaropoulos]{Alestas:2020zol}
Alestas, G.; Kazantzidis, L.; Perivolaropoulos, L.
\newblock {A $w$ phantom transition at $z_t<0.1$ as a resolution of the Hubble
  tension}  \emph{Phys. Rev.  D} {\bf 2020}, 103, 083517. 

\bibitem[Cooray and Huterer(1999)]{Cooray:1999da}
Cooray, A.R.; Huterer, D.
\newblock {Gravitational lensing as a probe of quintessence}.
\newblock {\em Astrophys. J. Lett.} {\bf 1999}, {\em 513},~L95--L98, doi:10.1086/311927.

\bibitem[Barboza and Alcaniz(2008)]{Barboza:2008rh}
\textls[-25]{Barboza, Jr., E.M.; Alcaniz, J.S.
\newblock {A parametric model for dark energy}.
\newblock {\em Phys. Lett. B} {\bf 2008}, {\em 666},~415--419, doi:10.1016/j.physletb.2008.08.012.}

\bibitem[Lazkoz \em{et~al.}(2011)Lazkoz, Salzano, and Sendra]{Lazkoz:2010gz}
Lazkoz, R.; Salzano, V.; Sendra, I.
\newblock {Oscillations in the dark energy EoS: New MCMC lessons}.
\newblock {\em Phys. Lett. B} {\bf 2011}, {\em 694},~198--208,  doi:10.1016/j.physletb.2010.10.002.

\bibitem[Jassal \em{et~al.}(2005)Jassal, Bagla, and Padmanabhan]{Jassal:2004ej}
Jassal, H.K.; Bagla, J.S.; Padmanabhan, T.
\newblock {WMAP constraints on low redshift evolution of dark energy}.
\newblock {\em Mon. Not. R. Astron. Soc.} {\bf 2005}, {\em 356},~L11--L16,  doi:10.1111/j.1745-3933.2005.08577.x.

\bibitem[Perivolaropoulos(2005)]{Perivolaropoulos:2005yv}
Perivolaropoulos, L.
\newblock {Crossing the phantom divide barrier with scalar tensor theories}.
\newblock {\em JCAP} {\bf 2005}, {\em 10},~001, doi:10.1088/1475-7516/2005/10/001.

\bibitem[{D.M. Scolnic}({\natexlab{a}})]{Pantheon_data_01}
{D.M. Scolnic}.
\newblock Pantheon Data (github).
\newblock  Available online: \url{https://github.com/dscolnic/Pantheon} (accessed on 21 April 2020).

\bibitem[{D.M. Scolnic}({\natexlab{b}})]{Pantheon_data_02}
Scolnic, D.~M. ; Jones, D.~O. ; Rest, A. ; Pan, Y.~C. ; Chornock, R. ; Foley, R.~J. ; Huber, M.~E. ; Kessler, R. ; Narayan, G. ; Riess, A.~G.;~{et~al.} 
Supernova Catalog.
\newblock
  Available online:  \url{https://archive.stsci.edu/prepds/ps1cosmo/scolnic_datatable.html} (accessed on 21 April 2020).

\bibitem[Escamilla-Rivera(2016)]{Escamilla-Rivera:2016qwv}
Escamilla-Rivera, C.
\newblock {Status on bidimensional dark energy parameterizations using SNe Ia
  JLA and BAO datasets}.
\newblock {\em Galaxies} {\bf 2016}, {\em 4},~8, doi:10.3390/galaxies4030008.

\bibitem[Ross \em{et~al.}(2015)Ross, Samushia, Howlett, Percival, Burden, and
  Manera]{Ross:2014qpa}
Ross, A.J.; Samushia, L.; Howlett, C.; Percival, W.J.; Burden, A.; Manera, M.
\newblock {The clustering of the SDSS DR7 main Galaxy sample – I. A 4 per
  cent distance measure at $z = 0.15$}.
\newblock {\em Mon. Not. R. Astron. Soc.} {\bf 2015}, {\em 449},~835--847, doi:10.1093/mnras/stv154.

\bibitem[Anderson \em{et~al.}(2014)Anderson et~al.]{Anderson:2013zyy}
Anderson, L.; Aubourg, Éric; Bailey, S.; Beutler, F.; Bhardwaj, V.; Blanton, M.; Bolton, A.S.; Brinkmann, J.; Brownstein, J.R.; \mbox{Burden, A.;} et al. 
\newblock {The clustering of galaxies in the SDSS-III Baryon Oscillation
  Spectroscopic Survey: Baryon acoustic oscillations in the Data Releases 10
  and 11 Galaxy samples}.
\newblock {\em Mon. Not. R. Astron. Soc.} {\bf 2014}, {\em 441},~24--62 , doi:10.1093/mnras/stu523.

\bibitem[de~Sainte~Agathe \em{et~al.}(2019)de~Sainte~Agathe
  et~al.]{Agathe:2019vsu}
Agathe, V.D.S.; Balland, C.; Bourboux, H.D.M.D.; Busca, N.G.; Blomqvist, M.; Guy, J.; Rich, J.; Font-Ribera, A.; Pieri, M.M.; Bautista, J.E.; et al. 
\newblock {Baryon acoustic oscillations at z = 2.34 from the correlations of
  Ly$\alpha$ absorption in eBOSS DR14}.
\newblock {\em Astron. Astrophys.} {\bf 2019}, {\em 629},~A85, doi:10.1051/0004-6361/201935638.

\bibitem[{Daniel Baumann}()]{Baumann_cosmology}
{Baumann, D}.
\newblock Cosmology Lecture Notes.
\newblock  Available online: \url{http://cosmology.amsterdam/education/cosmology/}   (accessed on 9 April 2020).

\bibitem[Hobson \em{et~al.}(2006)Hobson, Efstathiou, and
  Lasenby]{Hobson:2006se}
Hobson, M.P.; Efstathiou, G.P.; Lasenby, A.N.
\newblock {\em {General relativity: An Introduction for Physicists}};  Cambridge University Press: Cambridge, UK, 2006.

\bibitem[Tong()]{hubble_law}
Tong, D.
\newblock Cosmology.
\newblock  Available online: \url{http://www.damtp.cam.ac.uk/user/tong/cosmo/one.pdf} 
\newblock  (accessed on 7  July 2020).

\bibitem[hub({\natexlab{a}})]{hubble_rsrup}
Hubble Image of Variable Star RS Puppis.
\newblock Available online:  \url{https://esahubble.org/images/heic1323a/} 
\newblock  (accessed on 7  July 2020).

\bibitem[hub({\natexlab{b}})]{hubble_ceplig}
 Available online: \url{https://en.wikipedia.org/wiki/File:Delta_Cephei_lightcurve.jpg} 
\newblock  (accessed on 7  July 2020).

\bibitem[Blair()]{kepler_sup}
Blair, B.
\newblock Bill Blair's Kepler's Supernova Remnant Page.
\newblock
   Available online: \url{https://web.archive.org/web/20160316154134/http://fuse.pha.jhu.edu/~wpb/Kepler/kepler.html} 
\newblock  (accessed on 7  July 2020).

\bibitem[Perivolaropoulos(2006)]{Perivolaropoulos:2006ce}
Perivolaropoulos, L.
\newblock {Accelerating universe: Observational status and theoretical
  implications}.
\newblock {\em AIP Conf. Proc.} {\bf 2006}, {\em 848},~698--712, doi:10.1063/1.2348048.

\bibitem[gam()]{gamma_func}
Gamma Function.
\newblock  Available online: \url{https://en.wikipedia.org/wiki/Gamma_function} 
\newblock  (accessed on 7  July 2020).

\bibitem[err()]{error_func}
Error Function.
\newblock Available online:  \url{https://en.wikipedia.org/wiki/Error_function} 
\newblock  (accessed on 7  July 2020).

\bibitem[Chandler()]{sigma_MIT}
Chandler, D.L.
\newblock Explained: Sigma.
\newblock  Available online: \url{https://news.mit.edu/2012/explained-sigma-0209} 
\newblock  (accessed on 7  July 2020).

\bibitem[chi()]{chisq_dist}
Chi-Square Distribution.
\newblock Available online: \url{https://en.wikipedia.org/wiki/Chi-square_distribution} 
\newblock  (accessed on 7  July 2020).

\end{thebibliography}
\end{document}